\documentclass[pdflatex,sn-mathphys-num]{sn-jnl}% Math and Physical Sciences Numbered Reference Style 
\usepackage{graphicx}%
\usepackage{multirow}%
\usepackage{amsmath,amssymb,amsfonts}%
\usepackage{amsthm}%
\usepackage{mathrsfs}%
\usepackage[title]{appendix}%
\usepackage{xcolor}%
\usepackage{textcomp}%
\usepackage{manyfoot}%
\usepackage{booktabs}%
\usepackage{algorithm}%
\usepackage{algorithmicx}%
\usepackage{algpseudocode}%
\usepackage{listings}%
\usepackage{makecell}
\usepackage{tikz}%
\usepackage{bookmark}
\usepackage{siunitx}
\usepackage[version=4]{mhchem}
\usepackage[mathlines]{lineno}
\usetikzlibrary{arrows.meta,positioning,calc,fit,backgrounds,shapes.geometric}%

\theoremstyle{thmstyleone}%
\theoremstyle{thmstyletwo}%

\theoremstyle{thmstylethree}%

\begin{document}

\title[Article Title]{Symmetry- and Property-Aware Crystal Generation with Reinforcement Learning for Inverse Materials Design}

\author*[1,2]{\fnm{Ting-Wei} \sur{Hsu}}\email{hsu.ting@northeastern.edu}

\author[1,2]{\fnm{Arun} \sur{Bansil}} 
\author*[1,2]{\fnm{Qimin} \sur{Yan}}\email{q.yan@northeastern.edu}

\affil[1]{\orgdiv{Department of Physics}, \orgname{Northeastern University}, \city{Boston}, \postcode{02155}, \state{Massachusetts}, \country{USA}}

\affil[2]{\orgdiv{Quantum Materials and Sensing Institute}, \orgname{Northeastern University}, \city{Burlington}, \postcode{01803}, \state{Massachusetts}, \country{USA}}

\abstract{
The inverse design of crystalline materials ultimately seeks structures with desired physical properties. However, for many functional responses, a favorable numerical value is meaningful only when supported by the symmetry of the underlying crystal. Without the appropriate crystallographic constraints, an apparent response may be ill defined, accidental, or not symmetry protected. Here we introduce SPARC, a symmetry- and property-aware reinforcement learning framework that optimizes physical objectives while preserving the structural conditions required for their realization. We demonstrate SPARC on two complementary tasks. The first targets strong uniaxial dielectric anisotropy, a tensorial response that is well defined only within appropriate crystal classes. The second maximizes the spectroscopic limited maximum efficiency, a scalar device-level objective without a prescribed symmetry class, allowing the framework to identify favorable crystallographic motifs. These results show that symmetry is not merely an additional design constraint, but a physical foundation for generating candidates with meaningful, robust, and realizable functional properties.
}


\maketitle

\section{Introduction}\label{sec1}
The physical properties of functional materials are deeply constrained by the symmetry of their underlying crystal structures. Many technologically relevant responses are not determined by composition alone, but emerge from the coupled effects of atomic species, atomic positions, and lattice geometry. In the inverse design of functional materials, the central goal is therefore to identify crystal structures that satisfy a desired physical objective while remaining chemically and crystallographically plausible.

Symmetry plays a particularly important role in this process. From tensorial responses such as dielectric anisotropy, to nonlinear optical effects such as second-harmonic generation, and symmetry-protected electronic phenomena all depend on the allowed crystal symmetry. In some cases, the target response is permitted only within specific symmetry classes. For example, electric-dipole second-order nonlinear optical responses require the absence of inversion symmetry, while chiral charge-density-wave states can depend on layer-resolved charge-order patterns and their interlayer phase or stacking relationships. Thus, symmetry is not merely a descriptor of a generated structure, but a fundamental constraint that determines which physical properties can occur.

Recent advances in generative modeling have opened new opportunities for crystal structure design. Diffusion-based crystal generators can sample periodic structures across large chemical and configurational spaces~\cite{xieCrystalDiffusionVariational2022,jiao2023diffcsp,zeniGenerativeModelInorganic2025a}, while steering strategies such as classifier-free guidance~\cite{hoClassifierFreeDiffusionGuidance2022,guoInitioStructureSolutions2025a} and reinforcement learning (RL)~\cite{chenAcceleratingInverseMaterials2025,Park2026Guiding,govindarajan_crystalgym_2025,cao_reinforcement_2026} allow the generation process to be biased toward user-specified objectives. Most demonstrations focus on scalar targets such as formation energy, band gap, or magnetic density, whereas many functional materials are defined by tensorial, spectral, or device-level responses. A key distinction among these approaches lies in the structural representation on which the steering is applied. Recent work, for example, applies group-relative policy optimization to a latent diffusion model and combines novelty, stability, and diversity rewards to direct sampling toward underexplored regions of chemical space~\cite{Park2026Guiding}. This strategy controls where the generator samples in property and chemical space, but the latent and free-coordinate representations being optimized do not preserve crystallographic symmetry by construction. As a result, a generated crystal may appear successful in property space while losing the crystallographic symmetry required for the intended functionality, often collapsing toward the ${P}1$ space group when symmetry is not explicitly enforced during generation. This limitation reflects a deeper mismatch between local, continuous structure updates and global crystallographic constraints. Closing this gap requires generative frameworks that optimize functional objectives while preserving the symmetry conditions under which those objectives are physically admissible.

Here we introduce \textbf{SPARC} (\textbf{S}ymmetry- and \textbf{P}roperty-\textbf{A}ware \textbf{R}einforcement Learning for \textbf{C}rystal Generation), a framework for symmetry-aware, property-driven inverse design of crystalline materials. SPARC couples complex property objectives with crystallographic constraints during generation, enabling the search for structures that are simultaneously optimized for target functionality and consistent with physically meaningful symmetry. By integrating symmetry into the design loop, SPARC provides a route to generate candidates for functional materials whose desired properties cannot be specified by scalar objectives alone.

We evaluate SPARC on two representative inverse-design tasks that test complementary aspects of symmetry-aware property optimization. First, we consider a symmetry-sensitive optical target, the strong uniaxial dielectric anisotropy. This response is defined by a large contrast between the in-plane and out-of-plane components of the static-limit dielectric tensor (hereafter the dielectric tensor), as commonly found in layered van der Waals crystals~\cite{liuVanWaalsHeterostructures2016}. Because the special form of the dielectric tensor is constrained by crystal symmetry, this task tests whether SPARC can access and preserve the symmetry classes required for directional optical response. Second, we consider a device-level optical-efficiency target by maximizing the spectroscopic limited maximum efficiency (SLME) $\eta$. Unlike the anisotropy objective, SLME is a scalar figure of merit and does not prescribe a known symmetry class. It depends on the full frequency-resolved dielectric response and implicitly rewards an optimal band gap near the Shockley--Queisser maximum~\cite{yuIdentificationPotentialPhotovoltaic2012}. This task therefore provides a complementary test of whether symmetry-aware generation can discover preferential crystallographic motifs or space-group distributions for a complex device-level objective. Together, these case studies show that incorporating symmetry into complex-property-guided crystal generation is essential for producing realistic candidates and suggest a broader route toward symmetry-aware generative modeling for inverse materials design.

\section{Results}
\subsection{SPARC pipeline}
\begin{figure}[H]
    \centering
    \includegraphics[width=\textwidth]{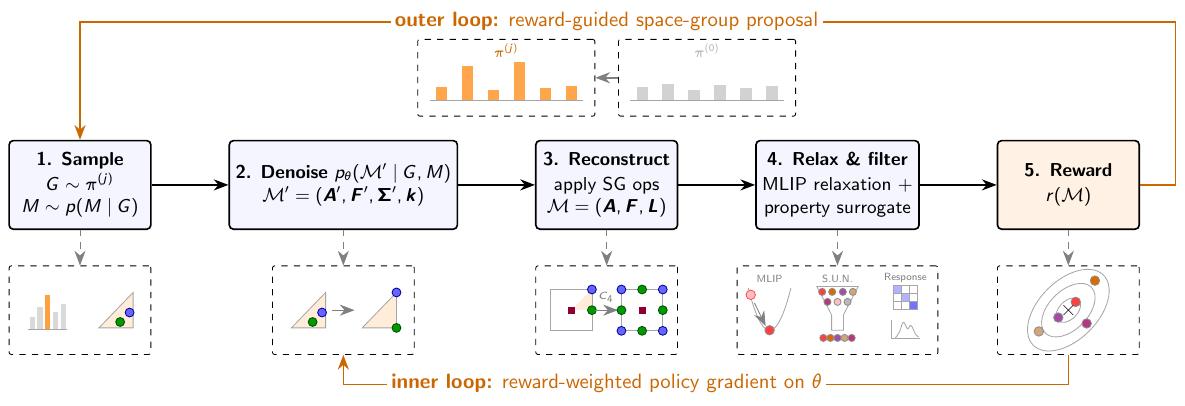}
    \caption[SPARC pipeline schematic]{
    \textbf{Schematic of the SPARC pipeline.}
    Each reinforcement-learning iteration consists of five stages.
    (1) The policy samples a space group $G$ from the proposal distribution
    $\pi^{(j)}$ and the number of representative asymmetric-unit sites $M$
    from the space-group-dependent prior $p(M\mid G)$.
    (2) The diffusion policy $p_\theta$ denoises the asymmetric-unit representation
    $\mathcal{M}'=(\boldsymbol{A}',\boldsymbol{F}',\boldsymbol{\Sigma}',\boldsymbol{k})$.
    (3) The full crystal
    $\mathcal{M}=(\boldsymbol{A},\boldsymbol{F},\boldsymbol{L})$
    is reconstructed by applying the symmetry operations of $G$.
    (4) The reconstructed structures are relaxed, filtered, and evaluated for the
    target property using a surrogate model.
    (5) Each resulting structure is assigned a terminal reward $r(\mathcal{M})$.
    The terminal reward drives two decoupled optimization processes.
    In the inner loop, the diffusion policy parameters $\theta$ are fine-tuned
    over multiple steps using a reward-weighted policy gradient together with a
    KL-divergence penalty that anchors the policy to the pretrained generator
    $p_{\mathrm{pre}}$.
    In the outer loop, the space-group proposal $\pi^{(j)}(G)$ is updated
    using rewards aggregated according to the realized space group, thereby
    increasing the sampling probability of productive symmetries relative to the
    dataset prior $\pi^{(0)}$.
    Throughout the figure panels, SG abbreviates space group.
    }
    \label{fig:sparc-schematic}
\end{figure}

SPARC couples a symmetry-constrained SymmCD \cite{levysymmcd} generator with reward optimization for property-directed crystal generation. 
Figure~\ref{fig:sparc-schematic} traces this pipeline as a left-to-right sequence of five stages, from sampling a crystal structure through denoising, reconstruction, relaxation and scoring, to the terminal reward, and the reward then closes the two feedback loops to update the diffusion model and the space-group proposal. At generation time the policy first samples a space group $G$ from the proposal $\pi^{(j)}$, then draws the number of representative asymmetric-unit sites $M$ from a per-space-group prior $p(M\mid G)$ estimated from Materials Project (MP-20 dataset) \cite{jainCommentaryMaterialsProject2013, xieCrystalDiffusionVariational2022}. 
Conditioned on $(G,M)$, it denoises the asymmetric unit $\mathcal{M}'=(\boldsymbol{A}',\boldsymbol{F}',\boldsymbol{\Sigma}',\boldsymbol{k})$ over $T$ reverse steps, jointly generating the atom types $\boldsymbol{A}'$, fractional coordinates $\boldsymbol{F}'$, site-symmetry labels $\boldsymbol{\Sigma}'$, and symmetric-matrix lattice basis of its representative sites $\boldsymbol{k}$. Applying the group operations of $G$ to these representatives then regenerates the full crystal. The reward is therefore optimized over symmetry-constrained crystals rather than arbitrary coordinate representations. Each decoded structure is first relaxed using MatterSim, a machine-learning interatomic potential (MLIP)~\cite{yang2024mattersim}. The relaxed structure is then evaluated according to the stable, unique, and novel (S.U.N.) criteria following MatterGen~\cite{zeniGenerativeModelInorganic2025a}, and its target property is evaluated by the surrogate model to assign the terminal reward.

Each generated batch drives two decoupled updates, both derived in full details in Methods. In simple words, the inner loop fine-tunes the diffusion policy $p_\theta(\boldsymbol{A}',\boldsymbol{F}',\boldsymbol{\Sigma}',\boldsymbol{k})$ to adjust the asymmetric-unit generation. Treating the reverse trajectory as a Markov decision process, it applies a reward-weighted policy gradient anchored to the pretrained generator $p_{\mathrm{pre}}$ by an explicit Kullback--Leibler (KL) term to discourage reward hacking and preserves the broad chemical and structural coverage learned during pretraining \cite{chenAcceleratingInverseMaterials2025}. The experience replay and a diversity filter to improve sample efficiency and downweight near-duplicate candidates. The outer loop updates the space-group proposal itself. Because SymmCD treats $G$ as a fixed conditioning input that the policy gradient cannot reshape, we accumulate reward feedback across the realized space groups and tilt the proposal $\pi^{(j)}$ toward desirable space groups. At every round, we mix the proposal back with the initial base distribution $\pi^{(0)}$ so that multiple symmetry classes remain active and the optimization does not collapse onto a few low-symmetry structures.

\subsection{Inverse design of uniaxial dielectric responses}
We first examine whether SPARC can generate crystals whose symmetry is dictated by a target physical response. We consider a uniaxial dielectric tensor,
\begin{equation}
\boldsymbol{\varepsilon}
=
\begin{pmatrix}
\varepsilon_{\parallel} & 0 & 0 \\
0 & \varepsilon_{\parallel} & 0 \\
0 & 0 & \varepsilon_{\perp}
\end{pmatrix},
\end{equation}
where $\varepsilon_{\parallel}$ is the response within the basal plane and $\varepsilon_{\perp}$ is the response along the unique axis. The target requires both in-plane isotropy, $\varepsilon_{xx}=\varepsilon_{yy}$, and a finite out-of-plane contrast, $\varepsilon_{zz}\neq\varepsilon_{xx}$. It therefore defines one distinguished dielectric axis without selecting a preferred direction within the perpendicular plane.

Crucially, this target is not merely a numerical relation among tensor components. Under Neumann's principle, the dielectric tensor must reflect the point-group symmetry of the crystal. Three-, four-, and six-fold rotational symmetries enforce $\varepsilon_{xx}=\varepsilon_{yy}$, making the uniaxial form symmetry-compatible with trigonal, tetragonal, and hexagonal crystals. Cubic symmetry instead removes the desired anisotropy by requiring $\varepsilon_{xx}=\varepsilon_{yy}=\varepsilon_{zz}$. Lower-symmetry crystals do not generally enforce the in-plane degeneracy, so any observed $\varepsilon_{xx}\approx\varepsilon_{yy}$ is accidental rather than symmetry-protected. We therefore seek structures in which the target dielectric response follows naturally from the underlying crystal symmetry.

A large contrast $|\varepsilon_{\perp}-\varepsilon_{\parallel}|$ is useful when a device requires different in-plane and out-of-plane responses without sensitivity to the in-plane orientation of the crystal. This tensor form is relevant to layered dielectric environments, gate dielectrics, and van der Waals heterostructures ~\cite{deanBoronNitrideSubstrates2010, laturiaDielectricPropertiesHexagonal2018, illarionovInsulators2DNanoelectronics2020}. At finite frequencies, uniaxial anisotropy also supports birefringence and polarization control. It can produce hyperbolic light propagation when the in-plane and out-of-plane permittivities have opposite signs ~\cite{niuGiantOpticalAnisotropy2018, ermolaevGiantOpticalAnisotropy2021}. Although the present task uses the static-limit dielectric tensor, it provides a direct test of whether SPARC can generate crystal symmetries that support a prescribed tensorial response.

Black phosphorus (BP) provides a useful example of the technological value of dielectric anisotropy. Its strong in-plane optical anisotropy has enabled polarization-sensitive photodetectors and birefringent optical devices~\cite{yuanPolarizationsensitiveBroadbandPhotodetector2015a, qiaoHighmobilityTransportAnisotropy2014, xiaRediscoveringBlackPhosphorus2014}. BP is orthorhombic and therefore exhibits a biaxial rather than uniaxial dielectric response. Its $\varepsilon_{xx}$ and $\varepsilon_{yy}$ components are not symmetry-equivalent. Exfoliated BP can also undergo oxidation and progressive degradation under ambient conditions~\cite{islandEnvironmentalInstabilityFewlayer2015}. These characteristics motivate the search for complementary anisotropic materials with uniaxial symmetry, thermodynamic accessibility, and synthetic feasibility. SPARC provides a route to explore this broader multi-constraint materials space.

To express the target response as a scalar objective, we evaluate the dielectric tensor of each generated structure using a modified Tensorial Spectra Equivariant Neural Network (TSENN) surrogate~\cite{hsuAccuratePredictionTensorial2026}. The tensor is evaluated in the standardized conventional cell. For trigonal, tetragonal, and hexagonal structures, this convention aligns the unique axis with $z$. We then assign a uniaxial anisotropy reward $r_{\mathrm{uni}}\in[0,1]$. The reward combines an in-plane equality term with an out-of-plane contrast gate $g_z$. The first favors $\varepsilon_{xx}=\varepsilon_{yy}$. The second requires a finite difference between $\varepsilon_{zz}$ and the mean in-plane response, thereby excluding the fully isotropic cubic limit. Because the target materials are insulating or semiconducting, the reward additionally includes a band-gap term that suppresses metallic structures. The full reward definition is provided in Methods, and the surrogate benchmark is reported in the Supplementary Information (SI).

Figure~\ref{fig:sparc-dielectric-story-composite}(a) shows the SPARC loop specialized to this task. The initial space-group proposal $\pi^{(0)}$ is obtained by temperature reweighting the MP-20 distribution \cite{kirkpatrickOptimizationSimulatedAnnealing1983}, yielding a nearly uniform distribution that reduces the strong space-group frequency bias of the training data. Starting from this broad prior, SPARC iteratively updates the proposal using reward feedback, allowing the optimization to identify which symmetry classes best support the target dielectric response. During RL fine-tuning, the mean uniaxial-anisotropy reward increases from $0.17$ to $0.56$ over 120 RL steps (Figure~\ref{fig:sparc-dielectric-story-composite}(b)). At each step, 30 structures that pass the S.U.N. filter are scored, giving a total of 3,600 evaluated candidates. Over the same optimization trajectory, the realized space-group distribution becomes increasingly enriched in trigonal, tetragonal, and hexagonal groups that symmetry-enforce the target tensor form (Figure~\ref{fig:sparc-dielectric-story-composite}(c)). Importantly, these symmetry classes are not prescribed in advance, but emerge through the reward-guided redistribution of the space-group proposal. Among the 3,600 scored structures, $P\bar{3}1m$, $P4_2/mnm$, $R\bar{3}$, $P\bar{3}$ and $P{3}_1$ together account for approximately $44\%$ of the generated population. The generated \ce{Cd2SnBr6} structure in Figure~\ref{fig:sparc-dielectric-story-composite}(a) provides a representative symmetry-enforced result. It belongs to the tetragonal space group $P\bar{3}1m$ and has the TSENN-predicted dielectric tensor
\begin{equation}
\boldsymbol{\varepsilon}
=
\begin{pmatrix}
6.2 & 0 & 0 \\
0 & 6.2 & 0 \\
0 & 0 & 4.7
\end{pmatrix},
\end{equation}
Its equality $\varepsilon_{xx}=\varepsilon_{yy}$ and vanishing off-diagonal components follow directly from its tetragonal symmetry. The finite difference between the in-plane and out-of-plane components provides the desired uniaxial contrast. The structure retains this response after MLIP relaxation and is selected for subsequent density function theory (DFT) validation. 

\begin{figure}[H]
    \centering
    \includegraphics[width=0.9\linewidth]{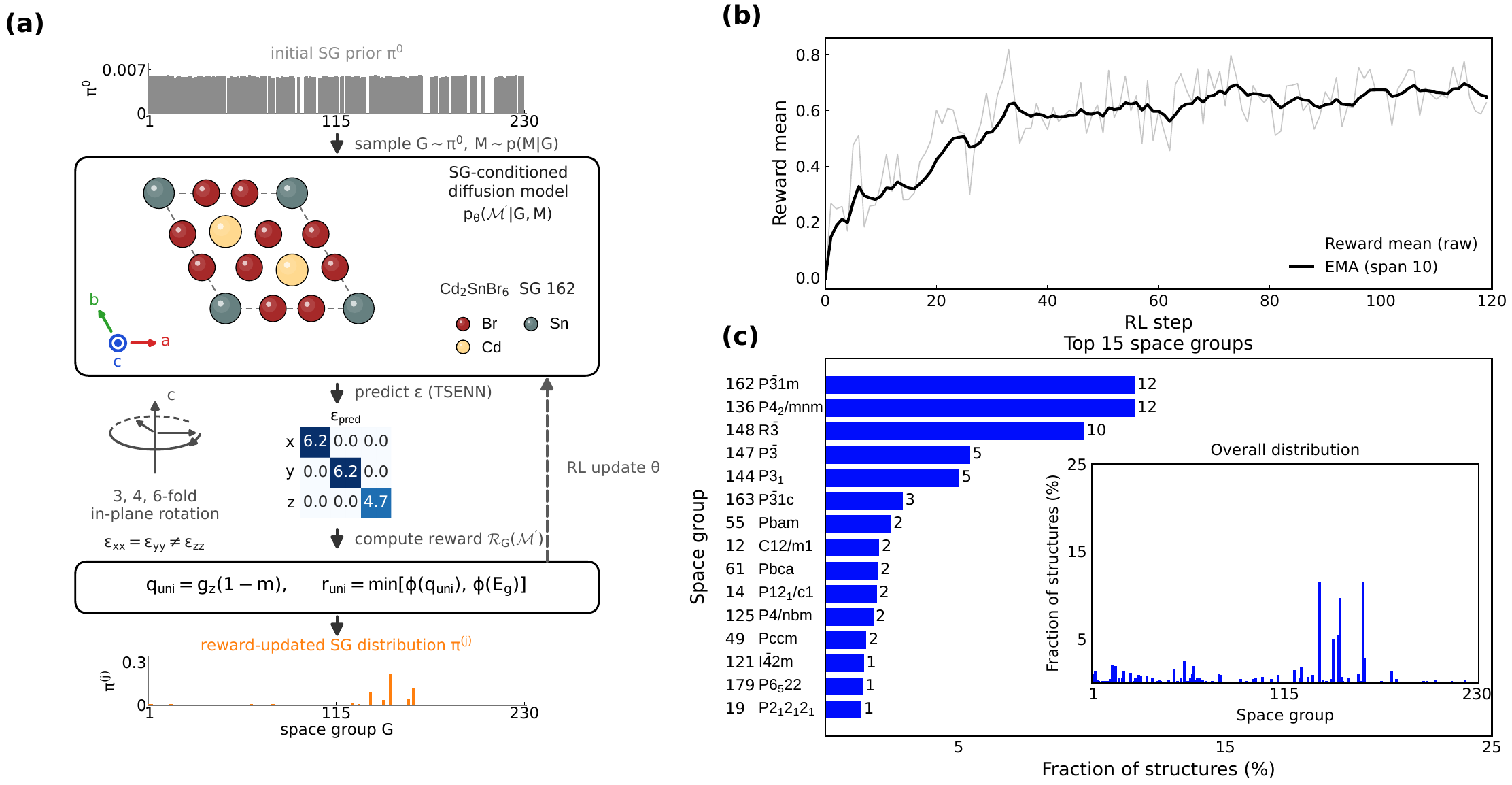}
    \caption[Dielectric]{
    \textbf{Steering generation toward uniaxial dielectric anisotropy.}
    \textbf{(a)}~The SPARC loop specialized to the dielectric task. A space group $G$ is drawn from the current proposal $\pi^{(j)}$, and the space-group-conditioned diffusion model $p_{\theta}(\mathcal{M}'|G,M)$ generates a crystal. The TSENN surrogate predicts the static dielectric tensor $\boldsymbol{\varepsilon}$ and band gap $E_g$. The uniaxial reward is defined as $r_{\mathrm{uni}}=\min[\Phi(q_{\mathrm{uni}}),\Phi(E_g)]$, where $q_{\mathrm{uni}}=g_z(1-m)$, $m=|\varepsilon_{xx}-\varepsilon_{yy}|/(|\varepsilon_{xx}|+|\varepsilon_{yy}|)$, and $g_z$ excludes cubic tensors. The band-gap term favors $E_g=0.3$--$0.8$~eV. The resulting reward updates both the policy $p_{\theta}$ and the space-group proposal $\pi^{(j)}$. Generated structures are filtered to stable, unique, and novel (S.U.N.) candidates before scoring.
    The worked example is a generated \ce{Cd2SnBr6} with $P\bar{3}1m$ symmetry whose
    predicted tensor is uniaxial, $\boldsymbol{\varepsilon}=\mathrm{diag}(6.2,6.2,4.7)$ with
    $\varepsilon_{xx}=\varepsilon_{yy}\neq\varepsilon_{zz}$ and vanishing off-diagonal entries.
    \textbf{(b)}~Mean reward versus RL step during fine-tuning (raw per-step mean in grey, exponential
    moving average with span 10 in black), and the uniaxial-anisotropy score $r_{\mathrm{uni}}$ climbs from
    {$0.17$ to $0.56$.}
    \textbf{(c)}~The fifteen most frequent realized space groups among the generated structures (inset, full distribution over all 230 groups). Sampling concentrates on the hexagonal, trigonal, and tetragonal groups that admit a uniaxial dielectric tensor, led by $P\bar{3}1m$ and $P4_2/mnm$, followed by $R\bar{3}$, $P\bar{3}$, and $P{3}_1$. Together, these five groups account for approximately 44\% of the generated structures.
    }
    \label{fig:sparc-dielectric-story-composite}
\end{figure}
Lower-symmetry groups also remain present in Figure~\ref{fig:sparc-dielectric-story-composite}(c), including the orthorhombic $Pbca$, and $Pbam$ group. These structures may approximately satisfy the component-level reward, but the in-plane equality is not symmetry-enforced. This motivates the backbone ablation below.

\begin{figure}[H]
\centering
\includegraphics[width=0.9\linewidth]{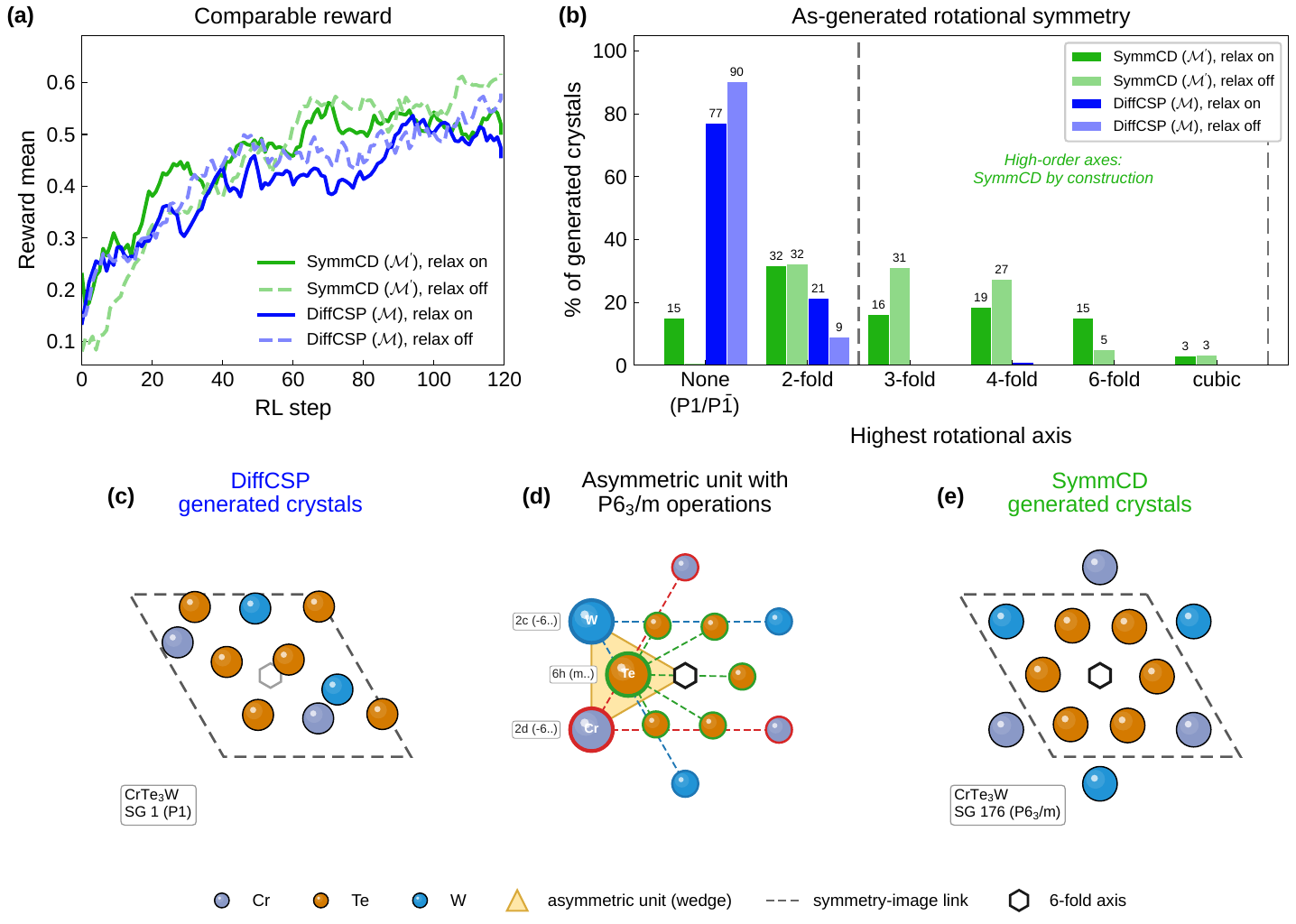}
\caption[Symmetry-aware backbone and relaxation ablation]{
\textbf{Ablation of symmetry enforcement and pre-reward relaxation.}
SymmCD and DiffCSP are fine-tuned using the same uniaxial-anisotropy reward. For each backbone, MLIP relaxation is either enabled or disabled before reward evaluation.
\textbf{(a)}~Mean reward as a function of RL step. All runs improve during fine-tuning, and the reward trajectories overlap substantially. Similar reward values therefore do not indicate whether the targeted tensor relation is symmetry-enforced.
\textbf{(b)}~Distribution of the highest rotational axis among generated crystals for the four ablation settings. Without relaxation, DiffCSP generates almost exclusively $P1$ or $P\bar{1}$ structures. Relaxation recovers mainly twofold symmetry, with only rare threefold and fourfold structures and no sixfold structures. SymmCD retains substantial populations with threefold, fourfold, and sixfold axes because symmetry is imposed during generation.
\textbf{(c)}~Representative DiffCSP-generated \ce{CrTe3W} structure in space group 1 ($P1$).
\textbf{(d)}~SymmCD asymmetric unit for the same composition under the $P6_3/m$ space-group operations. Dashed lines connect representative atoms to their symmetry images.
\textbf{(e)}~Full SymmCD crystal reconstructed from the asymmetric unit. The resulting structure belongs to space group 176 ($P6_3/m$) and contains a sixfold rotational axis.
}
\label{fig:symmcd-diffcsp-ablation}
\end{figure}

To test the necessity of a symmetry-aware diffusion backbone, we performed a controlled ablation by replacing SymmCD with DiffCSP~\cite{jiao2023diffcsp}, which diffuses atoms freely in continuous coordinate space without enforcing crystallographic symmetry. Both backbones were fine-tuned under the same uniaxial-anisotropy objective. For each backbone, we performed separate runs with MLIP relaxation either enabled or disabled before reward evaluation.
Figure~\ref{fig:symmcd-diffcsp-ablation}(a) shows that the four runs reach comparable mean rewards during fine-tuning. Their symmetry distributions, however, differ sharply (Figure~\ref{fig:symmcd-diffcsp-ablation}(b)). Without relaxation, approximately 90\% of DiffCSP structures have no nontrivial rotational axis and belong to $P1$ or $P\bar{1}$. MLIP relaxation recovers mainly twofold symmetry, with only a small fraction acquiring threefold or fourfold axes and almost no sixfold symmetry. SymmCD shows the opposite behavior. Because each structure is generated under explicit symmetry constraints, substantial populations with threefold, fourfold, and sixfold axes remain present both with and without relaxation.

The quantitative contrast is even more pronounced. With relaxation enabled, $50.1\%$ of SymmCD structures contain a threefold, fourfold, or sixfold axis, compared with only $1.4\%$ of DiffCSP structures. This corresponds to one uniaxially symmetric candidate every $2.0$ generations for SymmCD, versus every $70.6$ generations for DiffCSP, giving a $35\times$ higher symmetry yield. Sixfold symmetry appears in $546$ SymmCD structures but only once among $3,600$ DiffCSP samples. Because the two backbones required comparable computational wall-clock time for the same number of fine-tuning loops, this yield advantage reduced the computational cost per uniaxially symmetric candidate by approximately $34\times$ in our runs ($43$~s versus $1465$~s per candidate).

The structural examples in Figure~\ref{fig:symmcd-diffcsp-ablation}(c)--(e) illustrate the origin of this contrast. DiffCSP generates \ce{CrTe3W} in space group 1 ($P1$), with no imposed rotational symmetry. Because the generated atoms lie far from a high-symmetry arrangement, local relaxation cannot reliably recover the missing rotational symmetry afterward. For the same composition, SymmCD generates an asymmetric unit under $P6_3/m$ operations. Applying these operations reconstructs the full crystal in space group 176 ($P6_3/m$) and produces the required sixfold axis. These results show that reward optimization can satisfy the targeted dielectric relation, but a symmetry-aware backbone is needed when that relation must also be enforced by the crystal symmetry. We therefore adopt SymmCD as a central component of SPARC.

\begin{figure}[H]
    \centering
    \includegraphics[width=\linewidth]{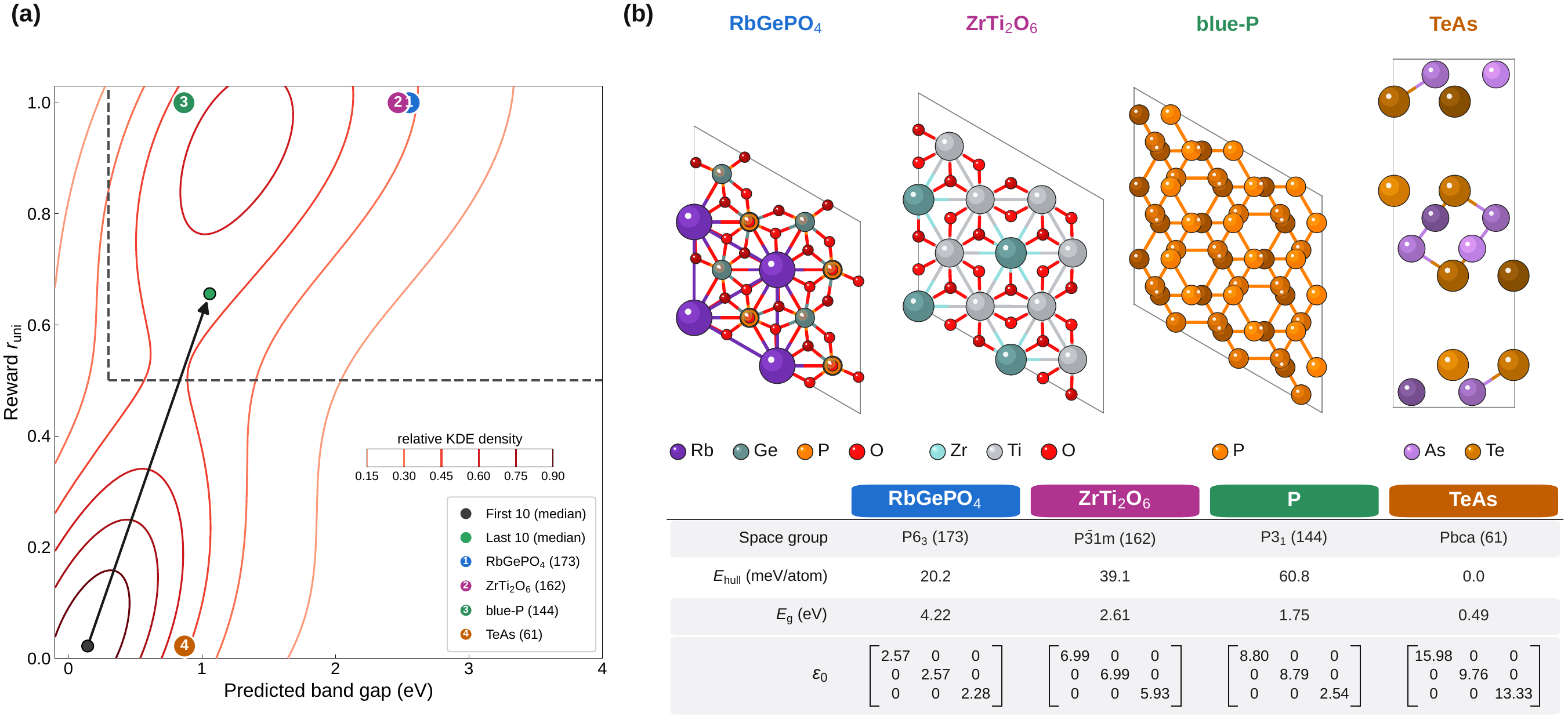}
    \caption[RL steering and DFT verification of dielectric-anisotropy candidates]{
    \textbf{RL-guided generation and DFT verification of dielectric-anisotropy candidates.}
    \textbf{(a)}~Distribution of generated structures in the plane of surrogate-predicted band gap $E_g$ and uniaxial dielectric reward $r_{\mathrm{uni}}$. Kernel-density contours compare structures generated during the first 10 RL rounds with those generated during the last 10 rounds. The black and green circles indicate the corresponding population medians, and the arrow highlights the shift toward the finite-gap, high-reward region during fine-tuning. The dashed lines indicate the band-gap and reward thresholds defining the target region. The four numbered markers denote the candidates selected for first-principles verification: \ce{RbGePO4}, \ce{ZrTi2O6}, blue phosphorus, and \ce{TeAs}.
    \textbf{(b)}~DFT-relaxed structures and calculated properties of the four selected candidates. The table reports the realized space group, $E_{\mathrm{hull}}$, PBE band gap $E_g$, and dielectric tensor $\boldsymbol{\varepsilon}_{0}$. The quantities in panel (a) are surrogate predictions before DFT validation, whereas those in panel (b) are obtained after DFT relaxation and may therefore differ. The three high-reward candidates, \ce{RbGePO4} ($P6_3$), \ce{ZrTi2O6} ($P\bar{3}1m$), and blue phosphorus ($P3_1$), retain the symmetry-constrained form $\varepsilon_{xx}=\varepsilon_{yy}\neq\varepsilon_{zz}$. Blue phosphorus shows the largest dielectric anisotropy, with $\varepsilon_{\parallel}/\varepsilon_{\perp}\approx3.46$. In contrast, orthorhombic \ce{TeAs} ($Pbca$) has three distinct diagonal components and serves as a low-reward biaxial comparison.
    }
    \label{fig:dft-gallery}
\end{figure}

To validate the reward-guided predictions with first-principles calculations, we performed DFT relaxations and computed dielectric tensors using density functional perturbation theory (DFPT)~\cite{petousisBenchmarkingDensityFunctional2016a} for a broader set of generated candidates. Figure~\ref{fig:dft-gallery} highlights four representative examples and summarizes the evolution of the generated population during fine-tuning.

As fine-tuning progresses, the reward function shifts the generated population toward the finite-gap, high-reward region. Figure~\ref{fig:dft-gallery}(a) compares the distributions obtained during the first and last 10 RL steps. The displacement between their population medians illustrates the corresponding shift in the sampled property space. The four highlighted candidates probe different regions of this distribution. \ce{RbGePO4}, \ce{ZrTi2O6}, and blue phosphorus occupy the predicted high-reward region, whereas \ce{TeAs} provides a low-reward biaxial comparison. The positions of the candidates in Figure~\ref{fig:dft-gallery}(a) are determined by the surrogate-predicted band gaps and rewards before first-principles validation, whereas the values reported in panel (b) are obtained after DFT relaxation. Differences between the two panels are therefore expected and directly reveal the remaining surrogate error.

Figure~\ref{fig:dft-gallery}(b) presents the DFT-relaxed structures together with their realized space groups, energies above the convex hull $E_{\mathrm{hull}}$, Perdew--Burke--Ernzerhof (PBE) band gaps $E_g$, and dielectric tensors $\boldsymbol{\varepsilon}_{0}$. The DFT band gaps range from $0.49$ to $4.22$~eV, while $E_{\mathrm{hull}}$ ranges from $0.0$ to $60.8$~meV/atom. The three candidates selected from the high-reward region retain the symmetry-constrained uniaxial form $\varepsilon_{xx}=\varepsilon_{yy}\neq\varepsilon_{zz}$ after DFT relaxation.
Among these candidates, blue phosphorus exhibits the strongest dielectric anisotropy, with $\varepsilon_{xx}=8.80$, $\varepsilon_{yy}=8.79$, and $\varepsilon_{zz}=2.54$. This corresponds to an anisotropy ratio $\varepsilon_{\parallel}/\varepsilon_{\perp}\approx3.46$. \ce{ZrTi2O6} and \ce{RbGePO4} retain weaker but symmetry-consistent uniaxial responses, with ratios of approximately $1.18$ and $1.13$, respectively. By contrast, orthorhombic \ce{TeAs} has three distinct diagonal components, $\varepsilon_{xx}=15.98$, $\varepsilon_{yy}=9.76$, and $\varepsilon_{zz}=13.33$, consistent with its low uniaxial reward.

\subsection{Inverse design of high-efficiency photovoltaic absorbers}

We next consider a second optical inverse-design task. Our objective is the SLME $\eta$ of a single-junction photovoltaic absorber~\cite{choudharyAcceleratedDiscoveryEfficient2019,yuIdentificationPotentialPhotovoltaic2012}. Although SLME is reported as a single percentage, it depends on the material's full optical response. The complex dielectric function $\varepsilon(\omega)$ determines the absorption coefficient $\alpha(\omega)$. This absorption spectrum controls the short-circuit current under the AM1.5G solar spectrum and the radiative dark current associated with black-body emission. Optimizing $\eta$ therefore requires control over the absorption spectrum above the band edge, rather than the response at a single frequency.

This objective complements the uniaxial dielectric task considered above. The uniaxial reward depends only on the dielectric tensor in the $\omega\rightarrow0$ limit. It is also restricted to crystal classes that permit the target tensor form. In contrast, SLME depends on the full frequency-dependent optical response and does not prescribe a particular crystal class. This task therefore tests whether SPARC can discover favorable chemical and structural distributions when the target property does not impose a preferred symmetry.

For each generated crystal, we evaluate $\eta$ under fixed device conditions. We use an absorber thickness of $L_{\mathrm{abs}}=\SI{0.3}{\micro\meter}$, a cell temperature of $T_{\mathrm{cell}}=\SI{300}{\kelvin}$, and a radiative recombination fraction of $f_r=1$. The modified TSENN surrogate predicts the imaginary dielectric response $\varepsilon_2(\omega)$. We recover the real part $\varepsilon_1(\omega)$ using the Kramers--Kronig relation. The resulting complex dielectric function is then converted to the absorption coefficient through
\begin{equation}
    \alpha(\omega)=\frac{2\omega k(\omega)}{c},
\end{equation}
where $k(\omega)$ is the extinction coefficient~\cite{fox2010optical,gajdosLinearOpticalProperties2006}. The architecture and performance of the modified surrogate are benchmarked in the SI.

SLME depends jointly on the band gap $E_g$ and the spectral response above the gap. The band-gap term in the reward is therefore aligned with the photovoltaic objective rather than acting as a separate competing target. It steers the population away from the near-metallic regime, where the predicted efficiency is low and the optical surrogate is less reliable. We center the search near $E_g\approx1.3$~eV, close to the optimal single-junction band gap under standard assumptions. We define a broader target region of $E_g\approx0.8$--$1.9$~eV and $\eta\geq25\%$. Reaching this region requires more than placing the band edge at an appropriate energy. The absorption above the edge must also be strong enough to convert a large fraction of the incident photon flux into photocurrent.

\begin{figure}[H]
    \centering
    \includegraphics[width=\linewidth]{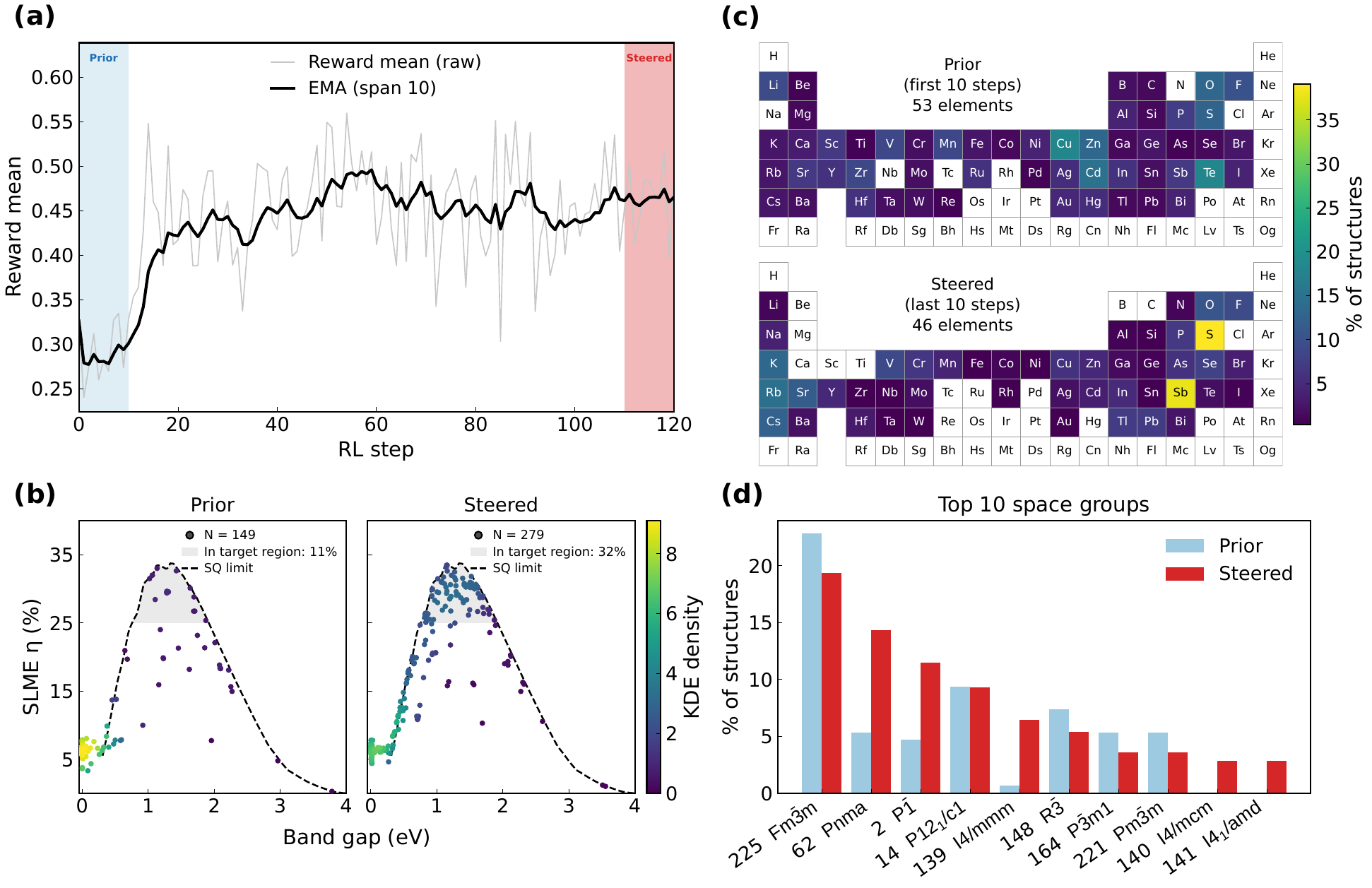}
    \caption[Inverse design of high-SLME photovoltaic absorbers]{
    \textbf{Steering crystal generation toward high-efficiency single-junction absorbers.}
    \textbf{(a)}~Mean reward as a function of RL step during SPARC fine-tuning. The gray line shows the raw mean reward at each step, while the black line shows the exponential moving average with a span of 10 steps. The shaded regions indicate the prior window, defined by the first 10 RL steps, and the steered window, defined by the last 10 RL steps.
    \textbf{(b)}~Spectroscopic limited maximum efficiency (SLME), $\eta$, as a function of band gap $E_g$ for structures generated in the prior window (left, 149 structures) and the steered window (right, 279 structures). The points are colored by their kernel-density estimate, and the dashed curve indicates the Shockley--Queisser limit. The light-gray area marks the target region, defined by $E_g\approx0.8$--$1.9$~eV and $\eta\geq25\%$. The fraction of structures within this region increases from $11\%$ to $32\%$ after steering.
    \textbf{(c)}~Periodic-table heatmaps showing the percentage of generated structures containing each element. The number of represented elements decreases from 53 in the prior population to 46 in the steered population. Steering redistributes the elemental frequencies and enriches sulfur- and antimony-containing structures.
    \textbf{(d)}~Relative frequencies of the ten most commonly realized space groups in the prior and steered populations. The cubic $Fm\bar{3}m$ space group remains the most frequent symmetry after steering. The orthorhombic $Pnma$ and triclinic $P\bar{1}$ space groups become substantially more prevalent.
    }
    \label{fig:sparc-schematic-slme}
\end{figure}

Figure~\ref{fig:sparc-schematic-slme}(a) shows that fine-tuning produces a clear increase in the reward. The mean reward rises rapidly during the early RL steps and remains elevated throughout the optimization process. Its exponential moving average increases from approximately $0.28$ in the prior window to approximately $0.46$ in the steered window. Figure~\ref{fig:sparc-schematic-slme}(b) shows the corresponding redistribution in property space. Before steering, the generated structures are concentrated mainly in the low-gap, low-efficiency region. Fine-tuning moves the population toward intermediate band gaps and higher SLME values. Many of the steered candidates approach the Shockley--Queisser limit within the target band-gap range. As a result, the fraction of structures satisfying both $E_g\approx0.8$--$1.9$~eV and $\eta\geq25\%$ increases from $11\%$ in the prior population (149 structures) to $32\%$ in the steered population (279 structures).

The improved photovoltaic response is accompanied by changes in the sampled chemistry. Figure~\ref{fig:sparc-schematic-slme}(c) shows the number of represented elements decreases only modestly, from 53 in the prior population to 46 after steering. The optimization therefore does not collapse the search onto a single chemical family. Instead, it redistributes the elemental frequencies while preserving substantial compositional diversity. Sulfur becomes particularly prominent after steering, and antimony is also enriched in the high-SLME population. This trend is consistent with the broader relevance of sulfide semiconductors to photovoltaics, such as kesterite \ce{Cu2ZnSnS4} and \ce{Sb2S3}, which have been extensively investigated as thin-film solar absorbers~\cite{yanCu2ZnSnS4SolarCells2018,yangUltrafastSelftrappingPhotoexcited2019}.

The steered population also remains crystallographically diverse. Figure~\ref{fig:sparc-schematic-slme}(d) compares the ten most frequently realized space groups before and after steering. Since SLME does not favor a specific crystal class, fine-tuning shifts the symmetry distribution without collapsing it to a single family. The cubic $Fm\bar{3}m$ group remains the most common after steering. The orthorhombic $Pnma$ and triclinic $P\bar{1}$ groups also become more prevalent. This shows that high SLME can be reached through several distinct crystallographic families. This behavior differs from the dielectric task, where the reward favors a much narrower set of rotational symmetries.

Taken together, these results establish SLME optimization as a multi-constraint inverse-design task. A successful candidate cannot be identified by optimizing the band gap or a single spectral feature alone. It must combine a suitable band gap with strong absorption across the relevant solar spectrum while remaining structurally valid and crystallographically plausible. SPARC addresses these coupled requirements by concentrating the generated population in favorable regions of property, compositional, and crystallographic space, while still preserving multiple structural routes to high efficiency. The increase in the target-region population from $11\%$ to $32\%$ therefore reflects more than an improvement in a single predicted quantity. It demonstrates a coordinated redistribution of the generated population toward candidates that simultaneously satisfy electronic, optical, and structural constraints. Additional candidates sampled during the RL runs, together with their first-principles validation, are reported in the SI. Overall, this task shows that SPARC can navigate a complex photovoltaic design space governed by the full optical response rather than by a single target descriptor.

\section{Discussion}\label{sec3}

The central result of this work is that symmetry and functional properties must be optimized together in crystal inverse design. They are not independent design objectives. Crystal symmetry constrains the form of many physical responses and determines whether a desired behavior is allowed, enforced, or forbidden, while the targeted property in turn determines which symmetry classes are relevant to the design problem. A lower-symmetry structure may still achieve the targeted dielectric response and therefore receive a high reward, but the required tensor relation is not enforced by symmetry and may occur only for a particular atomic configuration. Reward optimization alone therefore cannot guarantee that the desired response is realized in a symmetry-protected form. It must operate within a structural representation that can access the appropriate symmetry while simultaneously improving the target property. The two optical tasks illustrate complementary forms of this coupling. For the uniaxial dielectric objective, the target tensor directly restricts the admissible crystal classes because the required in-plane equivalence is associated with threefold, fourfold, or sixfold rotational symmetry. For SLME, the property depends on the band gap and the full optical spectrum but does not prescribe a unique crystal class, allowing multiple chemical and structural families to achieve favorable photovoltaic performance. Together, these results show that symmetry and property cannot generally be separated into independent stages of crystal design and should instead be treated within the same optimization problem.

The backbone ablation makes the physical role of symmetry explicit. SymmCD and DiffCSP reach similar reward values under the same dielectric objective, yet produce markedly different crystallographic distributions. Both can satisfy the targeted tensor relation numerically, but in a low-symmetry structure the equality $\varepsilon_{xx}\approx\varepsilon_{yy}$ is accidental rather than symmetry-enforced. This relation can be lost under small structural distortions, while local relaxation cannot reliably recover the missing global symmetry. The same reasoning extends to properties with known symmetry requirements. For example, inversion symmetry forbids piezoelectricity and electric-dipole second-harmonic generation, so the space-group proposal can be restricted to classes that permit the desired response. When the relevant symmetry class is not known in advance, SPARC can instead keep all symmetry channels available and use the reward-guided outer loop to identify which crystallographic families best support the target property, as demonstrated in the SLME task.

Blue phosphorus provides a useful benchmark for whether symmetry-aware steering can recover physically meaningful candidates. SPARC generates a layered phosphorus structure with trigonal $P3_1$ symmetry that remains gapped after first-principles relaxation and exhibits strong uniaxial dielectric anisotropy, with a band gap of $1.75$ eV. The structure is absent from the MP-20 training set, although a blue-phosphorus allotrope was proposed independently in prior work~\cite{zhuSemiconductingLayeredBlue2014}. No phosphorus-specific target or structural template was supplied during optimization. Its recovery therefore provides a positive benchmark. It shows that SPARC can reach a known and physically plausible motif beyond the training distribution, even though blue phosphorus is not the most stable phosphorus allotrope.

The two tasks also show that property steering can concentrate upon different dimensions of materials space. The dielectric objective acts primarily in symmetry space because only specific crystal classes enforce the target tensor. SLME leaves several crystallographic routes available and instead produces a stronger redistribution in property and chemical space. This selective concentration is an important output of inverse design. The goal is not to preserve the full diversity of the pretrained model, but to identify a smaller region that is more likely to contain the desired response. Such behavior should be distinguished from uncontrolled mode collapse because diversity remains along directions that are less strongly constrained. In the SLME task, the generated population becomes concentrated in a smaller region of chemical space while retaining a broad distribution of crystal symmetries.

The same principle extends beyond optical properties. Symmetry-aware generation can benefit the design of magnetocrystalline anisotropy, elastic and piezoelectric response, thermal expansion, and other symmetry-constrained material properties~\cite{zhangMagneticAnisotropy2021, yanSpaceGroupSymmetry2024}. More broadly, SPARC is not tied to a specific surrogate and can be coupled to models that predict tensorial properties, transport quantities, or the electronic Hamiltonian directly~\cite{dongAccuratePiezoelectricTensor2025,zengLearningThermoelectricTransport2026,liDeepLearningDensityFunctional2022,gongGeneralFrameworkE3equivariant2023a, zhongUniversalSpinOrbit2026a}. Electronic Hamiltonians are particularly attractive because they provide a common starting point for evaluating band structures, Berry-phase and Berry-curvature quantities, and topological properties through standard electronic-structure post-processing~\cite{gongGeneralFrameworkE3equivariant2023a,zhongUniversalSpinOrbit2026a, wangFirstprinciplesCalculationOptical2019}. The reward may change across applications, but the generated structure must still possess the symmetry needed to support the intended physics.

Realizing this broader potential will also require improving the accuracy of the reward signal. Learned surrogates inherit both their own prediction errors and the approximations of the reference calculations used for training. This is particularly relevant for the SLME task, where the efficiency depends sensitively on the band gap and absorption edge. The present surrogate is trained on PBE-level data, for which semiconductor band gaps are expected to be systematically underestimated~\cite{shamDensityFunctionalTheoryEnergy1983,borlidoLargeScaleBenchmarkExchange2019}. Promising candidates could instead be evaluated using higher-level methods such as hybrid functionals \cite{heydErratumHybridFunctionals2006} or $GW$~\cite{govoniLargeScaleGW2015, deslippeBerkeleyGWMassivelyParallel2012}. Because such calculations are substantially more expensive, the resulting high-fidelity dataset will naturally be much smaller than the available PBE dataset. Multi-fidelity learning provides a natural way to address this imbalance by combining abundant low-fidelity data with a smaller number of high-fidelity calculations to improve predictions at the desired level of accuracy~\cite{chenLearningPropertiesOrdered2021}. This strategy could be incorporated into an active-learning extension of SPARC, where candidates with high reward or high predictive uncertainty are prioritized for higher-level electronic-structure calculations. The most promising candidates could then be validated experimentally, providing an additional level of fidelity. These theoretical and experimental data could be fed back into the surrogate to improve the reward signal for subsequent rounds of generation and fine-tuning. In this way, increasingly accurate theory and experiment become part of the optimization loop rather than serving only as final validation steps.

Taken together, these results establish SPARC as a framework for crystal inverse design in which symmetry and functional properties are optimized together rather than treated as separate stages. By keeping crystallographic symmetry explicit during reward-guided generation, SPARC can enforce known symmetry requirements or identify favorable symmetry classes when they are not known in advance. More broadly, this provides a route toward inverse design in which the search is guided not only toward high-performing materials, but toward structures that can physically support the targeted response.

\section{Methods}\label{sec4}
\subsection{Symmetry-constrained crystal representation}
A crystal $\mathcal{M}$ is the periodic arrangement of atoms specified by the triplet $\mathcal{M}=(\boldsymbol{A},\boldsymbol{F},\boldsymbol{L})$, where $\boldsymbol{A}=[\boldsymbol{a}_1,\dots,\boldsymbol{a}_N]\in\mathbb{R}^{h\times N}$ collects the one-hot atom types of the $N$ atoms in the cell over the $h=94$ supported chemical species, $\boldsymbol{L}=[\boldsymbol{l}_1,\boldsymbol{l}_2,\boldsymbol{l}_3]\in\mathbb{R}^{3\times3}$ is the lattice whose columns span one period of the crystal, and $\boldsymbol{F}= [\boldsymbol{f}_1,\dots,\boldsymbol{f}_N]\in[0,1)^{3\times N}$ collects the fractional coordinates, related to the Cartesian coordinates by $\boldsymbol{X}=\boldsymbol{L}\boldsymbol{F}$, from which physical quantities such as interatomic distances are computed.

The space group $G$ is the set of Euclidean operations $g=(\boldsymbol{O},\boldsymbol{t})$, with $\boldsymbol{O}\in O(3)$ and $\boldsymbol{t}\in\mathbb{R}^3$, that leave the periodic crystal invariant. In three dimensions, space groups are classified into $230$ types. The operations that leave a given site fixed form its site-symmetry group. Sites related by operations of $G$ belong to the same crystallographic orbit, and these orbits partition the atomic sites into symmetry-equivalent classes. Points whose site-symmetry groups are conjugate within $G$ belong to the same Wyckoff position.

SPARC adopts the symmetry-constrained asymmetric-unit representation used in SymmCD~\cite{levysymmcd}. For a fixed space group $G$, a crystal can be reconstructed from a symmetry-compatible lattice and one representative atom from each occupied orbit. Applying the operations of $G$ to these representatives generates the remaining symmetry-equivalent sites. The representatives form the asymmetric unit, and their number is denoted by $M$. Conditioned on the sampled pair $(G,M)$, the asymmetric-unit representation is encoded as
\begin{equation}
\mathcal{M}'=(\boldsymbol{A}',\boldsymbol{F}',\boldsymbol{\Sigma}',\boldsymbol{k}),
\end{equation}
where $\boldsymbol{A}'$ and $\boldsymbol{F}'$ contain the atom types and fractional coordinates of the $M$ representative sites. The labels $\boldsymbol{\Sigma}'$ specify their site symmetries within $G$, and $\boldsymbol{k}\in\mathbb{R}^6$ parameterizes the lattice subject to the constraints imposed by $G$. The space group $G$ and the number of representatives $M$ are supplied as conditioning variables and are therefore not included among the generated components of $\mathcal{M}'$. The model generates $\mathcal{M}'$ conditioned on $(G,M)$, after which the full crystal $\mathcal{M}$ is reconstructed by applying the space-group operations.

Following DiffCSP++~\cite{jiaoSpaceGroupConstrained2024}, the lattice is parameterized in a rotation-invariant way rather than through the raw matrix $\boldsymbol{L}$. The polar decomposition $\boldsymbol{L}=\boldsymbol{Q}\exp(\boldsymbol{S})$ separates a rigid rotation $\boldsymbol{Q}\in SO(3)$ from a symmetric matrix $\boldsymbol{S}$, and $\boldsymbol{S}$ expands uniquely in a fixed six-dimensional symmetric basis $\{\boldsymbol{B}_i\}_{i=1}^{6}$ as $\boldsymbol{S}=\sum_{i=1}^{6}k_i\boldsymbol{B}_i$, yielding the coordinate vector $\boldsymbol{k}=(k_1,\dots,k_6)$. For each crystal family a subset of the $k_i$ is fixed by symmetry while the remainder are free, so a space group is enforced simply by holding the constrained entries at their symmetry-determined values. The explicit basis $\{\boldsymbol{B}_i\}$ and the per-family constraints on $\boldsymbol{k}$ are reproduced in the SI.

\subsection{Symmetry-aware generation}
We use the symmetry-constrained diffusion model SymmCD~\cite{levysymmcd}, which builds on DiffCSP++~\cite{jiaoSpaceGroupConstrained2024}. Given a space group $G$ and the number of representative sites $M$, the model jointly generates the four components of the asymmetric-unit representation,
\begin{equation}
p_{\theta}(\mathcal{M}'\mid G,M)=p_{\theta}(\boldsymbol{A}',\boldsymbol{F}',\boldsymbol{\Sigma}',\boldsymbol{k}\mid G,M).
\end{equation}
For simplicity, we do not write $G$ and $M$ in every equation in this work. But they are still used as inputs in all generation and denoising steps. Before RL fine-tuning, $G$ is sampled from the empirical MP-20 distribution $p_{\mathrm{marg}}(G)$, and $M$ is sampled from $p_{\mathrm{marg}}(M\mid G)$. These distributions provide the initial sampling rule used for property steering.

The reverse diffusion process jointly denoises the atomic species $\boldsymbol{A}'$, fractional coordinates $\boldsymbol{F}'$, site-symmetry descriptors $\boldsymbol{\Sigma}'$, and lattice coefficients $\boldsymbol{k}$. Crystal symmetry restricts the allowed values of $\boldsymbol{k}$. A binary mask $\boldsymbol{m}(G)\in\{0,1\}^{6}$ selects the coefficients that remain free under the chosen space group. Only these coefficients are denoised, while the symmetry-constrained coefficients remain fixed. The reconstructed lattice is therefore consistent with $G$ by construction.

The site-symmetry descriptors are also defined relative to $G$. For each representative site, $\boldsymbol{\Sigma}'$ is encoded as a $15\times13$ binary array. The 15 rows correspond to crystallographic axes such as $[001]$, $[010]$, $[100]$, $[110]$, and $[111]$. The 13 columns represent the possible site-symmetry operations associated with these axes, including rotations, reflections, and rotoinversions. Each entry indicates whether the corresponding operation is present, giving a 195-dimensional descriptor for each representative site. During the forward discrete diffusion process, these descriptors approach a marginal distribution defined for the selected space group. The complete encoding and transition process are provided in the SI.

The denoiser is trained to reconstruct all four components from their noisy representations. We use a score-matching loss for the fractional coordinates, a masked noise-prediction loss for the lattice coefficients, and categorical cross-entropy losses for the atomic species and site-symmetry descriptors. The total objective is
\begin{equation}
\mathcal{L}
=\lambda_{\boldsymbol{k}}\,\mathcal{L}_{\boldsymbol{k}}
+\lambda_{\boldsymbol{F}'}\,\mathcal{L}_{\boldsymbol{F}'}
+\lambda_{\boldsymbol{A}'}\,\mathcal{L}_{\boldsymbol{A}'}
+\lambda_{\boldsymbol{\Sigma}'}\,\mathcal{L}_{\boldsymbol{\Sigma}'},
\end{equation}
where the loss weights are $\lambda_{\boldsymbol{k}}=5$, $\lambda_{\boldsymbol{F}'}=1$, $\lambda_{\boldsymbol{A}'}=0.1$, and $\lambda_{\boldsymbol{\Sigma}'}=10$. The individual component losses and the transition matrices used for the discrete variables are given in the SI.

\subsection{Reward definition}

We define a separate reward for each inverse-design task. Both rewards depend on the target property and the predicted band gap, but the band gap plays a different role in each case. For the dielectric task, it penalizes near-metallic candidates for which the static dielectric response is not physically meaningful. For the SLME task, it guides the generated population toward the band-gap region favored by the Shockley--Queisser limit.

We use the clamped linear function
\begin{equation}
\Phi(x;a,b)=\operatorname{clip}\left(\frac{x-a}{b-a},0,1\right)
\end{equation}
to map selected quantities to the interval $[0,1]$.

\textbf{Uniaxial dielectric anisotropy.}
For the uniaxial dielectric task, we evaluate the dielectric tensor of each generated structure using the TSENN surrogate. Before evaluation, we standardize the refined cell so that the symmetry-determined unique axis is aligned with $z$.

We define the out-of-plane response as $\varepsilon_{\perp}\equiv\varepsilon_{zz}$ and the mean in-plane response as $\varepsilon_{\parallel}\equiv\frac{1}{2}\left(\varepsilon_{xx}+\varepsilon_{yy}\right)$. We then introduce an out-of-plane contrast factor
\begin{equation}
g_z=\operatorname{clip}\left(\frac{\left|\varepsilon_{\perp}-\varepsilon_{\parallel}\right|}{\delta_z},0,1\right),
\end{equation}
where $\delta_z=0.01$ sets the contrast required to receive full credit. The factor $g_z$ is zero for a fully isotropic tensor. It increases linearly with the dielectric contrast and reaches one when the contrast exceeds $\delta_z$.

The residual mismatch between the two in-plane components is
\begin{equation}
m=\frac{\left|\varepsilon_{xx}-\varepsilon_{yy}\right|}{\left|\varepsilon_{xx}\right|+\left|\varepsilon_{yy}\right|},
\end{equation}
The uniaxial-anisotropy score is then
\begin{equation}
q_{\mathrm{uni}}=g_z(1-m)\in[0,1].
\end{equation}
The factor $(1-m)$ penalizes differences between $\varepsilon_{xx}$ and $\varepsilon_{yy}$, while $g_z$ penalizes insufficient contrast between the in-plane and out-of-plane responses. A high score therefore requires both approximate in-plane equality and finite out-of-plane anisotropy.

We further penalize structures with vanishing band gaps by defining
\begin{equation}
r_{\mathrm{uni}}
=
\min\left[
\Phi\left(q_{\mathrm{uni}};a_{\mathrm{uni}},b_{\mathrm{uni}}\right),
\Phi\left(E_g;a_g,b_g\right)
\right].
\end{equation}
The band-gap score limits the final reward when $E_g$ approaches the metallic regime. A high anisotropy score therefore cannot compensate for a vanishing band gap. This keeps the generated population within the finite-gap regime where the static dielectric response is well defined and can be verified using DFPT.

\textbf{Spectroscopic limited maximum efficiency.}
For the photovoltaic task, the target property is the spectroscopic limited maximum efficiency $\eta$. It is evaluated under the AM1.5G solar spectrum using an absorber thickness $L_{\mathrm{abs}}=\SI{0.3}{\micro\meter}$, a cell temperature $T_{\mathrm{cell}}=\SI{300}{\kelvin}$, and a radiative recombination fraction $f_r=1$~\cite{yuIdentificationPotentialPhotovoltaic2012,hungUniversalEnsembleEmbeddingGraph2024}. The full calculation of $\eta$ from the predicted absorption spectrum is provided in the SI.

High SLME alone does not explicitly localize the generated population near the band-gap region favored for single-junction solar cells. We therefore combine an efficiency score with a band-gap localization score,
\begin{equation}
r_{\mathrm{SLME}}=w_{\eta}r_{\eta}+w_g r_{g,\mathrm{SLME}},
\end{equation}
where
\begin{equation}
r_{\eta}=\Phi(\eta;0,0.35),
\end{equation}
and
\begin{equation}
r_{g,\mathrm{SLME}}(E_g)=\operatorname{clip}\left(1-\frac{\left|E_g-E^*\right|}{\Delta},0,1\right).
\end{equation}
We use $w_{\eta}=0.8$, $w_g=0.2$, $E^*=1.3$~eV, and $\Delta=1.0$~eV. The band-gap score forms a symmetric tent centered at $E^*=1.3$~eV and becomes zero outside $E_g\in[0.3,2.3]$~eV. It therefore encourages the search to remain near the theoretically favorable single-junction region without acting as a hard constraint.

The larger value of $w_{\eta}$ keeps SLME as the dominant optimization objective. The smaller band-gap term discourages both near-metallic candidates and overly wide-gap materials with weak solar absorption. The reward therefore favors structures that combine high predicted photovoltaic efficiency with a band gap near the single-junction optimum. The ramp intervals and the resulting reward surfaces for both tasks are provided in the SI.

\subsection{Reward-guided fine-tuning on the asymmetric unit}

Reinforcement-learning fine-tuning shifts the reverse diffusion process toward high-reward crystals by weighting the likelihood of each denoising transition according to the terminal reward. Recent reinforcement-learning approaches to crystal generation optimize either the full crystal representation, including atom types, atomic coordinates, and lattice parameters~\cite{chenAcceleratingInverseMaterials2025}, or a learned continuous latent representation~\cite{Park2026Guiding}. SPARC instead fine-tunes the diffusion process directly in the symmetry-constrained asymmetric-unit representation $\mathcal{M}'=(\boldsymbol{A}',\boldsymbol{F}',\boldsymbol{\Sigma}',\boldsymbol{k})$. The generated variables are constrained by the selected space group, and the full crystal is reconstructed from the final asymmetric unit through the space-group operation $\mathcal{R}_G$. Because the reverse trajectory, policy, and KL anchor are all defined on $\mathcal{M}'$, symmetry is built into the optimization rather than imposed as a post hoc filter. We retain the reward-weighted objective with an explicit KL anchor introduced for crystal generation in MatInvent~\cite{chenAcceleratingInverseMaterials2025}. The choice of gradient estimator is independent of the representation, allowing other policy-gradient methods to be used without changing the symmetry constraints imposed during generation.

For a sampled space group $G$ and number of representative sites $M$, the reverse process generates a trajectory $\mathcal{M}'_{T},\ldots,\mathcal{M}'_{0}$. As above, we omit the explicit conditioning on $(G,M)$ for notational simplicity. At reverse step $t$, the current noisy asymmetric unit $\mathcal{M}'_t$ is the policy state, and the transition to $\mathcal{M}'_{t-1}$ is sampled from
\begin{equation}
p_\theta(\mathcal{M}'_{t-1}\mid\mathcal{M}'_t).
\end{equation}
The terminal sample $\mathcal{M}'_0$ is expanded into a full crystal $\mathcal{M}$ by applying the operations $\mathcal{R}_G$. The reward is evaluated only on this reconstructed crystal. Intermediate noisy states are not assigned separate rewards.

The reinforcement-learning objective maximizes the expected terminal reward,
\begin{equation}
\mathcal{J}_{\mathrm{RL}}(\theta)
=
\mathbb{E}_{p_\theta(\mathcal{M}'_0)}
\left[
r(\mathcal{M})
\right].
\end{equation}
Using the reverse trajectory as a policy rollout, the corresponding loss gradient can be written as
\begin{equation}
\nabla_\theta\mathcal{L}_{\mathrm{RL}}
=
\mathbb{E}_{p_\theta(\mathcal{M}'_{0:T})}
\left[
-r(\mathcal{M})
\sum_{t=1}^{T}
\nabla_\theta
\log p_\theta(\mathcal{M}'_{t-1}\mid\mathcal{M}'_t)
\right].
\end{equation}
The same terminal reward weights every transition along the trajectory that produced the crystal. This is the standard policy-gradient formulation for reinforcement-learning fine-tuning of diffusion models~\cite{blackTrainingDiffusionModels2023,fanDPOKReinforcementLearning2023} and follows its application to crystal generation in MatInvent~\cite{chenAcceleratingInverseMaterials2025}.

Optimizing the terminal reward alone can move the policy too far from the pretrained crystal distribution. The model may then generate implausible structures or exploit errors in the property surrogates. We therefore regularize the fine-tuned reverse process against the frozen pretrained process $p_{\mathrm{pre}}$. At each reverse step, we evaluate the KL divergence between the fine-tuned and pretrained transition distributions. The KL contribution for each trajectory is weighted by $\lambda-r(\mathcal{M})$, where $\lambda$ is chosen above the maximum attainable reward. This coefficient remains positive and decreases as the reward increases. Low-reward trajectories are therefore kept close to the pretrained process, while high-reward trajectories are allowed to deviate more strongly.

Combining the reward and KL contributions gives the gradient estimator used for fine-tuning,
\begin{equation}
\begin{aligned}
\nabla_\theta\mathcal{L}(\theta)=\mathbb{E}_{p_\theta(\mathcal{M}'_{0:T})}\Bigl[
&-\alpha\,r(\mathcal{M})\sum_{t=1}^{T}\nabla_\theta\log p_\theta(\mathcal{M}'_{t-1}\mid\mathcal{M}'_t)\\
&+\beta\bigl(\lambda-r(\mathcal{M})\bigr)\sum_{t=1}^{T}\nabla_\theta\,\mathrm{KL}\bigl(p_\theta(\mathcal{M}'_{t-1}\mid\mathcal{M}'_t)\,\|\,p_{\mathrm{pre}}(\mathcal{M}'_{t-1}\mid\mathcal{M}'_t)\bigr)
\Bigr],
\end{aligned}
\end{equation}
where $\alpha$ and $\beta$ control the reward and KL contributions, respectively. The sampled reward is treated as a fixed weight during each parameter update. The derivation of the policy-gradient term and the data-processing-inequality bound that replaces the intractable terminal-distribution KL with a sum of per-step conditional KL terms are provided in the SI.

Because the full crystal $\mathcal{M}$ is reconstructed deterministically from $\mathcal{M}'_0$ and $\mathcal{R}_G$, its terminal reward provides a learning signal for the asymmetric-unit trajectory that generated it. The policy update therefore increases the probability of asymmetric units whose reconstructed crystals achieve high rewards, while the KL term limits unsupported deviations from the pretrained generator. The space-group proposal is updated separately across reinforcement-learning rounds using the outer-loop procedure described next.

\subsection{Adaptive space-group sampling}

The reward-weighted policy gradient updates the conditional generator $p_\theta(\boldsymbol{A}',\boldsymbol{F}',\boldsymbol{\Sigma}',\boldsymbol{k}\mid G,M)$ but does not directly update the proposal distribution over $G$. At the beginning of each trajectory, a space group $G$ is sampled from the current proposal and the number of representative sites $M$ is sampled from $p_{\mathrm{marg}}(M\mid G)$. Both variables remain fixed throughout the $T$ reverse-diffusion steps. Although $G$ is provided to the denoising model as a conditioning variable, it is sampled outside the diffusion policy and is not parameterized by $\theta$. The policy-gradient update therefore changes the generated distribution conditioned on $(G,M)$ but not the probability of selecting each space group. We address this limitation using a discrete outer-loop update of the space-group proposal after each RL round.

Let $\pi^{(0)}$ denote the base proposal over the supported space groups. It is obtained from the empirical MP-20 marginal using an optional temperature reweighting \cite{kirkpatrickOptimizationSimulatedAnnealing1983},
\begin{equation}
\pi^{(0)}(G)
=
\frac{p_{\mathrm{marg}}(G)^{1/T_{\mathrm{sg}}}}
{\sum_{G'\in\mathcal{S}}p_{\mathrm{marg}}(G')^{1/T_{\mathrm{sg}}}},
\end{equation}
where $\mathcal{S}=\{G:p_{\mathrm{marg}}(G)>0\}$ is the support of the training distribution. The first RL round samples $G$ from $\pi^{(0)}$, while subsequent rounds use the updated proposal defined below.

For each generated crystal, we determine the realized space group from the reconstructed full structure using \texttt{SpacegroupAnalyzer} with $\texttt{symprec}=0.01$~\AA{}. The detected label can differ from the conditioning label because of accidental higher symmetry or small numerical deviations in the reconstructed structure. We associate the terminal reward with this realized space-group label. Let $\bar{r}^{(j)}(G)$ denote the mean terminal reward of structures assigned to space group $G$ during RL round $j$. We update a running reward estimate for each space group using
\begin{equation}
\mu^{(j)}(G)=
\begin{cases}
\gamma\,\mu^{(j-1)}(G)+(1-\gamma)\,\bar{r}^{(j)}(G), & G\ \text{observed in round }j,\\
\gamma\,\mu^{(j-1)}(G), & G\ \text{unobserved in round }j,
\end{cases}
\end{equation}
where $\gamma\in(0,1)$ controls the decay of previous reward estimates. The second branch gradually removes stale reward information for space groups that are not observed in the current round.

The running reward estimates are used to reweight the base proposal,
\begin{equation}
\widetilde{\pi}^{(j)}(G)
=\frac{\pi^{(0)}(G)\exp\left(\kappa\mu^{(j)}(G)\right)}
{\sum_{H\in\mathcal{S}}\pi^{(0)}(H)\exp\left(\kappa\mu^{(j)}(H)\right)},
\end{equation}
where $\kappa$ controls the sensitivity of the proposal to the accumulated reward. We then mix the reward-reweighted distribution with the base proposal,
\begin{equation}
\pi^{(j)}(G)
=\rho\,\pi^{(0)}(G)+(1-\rho)\,\widetilde{\pi}^{(j)}(G),
\end{equation}
where $\rho\in[0,1]$ controls the strength of the base-distribution anchor. A minimum-probability floor is subsequently applied to every $G\in\mathcal{S}$, followed by renormalization.

Both terms in the mixture retain the base proposal $\pi^{(0)}$. The adaptive component favors space groups with larger running rewards, while the anchored component continually restores probability mass across the original support. The probability floor provides an additional safeguard against eliminating rarely sampled space groups. Together, these terms limit premature concentration of the proposal and preserve exploration throughout fine-tuning.

We use different base proposals for the two design tasks,
\begin{equation}
\pi^{(0)}(G)\propto p_{\mathrm{marg}}(G)^{1/T_{\mathrm{sg}}}.
\end{equation}
For the uniaxial dielectric objective, we use $T_{\mathrm{sg}}=100$, $\gamma=0.9$, $\kappa=5.0$, and $\rho=0.15$. The large temperature produces a nearly uniform base proposal over $\mathcal{S}$, reducing the influence of the MP-20 space-group frequencies for this strongly symmetry-dependent objective. For the SLME objective, we use $T_{\mathrm{sg}}=1$, $\gamma=0.9$, $\kappa=3.0$, and $\rho=0.1$. This choice retains the empirical MP-20 marginal as the base proposal because SLME is not restricted to a narrow set of crystal symmetries.

\subsection{Training and fine-tuning details}
SPARC uses the SymmCD generator~\cite{levysymmcd} as its diffusion backbone. The generator was pretrained on the MP-20 dataset. We initialize the policy $p_\theta$ from the released checkpoint and retain a frozen copy as the KL reference policy $p_{\mathrm{pre}}$. Only the reward-guided fine-tuning stage is performed in this work. The original diffusion process is retained, with $T=1000$ reverse-diffusion steps.

Both objectives are fine-tuned for 120 RL rounds using an initial learning rate of $3\times10^{-5}$. The learning rate remains constant for the dielectric-anisotropy objective and is multiplied by $0.98$ after each round for the SLME objective. The policy-gradient coefficient is set to $\alpha=1$, and the reward-anchor threshold is set to $\lambda=1.1$. The KL coefficient is initialized as $\beta_0=0.025$ and decays with the fine-tuning update index $\nu$ according to
\begin{equation}
\beta_\nu=\frac{\beta_0}{\sqrt{1+\nu/50}}.
\end{equation}
Complete optimization, diffusion, and sampling settings are provided in SI.

\section*{Data Availability}
The data that support the findings of this study are available from the corresponding author upon
reasonable request. The datasets and trained model weights will be deposited in a publicly accessible
repository upon acceptance of the manuscript.

\section*{Code Availability}
The code that supports the findings of this study is available from the corresponding author upon
reasonable request. It will be released in a publicly accessible repository upon acceptance of the
manuscript.

\section*{Declarations}
\bmhead{Acknowledgements}
T.-W.H. and Q.Y. acknowledge support from the U.S. Department of Energy, Office of Science, Office of Basic Energy Sciences, under Award No. DE-SC0023664. T.-W.H. and A.B. acknowledge support from the U.S. National Science Foundation through the Expanding Capacity in Quantum Information Science and Engineering (ExpandQISE) program under Award No. OMA-2329067.

This research used resources of the National Energy Research Scientific Computing Center (NERSC), a U.S. Department of Energy Office of Science User Facility operated under Contract No. DE-AC02-05CH11231, through NERSC Award No. BES-ERCAP0029544. This work also benefited from Northeastern University's Quantum Materials and Sensing Institute (QMSI).

\bmhead{Author Contribution}
T.-W.H. and Q.Y. conceived the study. T.-W.H. developed the methodology, performed the calculations, trained the models, analyzed the results, and prepared the figures. Q.Y. and A.B. supervised the project. The manuscript was written with contributions from all authors, and all authors approved the final version.

\bmhead{Competing Interest}
The authors declare no competing interests.

\bibliography{sn-bibliography}% common bib file

@article{hsuAccuratePredictionTensorial2026,
  title = {Accurate Prediction of Tensorial Spectra Using Equivariant Graph Neural Network},
  author = {Hsu, Ting-Wei and Fang, Zhenyao and Bansil, Arun and Yan, Qimin},
  year = 2026,
  month = mar,
  journal = {Nature Communications},
  publisher = {Nature Publishing Group},
  issn = {2041-1723},
  doi = {10.1038/s41467-026-69159-9},
  urldate = {2026-03-08},
  copyright = {2026 The Author(s)},
  langid = {english}
}

@article{grunertDeepLearningSpectra2024,
  title = {Deep Learning of Spectra: {{Predicting}} the Dielectric Function of Semiconductors},
  shorttitle = {Deep Learning of Spectra},
  author = {Grunert, Malte and Gro{\ss}mann, Max and Runge, Erich},
  year = 2024,
  month = dec,
  journal = {Physical Review Materials},
  volume = {8},
  number = {12},
  pages = {L122201},
  publisher = {American Physical Society},
  doi = {10.1103/PhysRevMaterials.8.L122201},
  urldate = {2025-09-06},
}

@article{sangalliManybodyPerturbationTheory2019,
  title = {Many-Body Perturbation Theory Calculations Using the Yambo Code},
  author = {Sangalli, D and Ferretti, A and Miranda, H and Attaccalite, C and Marri, I and Cannuccia, E and Melo, P and Marsili, M and Paleari, F and Marrazzo, A and Prandini, G and Bonf{\`a}, P and Atambo, M O and Affinito, F and Palummo, M and {Molina-S{\'a}nchez}, A and Hogan, C and Gr{\"u}ning, M and Varsano, D and Marini, A},
  year = 2019,
  month = may,
  journal = {Journal of Physics: Condensed Matter},
  volume = {31},
  number = {32},
  pages = {325902},
  publisher = {IOP Publishing},
  issn = {0953-8984},
  doi = {10.1088/1361-648X/ab15d0},
  urldate = {2026-06-03},
  langid = {english},
}

@article{giannozziQUANTUMESPRESSOModular2009,
  title = {{{QUANTUM ESPRESSO}}: A Modular and Open-Source Software Project for Quantum Simulations of Materials},
  shorttitle = {{{QUANTUM ESPRESSO}}},
  author = {Giannozzi, Paolo and Baroni, Stefano and Bonini, Nicola and Calandra, Matteo and Car, Roberto and Cavazzoni, Carlo and Ceresoli, Davide and Chiarotti, Guido L and Cococcioni, Matteo and Dabo, Ismaila and Dal Corso, Andrea and {de Gironcoli}, Stefano and Fabris, Stefano and Fratesi, Guido and Gebauer, Ralph and Gerstmann, Uwe and Gougoussis, Christos and Kokalj, Anton and Lazzeri, Michele and {Martin-Samos}, Layla and Marzari, Nicola and Mauri, Francesco and Mazzarello, Riccardo and Paolini, Stefano and Pasquarello, Alfredo and Paulatto, Lorenzo and Sbraccia, Carlo and Scandolo, Sandro and Sclauzero, Gabriele and Seitsonen, Ari P and Smogunov, Alexander and Umari, Paolo and Wentzcovitch, Renata M},
  year = 2009,
  month = sep,
  journal = {Journal of Physics: Condensed Matter},
  volume = {21},
  number = {39},
  pages = {395502},
  issn = {0953-8984},
  doi = {10.1088/0953-8984/21/39/395502},
  urldate = {2026-06-03},
  langid = {english},
}

@article{schmidtMachineLearningAssistedDeterminationGlobal2023,
  title = {Machine-{{Learning-Assisted Determination}} of the {{Global Zero-Temperature Phase Diagram}} of {{Materials}}},
  author = {Schmidt, Jonathan and Hoffmann, Noah and Wang, Hai-Chen and Borlido, Pedro and Carri{\c c}o, Pedro J. M. A. and Cerqueira, Tiago F. T. and Botti, Silvana and Marques, Miguel A. L.},
  year = 2023,
  journal = {Advanced Materials},
  volume = {35},
  number = {22},
  pages = {2210788},
  issn = {1521-4095},
  doi = {10.1002/adma.202210788},
  urldate = {2026-06-03},
  copyright = {\copyright{} 2023 The Authors. Advanced Materials published by Wiley-VCH GmbH},
  langid = {english},
}

@article{jainCommentaryMaterialsProject2013,
  title = {Commentary: {{The Materials Project}}: {{A}} Materials Genome Approach to Accelerating Materials Innovation},
  shorttitle = {Commentary},
  author = {Jain, Anubhav and Ong, Shyue Ping and Hautier, Geoffroy and Chen, Wei and Richards, William Davidson and Dacek, Stephen and Cholia, Shreyas and Gunter, Dan and Skinner, David and Ceder, Gerbrand and Persson, Kristin A.},
  year = 2013,
  month = jul,
  journal = {APL Materials},
  volume = {1},
  number = {1},
  pages = {011002},
  issn = {2166-532X},
  doi = {10.1063/1.4812323},
  urldate = {2025-04-24},
}

@article{zeniGenerativeModelInorganic2025a,
  title = {A Generative Model for Inorganic Materials Design},
  author = {Zeni, Claudio and Pinsler, Robert and Z{\"u}gner, Daniel and Fowler, Andrew and Horton, Matthew and Fu, Xiang and Wang, Zilong and Shysheya, Aliaksandra and Crabb{\'e}, Jonathan and Ueda, Shoko and Sordillo, Roberto and Sun, Lixin and Smith, Jake and Nguyen, Bichlien and Schulz, Hannes and Lewis, Sarah and Huang, Chin-Wei and Lu, Ziheng and Zhou, Yichi and Yang, Han and Hao, Hongxia and Li, Jielan and Yang, Chunlei and Li, Wenjie and Tomioka, Ryota and Xie, Tian},
  year = 2025,
  month = mar,
  journal = {Nature},
  volume = {639},
  number = {8055},
  pages = {624--632},
  publisher = {Nature Publishing Group},
  issn = {1476-4687},
  doi = {10.1038/s41586-025-08628-5},
  urldate = {2026-06-03},
  copyright = {2025 The Author(s)},
  langid = {english}
}

@article{deanBoronNitrideSubstrates2010,
  title = {Boron Nitride Substrates for High-Quality Graphene Electronics},
  author = {Dean, C. R. and others},
  year = {2010},
  journal = {Nature Nanotechnology},
  volume = {5},
  pages = {722--726},
  doi = {10.1038/nnano.2010.172}
}

@article{laturiaDielectricPropertiesHexagonal2018,
  title = {Dielectric Properties of Hexagonal Boron Nitride and Transition Metal Dichalcogenides: From Monolayer to Bulk},
  author = {Laturia, Akash and Van de Put, Maarten L. and Vandenberghe, William G.},
  year = {2018},
  journal = {npj 2D Materials and Applications},
  volume = {2},
  pages = {6},
  doi = {10.1038/s41699-018-0050-x}
}

@article{illarionovInsulators2DNanoelectronics2020,
  title = {Insulators for 2D Nanoelectronics: The Gap to Bridge},
  author = {Illarionov, Yury Yu. and others},
  year = {2020},
  journal = {Nature Communications},
  volume = {11},
  pages = {3385},
  doi = {10.1038/s41467-020-16640-8}
}

@article{niuGiantOpticalAnisotropy2018,
  title = {Giant Optical Anisotropy in a Quasi-One-Dimensional Crystal},
  author = {Niu, Shanyuan and others},
  year = {2018},
  journal = {Nature Photonics},
  volume = {12},
  pages = {392--396},
  doi = {10.1038/s41566-018-0189-1}
}

@article{ermolaevGiantOpticalAnisotropy2021,
  title = {Giant Optical Anisotropy in Transition Metal Dichalcogenides for Next-Generation Photonics},
  author = {Ermolaev, G. A. and others},
  year = {2021},
  journal = {Nature Communications},
  volume = {12},
  pages = {854},
  doi = {10.1038/s41467-021-21139-x}
}

@inproceedings{levysymmcd,
  title={SymmCD: Symmetry-Preserving Crystal Generation with Diffusion Models},
  author={Levy, Daniel and Panigrahi, Siba Smarak and Kaba, S{\'e}kou-Oumar and Zhu, Qiang and Lee, Kin Long Kelvin and Galkin, Mikhail and Miret, Santiago and Ravanbakhsh, Siamak},
  booktitle={The Thirteenth International Conference on Learning Representations}
}

@article{yang2024mattersim,
      title={MatterSim: A Deep Learning Atomistic Model Across Elements, Temperatures and Pressures},
      author={Han Yang and Chenxi Hu and Yichi Zhou and Xixian Liu and Yu Shi and Jielan Li and Guanzhi Li and Zekun Chen and Shuizhou Chen and Claudio Zeni and Matthew Horton and Robert Pinsler and Andrew Fowler and Daniel Z{\"u}gner and Tian Xie and Jake Smith and Lixin Sun and Qian Wang and Lingyu Kong and Chang Liu and Hongxia Hao and Ziheng Lu},
      year={2024},
      eprint={2405.04967},
      archivePrefix={arXiv},
      primaryClass={cond-mat.mtrl-sci},
      url={https://arxiv.org/abs/2405.04967},
      journal={arXiv preprint arXiv:2405.04967}
}

@inproceedings{jiao2023diffcsp,
  title={Crystal Structure Prediction by Joint Equivariant Diffusion},
  author={Jiao, Rui and Huang, Wen-bing and Zhang, Yu and Rong, Jinliang and Liu, Yang},
  booktitle={Advances in Neural Information Processing Systems (NeurIPS)},
  year={2023}
}

@article{yuIdentificationPotentialPhotovoltaic2012,
  title = {Identification of {{Potential Photovoltaic Absorbers Based}} on {{First-Principles Spectroscopic Screening}} of {{Materials}}},
  author = {Yu, Liping and Zunger, Alex},
  year = 2012,
  month = feb,
  journal = {Physical Review Letters},
  volume = {108},
  number = {6},
  pages = {068701},
  publisher = {American Physical Society},
  doi = {10.1103/PhysRevLett.108.068701},
  url = {https://link.aps.org/doi/10.1103/PhysRevLett.108.068701},
  urldate = {2026-07-02},
}

@article{choudharyAcceleratedDiscoveryEfficient2019,
  title = {Accelerated {{Discovery}} of {{Efficient Solar Cell Materials Using Quantum}} and {{Machine-Learning Methods}}},
  author = {Choudhary, Kamal and Bercx, Marnik and Jiang, Jie and Pachter, Ruth and Lamoen, Dirk and Tavazza, Francesca},
  year = 2019,
  month = aug,
  journal = {Chemistry of Materials},
  volume = {31},
  number = {15},
  pages = {5900--5908},
  publisher = {American Chemical Society},
  issn = {0897-4756},
  doi = {10.1021/acs.chemmater.9b02166},
  url = {https://doi.org/10.1021/acs.chemmater.9b02166},
  urldate = {2026-07-02},
}

@article{zhuSemiconductingLayeredBlue2014,
  title = {Semiconducting Layered Blue Phosphorus: {{A}} Computational Study},
  author = {Zhu, Zhen and Tom{\'a}nek, David},
  year = 2014,
  month = apr,
  journal = {Physical Review Letters},
  volume = {112},
  number = {17},
  pages = {176802},
  publisher = {American Physical Society},
  doi = {10.1103/PhysRevLett.112.176802},
  url = {https://doi.org/10.1103/PhysRevLett.112.176802},
}

@inproceedings{xieCrystalDiffusionVariational2022,
  title = {Crystal Diffusion Variational Autoencoder for Periodic Material Generation},
  author = {Xie, Tian and Fu, Xiang and Ganea, Octavian-Eugen and Barzilay, Regina and Jaakkola, Tommi},
  booktitle = {International Conference on Learning Representations (ICLR)},
  year = 2022,
  eprint = {2110.06197},
  archiveprefix = {arXiv},
}

@inproceedings{jiaoSpaceGroupConstrained2024,
  title = {Space Group Constrained Crystal Generation},
  author = {Jiao, Rui and Huang, Wenbing and Liu, Yu and Zhao, Deli and Liu, Yang},
  booktitle = {International Conference on Learning Representations (ICLR)},
  year = 2024,
  eprint = {2402.03992},
  archiveprefix = {arXiv},
}

@article{blackTrainingDiffusionModels2023,
  title = {Training Diffusion Models with Reinforcement Learning},
  author = {Black, Kevin and Janner, Michael and Du, Yilun and Kostrikov, Ilya and Levine, Sergey},
  journal = {arXiv preprint arXiv:2305.13301},
  year = 2023,
  eprint = {2305.13301},
  archiveprefix = {arXiv},
}

@inproceedings{fanDPOKReinforcementLearning2023,
  title = {{DPOK}: Reinforcement Learning for Fine-Tuning Text-to-Image Diffusion Models},
  author = {Fan, Ying and Watkins, Olivia and Du, Yuqing and Liu, Hao and Ryu, Moonkyung and Boutilier, Craig and Abbeel, Pieter and Ghavamzadeh, Mohammad and Lee, Kangwook and Lee, Kimin},
  booktitle = {Advances in Neural Information Processing Systems (NeurIPS)},
  year = 2023,
  eprint = {2305.16381},
  archiveprefix = {arXiv},
}

@article{hoClassifierFreeDiffusionGuidance2022,
  title = {Classifier-Free Diffusion Guidance},
  author = {Ho, Jonathan and Salimans, Tim},
  journal = {arXiv preprint arXiv:2207.12598},
  year = 2022,
  eprint = {2207.12598},
  archiveprefix = {arXiv},
}

@article{chenAcceleratingInverseMaterials2025,
  title = {Accelerating Inverse Materials Design Using Generative Diffusion Models with Reinforcement Learning},
  author = {Chen, Junwu and Guo, Jeff and Fako, Edvin and Schwaller, Philippe},
  journal = {arXiv preprint arXiv:2511.03112},
  year = 2025,
  eprint = {2511.03112},
  archiveprefix = {arXiv},
}

@article{guoInitioStructureSolutions2025a,
  title = {Ab Initio Structure Solutions from Nanocrystalline Powder Diffraction Data via Diffusion Models},
  author = {Guo, Gabe and Saidi, Tristan Luca and Terban, Maxwell W. and Valsecchi, Michele and Billinge, Simon J. L. and Lipson, Hod},
  year = 2025,
  month = nov,
  journal = {Nature Materials},
  volume = {24},
  number = {11},
  pages = {1726--1734},
  publisher = {Nature Publishing Group},
  issn = {1476-4660},
  doi = {10.1038/s41563-025-02220-y},
  url = {https://www.nature.com/articles/s41563-025-02220-y},
  urldate = {2026-07-05},
  copyright = {2025 The Author(s), under exclusive licence to Springer Nature Limited},
  langid = {english},
}

@article{petousisBenchmarkingDensityFunctional2016a,
  title = {Benchmarking Density Functional Perturbation Theory to Enable High-Throughput Screening of Materials for Dielectric Constant and Refractive Index},
  author = {Petousis, Ioannis and Chen, Wei and Hautier, Geoffroy and Graf, Tanja and Schladt, Thomas D. and Persson, Kristin A. and Prinz, Fritz B.},
  year = 2016,
  month = mar,
  journal = {Physical Review B},
  volume = {93},
  number = {11},
  pages = {115151},
  publisher = {American Physical Society},
  doi = {10.1103/PhysRevB.93.115151},
  url = {https://link.aps.org/doi/10.1103/PhysRevB.93.115151},
  urldate = {2026-07-06},
}

@article{gajdosLinearOpticalProperties2006,
  title = {Linear Optical Properties in the Projector-Augmented Wave Methodology},
  author = {Gajdo{\v s}, M. and Hummer, K. and Kresse, G. and Furthm{\"u}ller, J. and Bechstedt, F.},
  year = 2006,
  month = jan,
  journal = {Physical Review B},
  volume = {73},
  number = {4},
  pages = {045112},
  publisher = {American Physical Society},
  doi = {10.1103/PhysRevB.73.045112},
  url = {https://link.aps.org/doi/10.1103/PhysRevB.73.045112},
  urldate = {2026-07-06},
}

@book{fox2010optical,
  author    = {Fox, Mark},
  title     = {Optical Properties of Solids},
  publisher = {Oxford University Press},
  address   = {Oxford, UK},
  year      = {2010}
}

@article{Park2026Guiding,
  author  = {Park, Hyunsoo and Walsh, Aron},
  title   = {Guiding generative models to uncover diverse and novel crystals via reinforcement learning},
  journal = {Nature Machine Intelligence},
  year    = {2026},
  doi     = {10.1038/s42256-026-01262-4},
  url     = {https://doi.org/10.1038/s42256-026-01262-4},
  note    = {Published online 6 July 2026}
}

@misc{govindarajan_crystalgym_2025,
	title = {{CrystalGym}: {A} {New} {Benchmark} for {Materials} {Discovery} {Using} {Reinforcement} {Learning}},
	shorttitle = {{CrystalGym}},
	url = {http://arxiv.org/abs/2509.23156},
	doi = {10.48550/arXiv.2509.23156},
	urldate = {2026-07-06},
	publisher = {arXiv},
	author = {Govindarajan, Prashant and Reymond, Mathieu and Clavaud, Antoine and Phielipp, Mariano and Miret, Santiago and Chandar, Sarath},
	month = sep,
	year = {2025},
	note = {arXiv:2509.23156 [cs.LG]},
}

@article{cao_reinforcement_2026,
	title = {Reinforcement {Fine}-{Tuning} for {Materials} {Design}},
	volume = {113},
	issn = {2469-9950, 2469-9969},
	url = {http://arxiv.org/abs/2504.02367},
	doi = {10.1103/45zh-44bg},
	number = {2},
	urldate = {2026-07-06},
	journal = {Physical Review B},
	author = {Cao, Zhendong and Wang, Lei},
	month = jan,
	year = {2026},
	note = {arXiv:2504.02367 [cond-mat.mtrl-sci]},
	pages = {024106},
}

@article{hungUniversalEnsembleEmbeddingGraph2024,
  title = {Universal {{Ensemble-Embedding Graph Neural Network}} for {{Direct Prediction}} of {{Optical Spectra}} from {{Crystal Structures}}},
  author = {Hung, Nguyen Tuan and Okabe, Ryotaro and Chotrattanapituk, Abhijatmedhi and Li, Mingda},
  year = 2024,
  journal = {Advanced Materials},
  volume = {36},
  number = {46},
  pages = {2409175},
  issn = {1521-4095},
  doi = {10.1002/adma.202409175},
  url = {https://onlinelibrary.wiley.com/doi/abs/10.1002/adma.202409175},
  urldate = {2025-11-24},
  copyright = {\copyright{} 2024 The Author(s). Advanced Materials published by Wiley-VCH GmbH},
  langid = {english},
}

@article{liuVanWaalsHeterostructures2016,
  title = {Van Der {{Waals}} Heterostructures and Devices},
  author = {Liu, Yuan and Weiss, Nathan O. and Duan, Xidong and Cheng, Hung-Chieh and Huang, Yu and Duan, Xiangfeng},
  year = 2016,
  month = jul,
  journal = {Nature Reviews Materials},
  volume = {1},
  number = {9},
  pages = {16042},
  publisher = {Nature Publishing Group},
  issn = {2058-8437},
  doi = {10.1038/natrevmats.2016.42},
  url = {https://www.nature.com/articles/natrevmats201642},
  urldate = {2026-07-07},
  copyright = {2016 Macmillan Publishers Limited},
  langid = {english},
}

@article{yuanPolarizationsensitiveBroadbandPhotodetector2015a,
  title = {Polarization-Sensitive Broadband Photodetector Using a Black Phosphorus Vertical p--n Junction},
  author = {Yuan, Hongtao and Liu, Xiaoge and Afshinmanesh, Farzaneh and Li, Wei and Xu, Gang and Sun, Jie and Lian, Biao and Curto, Alberto G. and Ye, Guojun and Hikita, Yasuyuki and Shen, Zhixun and Zhang, Shou-Cheng and Chen, Xianhui and Brongersma, Mark and Hwang, Harold Y. and Cui, Yi},
  year = 2015,
  month = aug,
  journal = {Nature Nanotechnology},
  volume = {10},
  number = {8},
  pages = {707--713},
  publisher = {Nature Publishing Group},
  issn = {1748-3395},
  doi = {10.1038/nnano.2015.112},
  url = {https://www.nature.com/articles/nnano.2015.112},
  urldate = {2026-07-26},
  copyright = {2015 Springer Nature Limited},
  langid = {english},
}

@article{islandEnvironmentalInstabilityFewlayer2015,
  title = {Environmental Instability of Few-Layer Black Phosphorus},
  author = {Island, Joshua O and Steele, Gary A and van der Zant, Herre S J and {Castellanos-Gomez}, Andres},
  year = 2015,
  month = jan,
  journal = {2D Materials},
  volume = {2},
  number = {1},
  pages = {011002},
  publisher = {IOP Publishing},
  issn = {2053-1583},
  doi = {10.1088/2053-1583/2/1/011002},
  url = {https://doi.org/10.1088/2053-1583/2/1/011002},
  urldate = {2026-07-26},
  langid = {english},
}

@article{xiaRediscoveringBlackPhosphorus2014,
  title = {Rediscovering Black Phosphorus as an Anisotropic Layered Material for Optoelectronics and Electronics},
  author = {Xia, Fengnian and Wang, Han and Jia, Yichen},
  date = {2014-07-21},
  journaltitle = {Nature Communications},
  shortjournal = {Nat. Commun.},
  volume = {5},
  number = {1},
  pages = {4458},
  publisher = {Nature Publishing Group},
  issn = {2041-1723},
  doi = {10.1038/ncomms5458},
  url = {https://www.nature.com/articles/ncomms5458},
  urldate = {2026-08-31},
  langid = {english},
}

@article{qiaoHighmobilityTransportAnisotropy2014,
  title = {High-Mobility Transport Anisotropy and Linear Dichroism in Few-Layer Black Phosphorus},
  author = {Qiao, Jingsi and Kong, Xianghua and Hu, Zhi-Xin and Yang, Feng and Ji, Wei},
  date = {2014-07-21},
  journaltitle = {Nature Communications},
  shortjournal = {Nat. Commun.},
  volume = {5},
  number = {1},
  pages = {4475},
  publisher = {Nature Publishing Group},
  issn = {2041-1723},
  doi = {10.1038/ncomms5475},
  url = {https://www.nature.com/articles/ncomms5475},
  urldate = {2026-08-31},
  langid = {english},
}

@article{yangUltrafastSelftrappingPhotoexcited2019,
  title = {Ultrafast Self-Trapping of Photoexcited Carriers Sets the Upper Limit on Antimony Trisulfide Photovoltaic Devices},
  author = {Yang, Zhaoliang and Wang, Xiaomin and Chen, Yuzhong and Zheng, Zhenfa and Chen, Zeng and Xu, Wenqi and Liu, Weimin and Yang, Yang (Michael) and Zhao, Jin and Chen, Tao and Zhu, Haiming},
  date = {2019-10-04},
  journaltitle = {Nature Communications},
  shortjournal = {Nat. Commun.},
  volume = {10},
  number = {1},
  pages = {4540},
  publisher = {Nature Publishing Group},
  issn = {2041-1723},
  doi = {10.1038/s41467-019-12445-6},
  url = {https://www.nature.com/articles/s41467-019-12445-6},
  urldate = {2026-08-31},
  langid = {english},
}

@article{yanCu2ZnSnS4SolarCells2018,
  title = {{{Cu2ZnSnS4}} Solar Cells with over 10\% Power Conversion Efficiency Enabled by Heterojunction Heat Treatment},
  author = {Yan, Chang and Huang, Jialiang and Sun, Kaiwen and Johnston, Steve and Zhang, Yuanfang and Sun, Heng and Pu, Aobo and He, Mingrui and Liu, Fangyang and Eder, Katja and Yang, Limei and Cairney, Julie M. and Ekins-Daukes, N. J. and Hameiri, Ziv and Stride, John A. and Chen, Shiyou and Green, Martin A. and Hao, Xiaojing},
  date = {2018-09},
  journaltitle = {Nature Energy},
  shortjournal = {Nat. Energy},
  volume = {3},
  number = {9},
  pages = {764--772},
  publisher = {Nature Publishing Group},
  issn = {2058-7546},
  doi = {10.1038/s41560-018-0206-0},
  url = {https://www.nature.com/articles/s41560-018-0206-0},
  urldate = {2026-08-31},
  langid = {english},
}

@article{shamDensityFunctionalTheoryEnergy1983,
  title = {Density-{{Functional Theory}} of the {{Energy Gap}}},
  author = {Sham, L. J. and Schl\"uter, M.},
  date = {1983-11-14},
  journaltitle = {Physical Review Letters},
  shortjournal = {Phys. Rev. Lett.},
  volume = {51},
  number = {20},
  pages = {1888--1891},
  publisher = {American Physical Society},
  doi = {10.1103/PhysRevLett.51.1888},
  url = {https://link.aps.org/doi/10.1103/PhysRevLett.51.1888},
  urldate = {2026-09-02},
}

@article{borlidoLargeScaleBenchmarkExchange2019,
  title = {Large-{{Scale Benchmark}} of {{Exchange}}--{{Correlation Functionals}} for the {{Determination}} of {{Electronic Band Gaps}} of {{Solids}}},
  author = {Borlido, Pedro and Aull, Thorsten and Huran, Ahmad W. and Tran, Fabien and Marques, Miguel A. L. and Botti, Silvana},
  date = {2019-07-15},
  journaltitle = {Journal of Chemical Theory and Computation},
  shortjournal = {J. Chem. Theory Comput.},
  volume = {15},
  number = {9},
  pages = {5069--5079},
  issn = {1549-9618},
  doi = {10.1021/acs.jctc.9b00322},
  url = {https://doi.org/10.1021/acs.jctc.9b00322},
  urldate = {2026-09-02},
}

@article{govoniLargeScaleGW2015,
  title = {Large {{Scale GW Calculations}}},
  author = {Govoni, Marco and Galli, Giulia},
  date = {2015-01-12},
  journaltitle = {Journal of Chemical Theory and Computation},
  shortjournal = {J. Chem. Theory Comput.},
  volume = {11},
  number = {6},
  pages = {2680--2696},
  issn = {1549-9618},
  doi = {10.1021/ct500958p},
  url = {https://doi.org/10.1021/ct500958p},
  urldate = {2026-09-02},
}

@article{kirkpatrickOptimizationSimulatedAnnealing1983,
  title = {Optimization by {{Simulated Annealing}}},
  author = {Kirkpatrick, S. and Gelatt, C. D. and Vecchi, M. P.},
  date = {1983},
  journaltitle = {Science},
  volume = {220},
  number = {4598},
  eprint = {1690046},
  eprinttype = {jstor},
  pages = {671--680},
  publisher = {American Association for the Advancement of Science},
  issn = {0036-8075},
  url = {https://www.jstor.org/stable/1690046},
  urldate = {2026-09-02},
}

@online{yanSpaceGroupSymmetry2024,
  title = {A {{Space Group Symmetry Informed Network}} for {{O}}(3) {{Equivariant Crystal Tensor Prediction}}},
  author = {Yan, Keqiang and Saxton, Alexandra and Qian, Xiaofeng and Qian, Xiaoning and Ji, Shuiwang},
  date = {2024-06-03},
  eprint = {2406.12888},
  eprinttype = {arXiv},
  eprintclass = {cond-mat},
  doi = {10.48550/arXiv.2406.12888},
  url = {http://arxiv.org/abs/2406.12888},
  urldate = {2025-02-11},
  pubstate = {prepublished},
}

@article{liDeepLearningDensityFunctional2022,
  title = {Deep-{{Learning Density Functional Theory Hamiltonian}} for {{Efficient}} Ab Initio {{Electronic-Structure Calculation}}},
  author = {Li, He and Wang, Zun and Zou, Nianlong and Ye, Meng and Xu, Runzhang and Gong, Xiaoxun and Duan, Wenhui and Xu, Yong},
  date = {2022-06-23},
  journaltitle = {Nature Computational Science},
  shortjournal = {Nat. Comput. Sci.},
  volume = {2},
  number = {6},
  eprint = {2104.03786},
  eprinttype = {arXiv},
  eprintclass = {cond-mat, physics:physics, physics:quant-ph},
  pages = {367--377},
  issn = {2662-8457},
  doi = {10.1038/s43588-022-00265-6},
  url = {http://arxiv.org/abs/2104.03786},
  urldate = {2024-06-03},
}

@article{gongGeneralFrameworkE3equivariant2023a,
  title = {General Framework for {{E}}(3)-Equivariant Neural Network Representation of Density Functional Theory {{Hamiltonian}}},
  author = {Gong, Xiaoxun and Li, He and Zou, Nianlong and Xu, Runzhang and Duan, Wenhui and Xu, Yong},
  date = {2023-05-18},
  journaltitle = {Nature Communications},
  shortjournal = {Nat. Commun.},
  volume = {14},
  number = {1},
  pages = {2848},
  publisher = {Nature Publishing Group},
  issn = {2041-1723},
  doi = {10.1038/s41467-023-38468-8},
  url = {https://www.nature.com/articles/s41467-023-38468-8},
  urldate = {2026-09-04},
  langid = {english},
}

@article{zhongUniversalSpinOrbit2026a,
  title = {A Universal Spin--Orbit-Coupled {{Hamiltonian}} Model for Accelerated Quantum Material Discovery},
  author = {Zhong, Yang and Wang, Rui and Gong, Xingao and Xiang, Hongjun},
  date = {2026-03},
  journaltitle = {Nature Machine Intelligence},
  shortjournal = {Nat. Mach. Intell.},
  volume = {8},
  number = {3},
  pages = {403--414},
  publisher = {Nature Publishing Group},
  issn = {2522-5839},
  doi = {10.1038/s42256-026-01196-x},
  url = {https://www.nature.com/articles/s42256-026-01196-x},
  urldate = {2026-09-04},
  langid = {english},
}

@article{dongAccuratePiezoelectricTensor2025,
  title = {Accurate Piezoelectric Tensor Prediction with Equivariant Attention Tensor Graph Neural Network},
  author = {Dong, Luqi and Zhang, Xuanlin and Yang, Ziduo and Shen, Lei and Lu, Yunhao},
  date = {2025-03-06},
  journaltitle = {npj Computational Materials},
  shortjournal = {Npj Comput. Mater.},
  volume = {11},
  number = {1},
  publisher = {{Springer Science and Business Media LLC}},
  issn = {2057-3960},
  doi = {10.1038/s41524-025-01546-0},
  url = {https://www.nature.com/articles/s41524-025-01546-0},
  urldate = {2025-07-22},
  langid = {english},
}

@article{zengLearningThermoelectricTransport2026,
  title = {Learning Thermoelectric Transport from Crystal Structures via Multiscale Graph Neural Network},
  author = {Zeng, Yuxuan and Cao, Wei and Zuo, Yijing and Lyu, Fang and Xie, Wenhao and Peng, Tan and Hou, Yue and Miao, Ling and Wang, Ziyu and Shi, Jing},
  date = {2026-07-07},
  journaltitle = {Physical Review Applied},
  shortjournal = {Phys. Rev. Appl.},
  volume = {26},
  number = {1},
  pages = {014019},
  publisher = {American Physical Society},
  doi = {10.1103/m8nb-bbp8},
  url = {https://link.aps.org/doi/10.1103/m8nb-bbp8},
  urldate = {2026-09-04},
}

@article{zhangMagneticAnisotropy2021,
  title   = {Magnetic anisotropy in the van der Waals ferromagnet VI3},
  journal = {Physical Review B},
  volume  = {103},
  pages   = {174401},
  year    = {2021},
  doi     = {10.1103/PhysRevB.103.174401}
}

@article{wangFirstprinciplesCalculationOptical2019,
  title = {First-Principles Calculation of Optical Responses Based on Nonorthogonal Localized Orbitals},
  author = {Wang, Chong and Zhao, Sibo and Guo, Xiaomi and Ren, Xinguo and Gu, Bing-Lin and Xu, Yong and Duan, Wenhui},
  date = {2019-09},
  journaltitle = {New Journal of Physics},
  shortjournal = {New J. Phys.},
  volume = {21},
  number = {9},
  pages = {093001},
  publisher = {IOP Publishing},
  issn = {1367-2630},
  doi = {10.1088/1367-2630/ab3c9c},
  url = {https://dx.doi.org/10.1088/1367-2630/ab3c9c},
  urldate = {2024-07-01},
  langid = {english},
}

@article{chenLearningPropertiesOrdered2021,
  title = {Learning Properties of Ordered and Disordered Materials from Multi-Fidelity Data},
  author = {Chen, Chi and Zuo, Yunxing and Ye, Weike and Li, Xiangguo and Ong, Shyue Ping},
  date = {2021-01},
  journaltitle = {Nature Computational Science},
  shortjournal = {Nat. Comput. Sci.},
  volume = {1},
  number = {1},
  pages = {46--53},
  publisher = {Nature Publishing Group},
  issn = {2662-8457},
  doi = {10.1038/s43588-020-00002-x},
  url = {https://www.nature.com/articles/s43588-020-00002-x},
  urldate = {2026-09-04},
  langid = {english},
}

@article{deslippeBerkeleyGWMassivelyParallel2012,
  title = {{{BerkeleyGW}}: {{A}} Massively Parallel Computer Package for the Calculation of the Quasiparticle and Optical Properties of Materials and Nanostructures},
  shorttitle = {{{BerkeleyGW}}},
  author = {Deslippe, Jack and Samsonidze, Georgy and Strubbe, David A. and Jain, Manish and Cohen, Marvin L. and Louie, Steven G.},
  date = {2012-06-01},
  journaltitle = {Computer Physics Communications},
  shortjournal = {Comput. Phys. Commun.},
  volume = {183},
  number = {6},
  pages = {1269--1289},
  issn = {0010-4655},
  doi = {10.1016/j.cpc.2011.12.006},
  url = {https://www.sciencedirect.com/science/article/pii/S0010465511003912},
  urldate = {2026-09-04},
}

@article{heydErratumHybridFunctionals2006,
  title = {Erratum: ``{{Hybrid}} Functionals Based on a Screened {{Coulomb}} Potential'' [{{J}}. {{Chem}}. {{Phys}}. 118, 8207 (2003)]},
  shorttitle = {Erratum},
  author = {Heyd, Jochen and Scuseria, Gustavo E. and Ernzerhof, Matthias},
  date = {2006-06-07},
  journaltitle = {The Journal of Chemical Physics},
  shortjournal = {J. Chem. Phys.},
  volume = {124},
  number = {21},
  pages = {219906},
  issn = {0021-9606},
  doi = {10.1063/1.2204597},
  url = {https://doi.org/10.1063/1.2204597},
  urldate = {2026-09-04},
}

\end{document}

\title[Article Title]{Supplementary Information for Symmetry- and Property-Aware Crystal Generation with Reinforcement Learning for Inverse Materials Design}

\author*[1,2]{\fnm{Ting-Wei} \sur{Hsu}}\email{hsu.ting@northeastern.edu}

\author[1,2]{\fnm{Arun} \sur{Bansil}} 
\author*[1,2]{\fnm{Qimin} \sur{Yan}}\email{q.yan@northeastern.edu}

\affil[1]{\orgdiv{Department of Physics}, \orgname{Northeastern University}, \city{Boston}, \postcode{02155}, \state{Massachusetts}, \country{USA}}

\affil[2]{\orgdiv{Quantum Materials and Sensing Institute}, \orgname{Northeastern University}, \city{Burlington}, \postcode{01803}, \state{Massachusetts}, \country{USA}}

\maketitle

\section{Reinforcement-learning experimental setup}

This section provides the complete experimental workflow and optimization settings used for reinforcement-learning (RL) fine-tuning. SPARC uses the publicly released SymmCD model~\cite{levysymmcd}, pretrained on MP-20, as the initial generative model and RL policy. A frozen copy of the same checkpoint is retained as the reference policy $p_{\mathrm{pre}}$ for Kullback--Leibler (KL) regularization. Both design campaigns are run for $120$ RL rounds. Each round consists of the following steps.

\begin{enumerate}
    \item For each generation trajectory, a space group $G$ is sampled from the current proposal distribution $\pi^{(j)}(G)$, and the number of representative asymmetric-unit sites $M$ is sampled from $p_{\mathrm{marg}}(M\mid G)$. The initial proposal $\pi^{(0)}(G)$ is obtained from the MP-20 space-group marginal using the task-specific temperature transformation described in the main text. The proposal over $G$ is updated between RL rounds, while $p_{\mathrm{marg}}(M\mid G)$ remains fixed.

    \item Conditioned on the sampled pair $(G,M)$, the diffusion model generates asymmetric-unit candidates $\mathcal{M}'$ through the reverse denoising process. The uniaxial dielectric task generates $96$ candidates per round, while the spectroscopic limited maximum efficiency (SLME) task generates $128$ candidates per round. Each asymmetric-unit candidate is expanded using the space-group reconstruction operator $\mathcal{R}_G$ to obtain the corresponding full crystal,
    \begin{equation}
    \mathcal{M}=\mathcal{R}_G(\mathcal{M}').
    \end{equation}
    Generation and policy updates are performed on the compact asymmetric-unit representation. Relaxation, filtering, property prediction, and reward evaluation are performed on the reconstructed full crystal.

    \item The reconstructed structures are relaxed using the \texttt{MatterSim-v1.0.0-5M} machine-learning interatomic potential (MLIP)~\cite{yang2024mattersim}. Relaxation is performed with the FIRE optimizer under full variable-cell relaxation, with the atomic positions and lattice optimized jointly using an exponential cell filter. We use a force-convergence threshold of $f_{\max}=\SI{0.05}{\electronvolt\per\angstrom}$ and a maximum of $5000$ optimization steps.

    \item The relaxed structures are first checked for structural validity and then evaluated using the stable, unique, and novel (S.U.N.) criteria following MatterGen~\cite{zeniGenerativeModelInorganic2025a}. A structure is considered stable when its MLIP-estimated energy above the full Materials Project convex hull satisfies $E_{\mathrm{hull}}\leq\SI{0.10}{\electronvolt}/atom$~\cite{jainCommentaryMaterialsProject2013}. Uniqueness is evaluated by comparing the generated structures with one another using the pymatgen \texttt{StructureMatcher}. Novelty is evaluated by comparing the generated structures with the Materials Project reference set using the same matcher. Only structurally valid candidates satisfying all three S.U.N. criteria proceed to property evaluation.

    \item Up to $30$ surviving structures are evaluated by the fixed property surrogates in each RL round. When more than $30$ structures pass the filters, $30$ are selected for surrogate evaluation. Otherwise, all surviving structures are evaluated. Each scored structure is assigned the task-specific terminal reward defined in the main Methods. These are $r_{\mathrm{uni}}$ for the uniaxial dielectric task and $r_{\mathrm{SLME}}$ for the photovoltaic task.

    \item The terminal reward $r(\mathcal{R}_G(\mathcal{M}'_0))$ is assigned to the asymmetric unit that generated the relaxed full crystal. For each policy update, a diffusion time $t$ is sampled and the clean asymmetric unit $\mathcal{M}'_0$ is forward-noised to $\mathcal{M}'_t$. The policy is then updated using the reward-weighted denoising loss together with the KL penalty relative to $p_{\mathrm{pre}}$. The noising and denoising operations act on the asymmetric-unit variables $\boldsymbol{A}'$, $\boldsymbol{F}'$, $\boldsymbol{\Sigma}'$, and $\boldsymbol{k}$. The full crystal is recovered through $\mathcal{R}_G$, allowing the policy to improve the target reward while preserving compatibility with the conditioning space group.

    \item After reward evaluation and policy fine-tuning, the realized space group of each reconstructed structure is determined using \texttt{SpacegroupAnalyzer} with $\texttt{symprec}=0.01$~\AA{}. The rewards are grouped by the realized space-group labels and used to update $\pi^{(j)}(G)$ according to the outer-loop rule defined in the main text. Space groups that repeatedly yield high-reward structures receive increased proposal probability. The base-distribution anchor and minimum-probability floor retain sampling probability across the original support. The outer-loop parameters are the proposal temperature $T_{\mathrm{sg}}$, the reward-decay factor $\gamma$, the tilt strength $\kappa$, and the base-proposal mixture $\rho$, set to $T_{\mathrm{sg}}=100$, $\gamma=0.9$, $\kappa=5.0$, $\rho=0.15$ for the dielectric task and $T_{\mathrm{sg}}=1$, $\gamma=0.9$, $\kappa=3.0$, $\rho=0.1$ for the SLME task. The updated proposal is used in the next RL round.
\end{enumerate}

The filtering stage can substantially reduce the number of structures available for property evaluation. Candidates may be removed because of invalid geometries, insufficient stability, duplication within the generated population, or matches to known Materials Project structures. We therefore generate more structures than are used in each policy update. The dielectric task generates $96$ candidates per round, while the SLME task generates $128$. At most $30$ surviving structures are evaluated and used for fine-tuning in each round. Over $120$ rounds, the SLME campaign generated $15{,}360$ structures. Of these, $4{,}825$ passed the validity checks and $3{,}522$ entered the capped per-round evaluation pool and were scored by the property surrogates.

To improve sample efficiency, high-reward structures from previous rounds are reused through an experience-replay buffer. The buffer stores up to $100$ structures, and $10$ stored structures are replayed in each RL round. Only structures with reward $r\geq0.1$ are eligible to enter or be sampled from the buffer. A diversity filter is also applied to reduce the influence of near-duplicate candidates. The filter uses the implementation parameters of tolerance $3$ and buffer size $6$.

Fine-tuning retains the diffusion settings of the pretrained SymmCD model. The diffusion process contains $T=1000$ steps. A cosine variance schedule with offset $s=0.008$ is used for the lattice coefficients $\boldsymbol{k}$ and the discrete atom-type and site-symmetry variables. A geometric noise-scale schedule from $\sigma_{\min}=0.005$ to $\sigma_{\max}=0.5$ is used for the fractional coordinates.

The initial learning rate is $3\times10^{-5}$ for both tasks. It remains constant during dielectric fine-tuning and is multiplied by $0.98$ after each RL round for the SLME task. The KL coefficient is initialized at $\beta_0=0.025$ and decays with the fine-tuning update index $\nu$ according to
\begin{equation}
\beta_\nu=\frac{\beta_0}{\sqrt{1+\nu/50}}.
\end{equation}
The policy-gradient coefficient is $\alpha=1$, and the reward-anchor threshold is $\lambda=1.1$. Each RL round contains three inner fine-tuning epochs. The training mini-batch size is $30$, and gradients are accumulated over $50$ steps before each optimizer update.

The property surrogates remain frozen throughout RL fine-tuning. The uniaxial dielectric reward uses the band-gap and dielectric-tensor surrogates, while the SLME reward uses the band-gap and spectral surrogates. Their architectures, training datasets, and predictive benchmarks are described separately in Section~\ref{sec:surrogate}.

\begin{table}[h]
\centering
\small
\caption{\textbf{Reinforcement-learning optimization settings.} Shared and task-specific settings used for SPARC fine-tuning. Both campaigns were run for $120$ RL rounds.}
\label{tab:rl-training}
\begin{tabular}{p{0.3\linewidth}p{0.25\linewidth}p{0.15\linewidth}p{0.15\linewidth}}
\toprule
Setting & Shared & Dielectric RL & SLME RL \\
\midrule
Initial policy & SymmCD MP-20 checkpoint & -- & -- \\
Reference policy & Frozen initial checkpoint & -- & -- \\
Optimizer & Adam & -- & -- \\
Initial learning rate & -- & $3\times10^{-5}$ & $3\times10^{-5}$ \\
Learning-rate multiplier per round & -- & $1.0$ & $0.98$ \\
Weight decay & $0$ & -- & -- \\
Diffusion steps $T$ & $1000$ & -- & -- \\
Cosine-schedule offset $s$ & $0.008$ & -- & -- \\
Coordinate noise range & $\sigma_{\min}=0.005$, $\sigma_{\max}=0.5$ & -- & -- \\
Denoising loss weights & $\lambda_{\boldsymbol{k}}{=}5$, $\lambda_{\boldsymbol{F}'}{=}1$, $\lambda_{\boldsymbol{A}'}{=}0.1$, $\lambda_{\boldsymbol{\Sigma}'}{=}10$ & -- & -- \\
Training mini-batch size & $30$ & -- & -- \\
Gradient accumulation steps & $50$ & -- & -- \\
Inner fine-tuning epochs per round & $3$ & -- & -- \\
RL rounds & $120$ & -- & -- \\
Generated candidates per round & -- & $96$ & $128$ \\
Total generated candidates & -- & $11{,}520$ & $15{,}360$ \\
Scored candidates per round & $30$ & -- & -- \\
Policy-gradient coefficient $\alpha$ & $1$ & -- & -- \\
Initial KL coefficient $\beta_0$ & $0.025$ & -- & -- \\
KL schedule & $\beta_\nu=\beta_0/\sqrt{1+\nu/50}$ & -- & -- \\
Reward-anchor threshold $\lambda$ & $1.1$ & -- & -- \\
Replay-buffer capacity & $100$ & -- & -- \\
Replayed structures per round & $10$ & -- & -- \\
Replay reward threshold & $0.1$ & -- & -- \\
\bottomrule
\end{tabular}
\end{table}

\newpage
\section{Symmetry-constrained generation}

SPARC inherits its symmetry-constrained generation from SymmCD~\cite{levysymmcd}. The central idea is to
diffuse not only over atom types, fractional coordinates, and the lattice, but also over the site
symmetries of the representative atoms, conditioned on the space group $G$. Because the asymmetric unit is
generated together with its site-symmetry assignments, applying the operations of $G$ regenerates a crystal
that respects $G$ by construction, rather than one whose symmetry must be recovered after generation. This
decomposition is illustrated on a two-dimensional toy example in Supplementary
Figure~\ref{fig:symmcd-decomposition}.

\begin{figure}[H]
    \centering
    \includegraphics[width=1.0\linewidth]{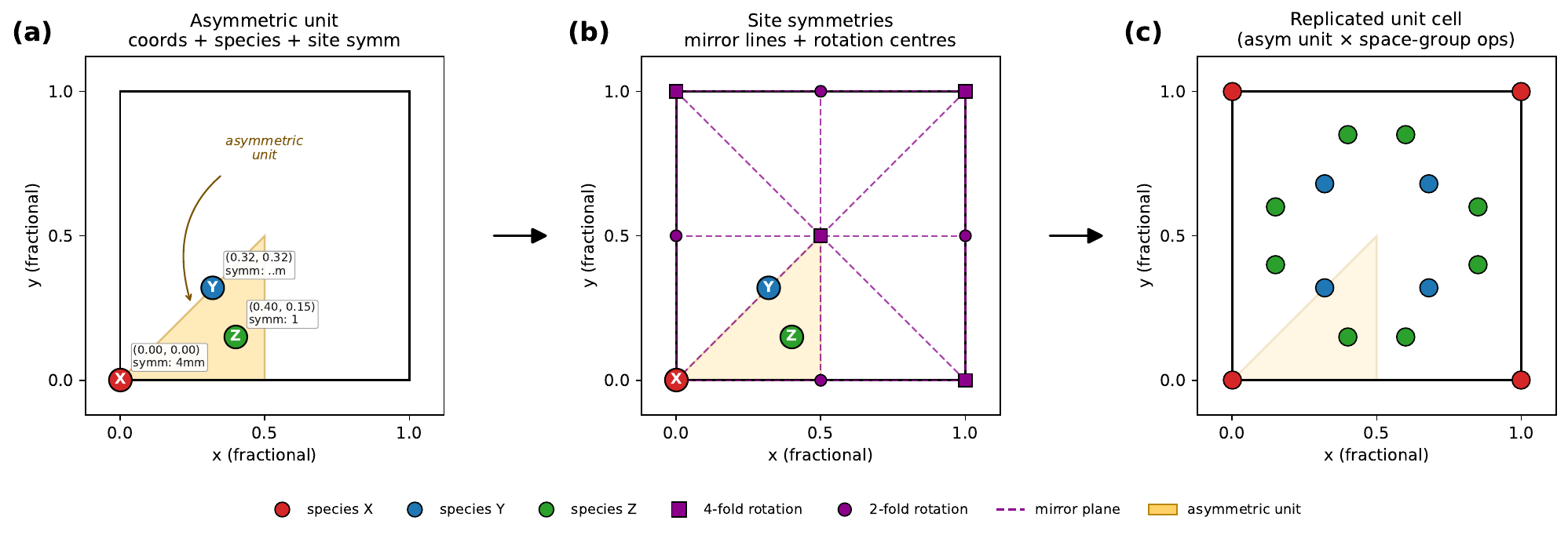}
    \caption[Symmetry-constrained decomposition of a crystal]{
    \textbf{Symmetry-constrained decomposition of a crystal}. On a two-dimensional toy example (wallpaper group $p4mm$, fictitious species X, Y, Z). \textbf{(a)}~The asymmetric unit, comprising the representative atoms with their fractional coordinates, species, and site symmetries (X on a $4mm$ site, Y on a mirror line, Z on a general site). \textbf{(b)}~The corresponding space-group symmetry elements, the mirror lines and rotation centres. \textbf{(c)}~Applying these operations to the asymmetric unit replicates it across the unit cell and recovers the full crystal. SPARC generates only the asymmetric unit in \textbf{(a)} and reconstructs the complete structure through \textbf{(b)} and \textbf{(c)}, so every generated crystal is symmetry-valid by construction.
    }
    \label{fig:symmcd-decomposition}
\end{figure}

\begin{figure}[H]
    \centering
    \includegraphics[width=0.5\linewidth]{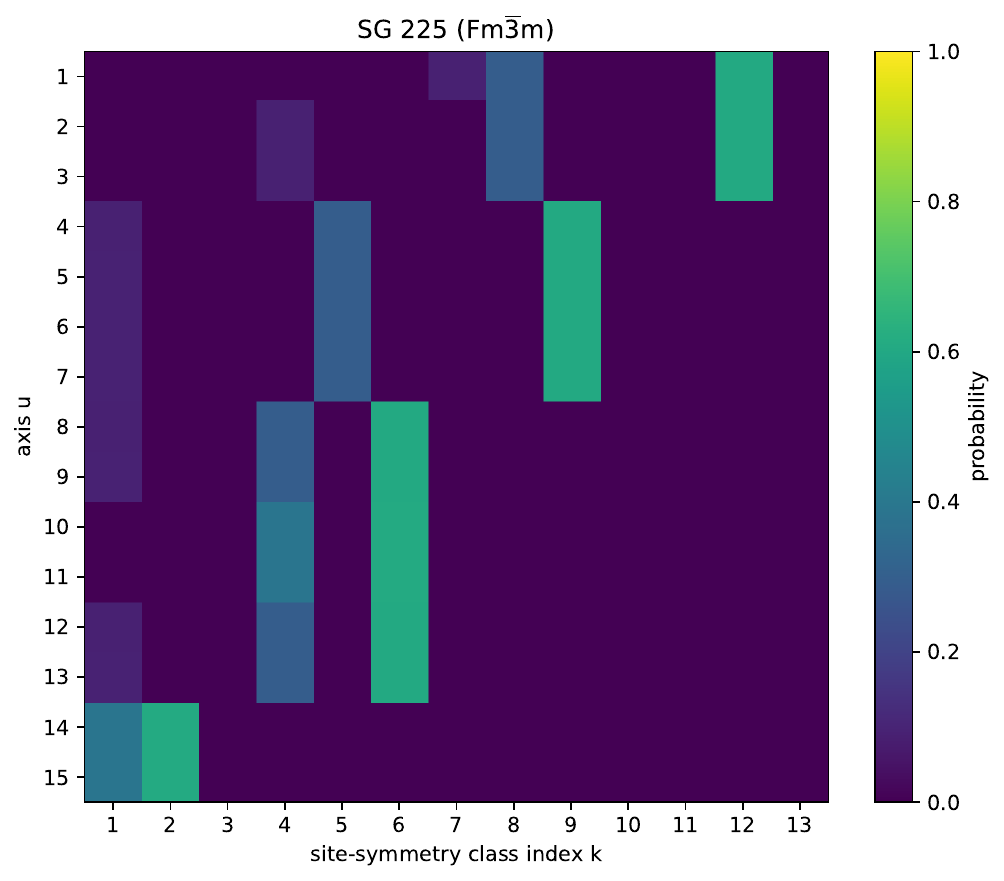}
    \caption[Site-symmetry marginal for space group 225]{
    \textbf{Site-symmetry marginal for space group 225 ($Fm\bar{3}m$).}
    Empirical site-symmetry marginal $p_{\mathrm{marg}}(\boldsymbol{\Sigma}'\mid G{=}225)$, aggregated over all training materials in $Fm\bar{3}m$ and shown as a $15\times13$ heatmap over the $15$ crystallographic axes and $13$ site-symmetry categories. Color encodes probability.
    }
    \label{fig:sg225-site-symm-marginal}
\end{figure}

The symmetry enters the generator through a data-driven prior. For a given space group $G$ we aggregate the site-symmetry assignments of every training material in $G$ to form the site-symmetry marginal $p_{\mathrm{marg}}(\boldsymbol{\Sigma}'\mid G)$. For example, $G=225$ ($Fm\bar{3}m$), the marginal $p_{\mathrm{marg}}(\boldsymbol{\Sigma}'\mid G=225)$ is naturally displayed as a $15\times13$ heatmap over the $15$ crystallographic axes and $13$ site-symmetry categories, with color encoding probability (Supplementary Figure~\ref{fig:sg225-site-symm-marginal}).

The site-symmetry representation $\boldsymbol{\Sigma}'$ is treated as a collection of categorical variables. 
For each representative atom $i$, the site symmetry is represented by 15 one-hot vectors 
$\boldsymbol{\Sigma}'_{i,u}\in\{0,1\}^{13}$, where $u\in\{1,\ldots,15\}$ indexes the crystallographic symmetry 
axis and the 13 categories correspond to the possible site-symmetry operations along that axis. 
We apply discrete diffusion independently to each axis. Given the initial site-symmetry representation 
$\boldsymbol{\Sigma}'_{0,i,u}$ and space group $G$, the noising process is defined as
$$
q(\boldsymbol{\Sigma}'_{t,i,u}\mid \boldsymbol{\Sigma}'_{0,i,u},G)
=
\mathrm{Cat}
\left(
\boldsymbol{\Sigma}'_{t,i,u};
\boldsymbol{p}
=
\boldsymbol{\Sigma}'_{0,i,u}{}^{\top}
\overline{\boldsymbol{Q}}_{t,u,G}
\right),
$$
with the cumulative transition matrix
$$
\overline{\boldsymbol{Q}}_{t,u,G}
=
\prod_{\tau=1}^{t}
\boldsymbol{Q}_{\tau,u,G},
\qquad
\boldsymbol{Q}_{t,u,G}
=
\alpha_t\boldsymbol{I}
+
\beta_t\mathbf{1}\boldsymbol{m}_{\Sigma,u,G}^{\top}.
$$
Here, $\boldsymbol{m}_{\Sigma,u,G}$ denotes the empirical marginal distribution of site-symmetry operations 
for axis $u$ within space group $G$, so the forward process drives each axis from its clean assignment
toward this space-group-specific marginal (Supplementary Figure~\ref{fig:sg225-site-symm-marginal}). The
denoising network predicts the clean site-symmetry distribution 
$\hat{\boldsymbol{\Sigma}}'_0=\phi_{\Sigma}(\mathcal{M}'_t,t,G)$ and is trained using the averaged cross-entropy loss
$$
\mathcal{L}_{\Sigma'}
=
\mathbb{E}_{t\sim\mathcal{U}(1,T)}
\mathbb{E}_{\boldsymbol{\Sigma}'_t
\sim q(\boldsymbol{\Sigma}'_t\mid\boldsymbol{\Sigma}'_0,G)}
\left[
\frac{1}{15M}
\sum_{i=1}^{M}
\sum_{u=1}^{15}
\operatorname{CrossEntropy}
\left(
\boldsymbol{\Sigma}'_{0,i,u},
\hat{\boldsymbol{\Sigma}}'_{0,i,u}
\right)
\right].
$$
where $M$ is the number of representative atoms in the asymmetric unit.

\begin{figure}[H]
    \centering
    \includegraphics[width=1.0\linewidth]{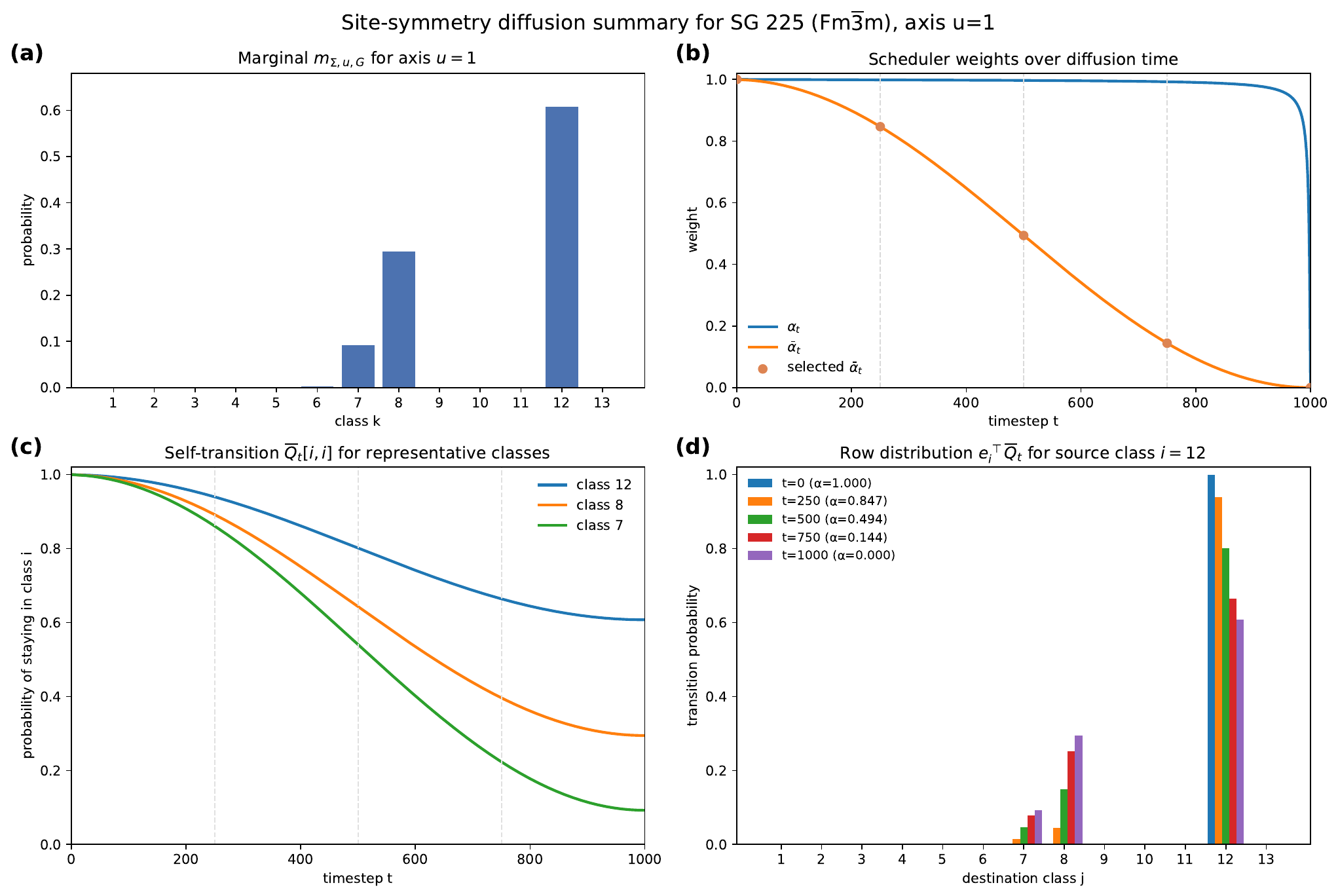}
    \caption[Discrete-diffusion transition toward the site-symmetry marginal]{
    \textbf{Discrete-diffusion transition toward the site-symmetry marginal (space group 225, $Fm\bar{3}m$, axis $u=1$).}
    The forward process for the site-symmetry labels along one crystallographic axis, shown through the
    cumulative transition $\overline{\boldsymbol{Q}}_{t,u,G}$ at diffusion steps $t=0,250,500,750,1000$.
    \textbf{(a)}~The space-group marginal $\boldsymbol{m}_{\Sigma,u,G}$ over the $13$ site-symmetry
    classes, the absorbing prior the forward process relaxes onto; for this axis the mass concentrates
    on a few classes.
    \textbf{(b)}~The noise-schedule weights $\alpha_t$ and the cumulative $\bar\alpha_t$ over diffusion
    time, with the five sampled timesteps marked; $\bar\alpha_t$ decreases from $1$ to $0$ as the clean
    signal is replaced by the marginal.
    \textbf{(c)}~The self-transition probability $\overline{\boldsymbol{Q}}_{t}[i,i]$, the probability of
    remaining in the initial class $i$, versus $t$ for three representative classes; each decays from $1$
    toward its marginal weight.
    \textbf{(d)}~The row distribution $\boldsymbol{e}_i^\top\overline{\boldsymbol{Q}}_{t}$ for source
    class $i=12$ at the five timesteps, showing how probability initially localized on class $12$ spreads
    toward the marginal as $t$ grows.
    }
    \label{fig:sg225-qt-evolution}
\end{figure}

Supplementary Figure~\ref{fig:sg225-qt-evolution} makes this forward process concrete for a single crystallographic axis of space group 225. Panel (a) showing the initial prior is the space-group marginal $\boldsymbol{m}_{\Sigma,u,G}$ at axis $u=1$, which for this axis places its mass on only site-symmetry classes number 7, 8, and 12. Panel (b) showing the noise schedule drives the cumulative weight $\bar\alpha_t$ from $1$ to $0$ over the trajectory, interpolating from the clean assignment at $t=0$ toward this marginal at $t=T$. 
Panels (c) and (d) trace the resulting relaxation. Panel (c) shows that the self-transition probability $\overline{\boldsymbol{Q}}_t[i,i]$ decays from 1 as $t$ increases, so an initially one-hot class progressively transfers probability to other allowed classes. Panel (d) shows the full row distribution $\boldsymbol{e}_i^\top \overline{\boldsymbol{Q}}_t$ for a representative source class, $i=12$. As time progresses, this distribution evolves from a sharp peak at the initial class toward the space-group marginal, with probability redistributed only among the symmetry classes supported by the marginal. Because this absorbing distribution is space-group-specific, the forward noise process is itself symmetry-aware. During sampling, the reverse denoising model learns to invert this process, mapping noisy site-symmetry assignments back toward configurations compatible with the target space group $G$. Thus, the model samples within a symmetry-conditioned space rather than an unconstrained categorical space. Results in the generated structures preserve local site-symmetry assignments consistent with the prescribed global symmetry.

\subsection{Lattice parameterization and space-group constraints}
SPARC inherits its lattice parameterization from DiffCSP++~\cite{jiaoSpaceGroupConstrained2024}.
Rather than denoising the raw $3\times3$ lattice matrix $\boldsymbol{L}$, we use the polar
decomposition $\boldsymbol{L}=\boldsymbol{Q}\exp(\boldsymbol{S})$, where $\boldsymbol{Q}\in SO(3)$ is
orthogonal, $\boldsymbol{S}=\boldsymbol{S}^{\top}$ is symmetric, and
$\exp(\boldsymbol{S})=\sum_{n=0}^{\infty}\boldsymbol{S}^{n}/n!$ is the matrix exponential. Any rigid
rotation of $\boldsymbol{L}$ is absorbed entirely by $\boldsymbol{Q}$ and leaves $\boldsymbol{S}$
unchanged, so $\boldsymbol{S}$ furnishes a rotation-invariant description of the lattice. Every
symmetric $\boldsymbol{S}$ expands uniquely in a fixed six-dimensional symmetric basis,
\begin{equation}
\boldsymbol{S}=\sum_{i=1}^{6}k_i\boldsymbol{B}_i,\qquad
\boldsymbol{k}=(k_1,\dots,k_6)\in\mathbb{R}^6,
\end{equation}
with basis matrices
\begin{equation}
\begin{aligned}
\boldsymbol{B}_1=\left(\begin{array}{lll}
0 & 1 & 0 \\ 1 & 0 & 0 \\ 0 & 0 & 0
\end{array}\right), &\boldsymbol{B}_2=\left(\begin{array}{lll}
0 & 0 & 1 \\ 0 & 0 & 0 \\ 1 & 0 & 0
\end{array}\right), \boldsymbol{B}_3=\left(\begin{array}{lll}
0 & 0 & 0 \\ 0 & 0 & 1 \\ 0 & 1 & 0
\end{array}\right), \\
\boldsymbol{B}_4=\left(\begin{array}{ccc}
1 & 0 & 0 \\ 0 & -1 & 0 \\ 0 & 0 & 0
\end{array}\right), &\boldsymbol{B}_5=\left(\begin{array}{ccc}
1 & 0 & 0 \\ 0 & 1 & 0 \\ 0 & 0 & -2
\end{array}\right), \boldsymbol{B}_6=\left(\begin{array}{lll}
1 & 0 & 0 \\ 0 & 1 & 0 \\ 0 & 0 & 1
\end{array}\right) .
\end{aligned}
\end{equation}
For a chosen space group $G$, the crystal family fixes a subset of the $k_i$ while the rest remain
free, as listed in Supplementary Table~\ref{tab:spacegroup-constraint}. During generation we hold the
constrained coefficients at their required values through a mask $\boldsymbol{m}\in\{0,1\}^6$ and
denoise only the free entries, guaranteeing that every generated lattice is consistent with $G$.

\begin{table}[t]
\centering
\caption{Relationship between the lattice shape and the constraint on the symmetric-basis
coefficients $\boldsymbol{k}$, where $a, b, c$ and $\alpha, \beta, \gamma$ denote the lengths and
angles of the lattice bases, respectively (after DiffCSP++~\cite{jiaoSpaceGroupConstrained2024}).}
\label{tab:spacegroup-constraint}
\begin{tabular}{llll}
\toprule
Crystal Family & Space Group No. & Lattice Shape & Constraint of Symmetric Bases \\
\midrule
Triclinic & $1 \sim 2$ & No Constraint & No Constraint \\
\midrule
Monoclinic & $3 \sim 15$ & $\alpha = \gamma = 90^\circ$ & $k_1 = k_3 = 0$ \\
\midrule
Orthorhombic & $16 \sim 74$ & $\alpha = \beta = \gamma = 90^\circ$ & $k_1 = k_2 = k_3 = 0$ \\
\midrule
Tetragonal & $75 \sim 142$
  & \makecell[l]{$\alpha = \beta = \gamma = 90^\circ$ \\ $a = b$}
  & \makecell[l]{$k_1 = k_2 = k_3 = 0$ \\ $k_4 = 0$} \\
\midrule
Trigonal/Hexagonal & $143 \sim 194$
  & \makecell[l]{$\alpha = \beta = 90^\circ,\ \gamma = 120^\circ$ \\ $a = b$}
  & \makecell[l]{$k_2 = k_3 = 0,\ k_1 = -\log(3)/4$ \\ $k_4 = 0$} \\
\midrule
Cubic & $195 \sim 230$
  & \makecell[l]{$\alpha = \beta = \gamma = 90^\circ$ \\ $a = b = c$}
  & \makecell[l]{$k_1 = k_2 = k_3 = 0$ \\ $k_4 = k_5 = 0$} \\
\bottomrule
\end{tabular}
\end{table}

\subsection{Denoising losses}
Following SymmCD~\cite{levysymmcd}, the denoiser reconstructs the four asymmetric-unit components
jointly, minimizing
$\mathcal{L}=\lambda_{\boldsymbol{k}}\mathcal{L}_{\boldsymbol{k}}
+\lambda_{\boldsymbol{F}'}\mathcal{L}_{\boldsymbol{F}'}
+\lambda_{\boldsymbol{A}'}\mathcal{L}_{\boldsymbol{A}'}
+\lambda_{\boldsymbol{\Sigma}'}\mathcal{L}_{\boldsymbol{\Sigma}'}$,
with the weights listed in Supplementary Table~\ref{tab:rl-training}.
The fractional coordinates use a denoising score-matching loss,
\begin{equation}
\mathcal{L}_{\boldsymbol{F}^{\prime}}=\mathbb{E}_{\boldsymbol{F}_t^{\prime} \sim q^{\prime}\left(\boldsymbol{F}_t^{\prime} \mid \boldsymbol{F}_0^{\prime}\right), t \sim \mathcal{U}(1, T)}\left[\lambda_t\left\|\nabla_{\boldsymbol{F}_t^{\prime}} \log q^{\prime}\left(\boldsymbol{F}_t^{\prime} \mid \boldsymbol{F}_0^{\prime}\right)-\hat{\boldsymbol{\epsilon}}_{\boldsymbol{F}^{\prime}}\left(\mathcal{M}_t^{\prime}, t\right)\right\|_2^2\right],
\end{equation}
the lattice coefficients a masked $\boldsymbol{\epsilon}$-prediction loss, in which the mask
$\boldsymbol{m}$ zeroes the symmetry-constrained entries,
\begin{equation}
\mathcal{L}_{\boldsymbol{k}}=\mathbb{E}_{\boldsymbol{\epsilon}_{\boldsymbol{k}} \sim \mathcal{N}(0, \boldsymbol{I}), t \sim \mathcal{U}(1, T)}\left[\left\|\boldsymbol{m} \odot \boldsymbol{\epsilon}_{\boldsymbol{k}}-\hat{\boldsymbol{\epsilon}}_{\boldsymbol{k}}\left(\mathcal{M}_t^{\prime}, t\right)\right\|_2^2\right],
\end{equation}
and the atom types a categorical cross-entropy under the discrete-diffusion transition
$\overline{\boldsymbol{Q}}_t$,
\begin{equation}
\mathcal{L}_{\boldsymbol{A}^{\prime}}=\mathbb{E}_{\boldsymbol{a}_t \sim \operatorname{Cat}\left(\boldsymbol{a}_0^{\top} \overline{\boldsymbol{Q}}_t\right), t \sim \mathcal{U}(1, T)} \sum_{i=1}^M \operatorname{CrossEntropy}\left(\boldsymbol{a}_i^{\prime}, \hat{\boldsymbol{a}}_i^{\prime}\right).
\end{equation}
The site-symmetry loss $\mathcal{L}_{\boldsymbol{\Sigma}'}$ has the same cross-entropy form and is
defined earlier in this section.

\subsection{Denoising-network architecture}

SPARC retains the denoising architecture of the released SymmCD model~\cite{levysymmcd}, which is adapted from the graph neural network used in DiffCSP~\cite{jiao2023diffcsp}. The network operates on a fully connected graph whose $M$ nodes correspond to the representative sites in the asymmetric unit. Following SymmCD, the denoiser is not explicitly rotation-equivariant because the conventional unit-cell axes define a canonical reference frame. Periodic translation invariance is incorporated through Fourier features of the relative fractional coordinates.

For representative site $i$, the initial node feature $\boldsymbol{h}_i$ is constructed from its atom type $\boldsymbol{a}_i$, fractional coordinate $\boldsymbol{x}_i$, site-symmetry descriptor $\boldsymbol{\Sigma}'_i$, the space group $G$, and the diffusion timestep $t$. The timestep is encoded using a sinusoidal embedding $\psi_t(t)$, while the space group is embedded using a multilayer perceptron (MLP) $\phi_G$. The site-symmetry descriptor is embedded separately along each of the 15 crystallographic axes using a shared MLP $\phi_U$, and the resulting axis-wise embeddings are combined using a second MLP $\phi_S$,
\begin{equation}
\boldsymbol{h}_i
\leftarrow
\phi_h\left(
\boldsymbol{a}_i,
\boldsymbol{x}_i,
\phi_S\left(
\bigoplus_{u=1}^{15}
\phi_U\left(\boldsymbol{\Sigma}'_{i,u}\right)
\right),
\phi_G(G),
\psi_t(t)
\right).
\end{equation}
At each message-passing layer, the message from representative $j$ to representative $i$ is computed as
\begin{equation}
\boldsymbol{m}_{ij}
=
\phi_m\left(
\boldsymbol{h}_i,
\boldsymbol{h}_j,
\boldsymbol{k},
\psi_x\left(\boldsymbol{x}_i-\boldsymbol{x}_j\right)
\right),
\end{equation}
where $\psi_x$ is a Fourier embedding of the relative fractional coordinate. The node features are then updated by aggregating the incoming messages,
\begin{equation}
\boldsymbol{h}_i
\leftarrow
\boldsymbol{h}_i
+
\phi_n\left(
\boldsymbol{h}_i,
\sum_{j=1}^{M}\boldsymbol{m}_{ij}
\right).
\end{equation}
The edge and node networks $\phi_m$ and $\phi_n$ are MLPs with SiLU activations, and layer normalization is applied at each message-passing layer.

After the final layer, node-level output heads predict the fractional-coordinate denoising term, atom-type logits, and site-symmetry logits. The site-symmetry output contains one categorical distribution over 13 possible operations for each of the 15 crystallographic axes. The lattice denoising term is predicted from the aggregated graph representation,
\begin{equation}
\boldsymbol{h}_{\mathrm{graph}}
=
\sum_{i=1}^{M}\boldsymbol{h}_i.
\end{equation}
The resulting outputs are denoted by
$\hat{\boldsymbol{\epsilon}}_{\boldsymbol{F}'},
\hat{\boldsymbol{A}}',
\hat{\boldsymbol{\Sigma}}',
\hat{\boldsymbol{\epsilon}}_{\boldsymbol{k}}$.
The architecture and diffusion hyperparameters are summarized in Supplementary Table~\ref{tab:arch}.
\begin{table}[h]
\centering
\caption{Denoising-network architecture hyperparameters (inherited from the SymmCD MP-20 model).}
\label{tab:arch}
\begin{tabular}{ll}
\toprule
Component & Value \\
\midrule
Backbone & CSPNet message-passing graph neural network (fully connected over representatives) \\
Message-passing layers & 8 \\
Hidden / representation dim. & 1024 \\
Timestep embedding dim. & 10 \\
Coordinate Fourier frequencies & 128 \\
Activation & SiLU \\
Normalization & LayerNorm (each layer) \\
Lattice parameterization & $\boldsymbol{k}\in\mathbb{R}^{6}$ \\
Site-symmetry dimension & $15\times13=195$ \\
Atom-type  & 94 \\
Prediction heads (linear) & $\boldsymbol{k}$ (6), $\boldsymbol{F}'$ (3), $\boldsymbol{A}'$ (94), $\boldsymbol{\Sigma}'$ (195) \\
Parameters & 61,010,350 \\
\bottomrule
\end{tabular}
\end{table}

\section{Reinforcement-learning fine-tuning objective}
This section reproduces the reward-weighted KL objective of
MatInvent~\cite{chenAcceleratingInverseMaterials2025}, see also the reinforcement-learning treatments
of diffusion samplers in \cite{blackTrainingDiffusionModels2023,fanDPOKReinforcementLearning2023}.
In SPARC the diffusion policy acts on the asymmetric unit rather than the full crystal, so we write the
objective directly in the asymmetric-unit representation
$\mathcal{M}'=(\boldsymbol{A}',\boldsymbol{F}',\boldsymbol{\Sigma}',\boldsymbol{k})$. Given a space group
$G$, a generated asymmetric unit is decoded into a full crystal by the deterministic space-group
replication map $\mathcal{R}_G$, $\mathcal{M}=\mathcal{R}_G(\mathcal{M}')$, and the reward is evaluated
on this decoded crystal. Because $\mathcal{R}_G$ is fixed and non-learned, a reward on $\mathcal{M}$ acts
as a reward on the $\mathcal{M}'$ that generated it, and the entire trajectory, policy, and
regularizer below live in $\mathcal{M}'$-space (all conditioned on $G$, which we suppress in the
notation for brevity).

We treat the length-$T$ reverse denoising trajectory $\mathcal{M}'_{0:T}$ as a Markov decision process
whose actions are the per-step transitions $p_\theta(\mathcal{M}'_{t-1}\mid\mathcal{M}'_t)$, and assign
the terminal reward $r(\mathcal{R}_G(\mathcal{M}'_0))$ to the decoded crystal. The fine-tuning objective is
\begin{equation}
\min_\theta\mathbb{E}_{p_\theta(\mathcal{M}'_0)}\left[-r(\mathcal{R}_G(\mathcal{M}'_0))\right].
\end{equation}
Since $\log p_\theta(\mathcal{M}'_0)$ is intractable, we differentiate through the trajectory
likelihood using the score-function identity
$\nabla_\theta p_\theta=p_\theta\,\nabla_\theta\log p_\theta$, giving the policy gradient
\begin{equation}
\nabla_\theta\,\mathbb{E}_{p_\theta(\mathcal{M}'_0)}\left[-r(\mathcal{R}_G(\mathcal{M}'_0))\right]
=\mathbb{E}_{p_\theta(\mathcal{M}'_{0:T})}\left[r(\mathcal{R}_G(\mathcal{M}'_0))\sum_{t=1}^{T}\bigl(-\nabla_\theta\log p_\theta(\mathcal{M}'_{t-1}\mid\mathcal{M}'_t)\bigr)\right],
\end{equation}
i.e.\ the per-step denoising-loss gradient weighted by the terminal reward.

To keep the fine-tuned policy near the pretrained generator we regularize the marginal divergence
$\mathrm{KL}(p_\theta(\mathcal{M}'_0)\,\|\,p_{\mathrm{pre}}(\mathcal{M}'_0))$. This marginal is
intractable, but the data-processing inequality bounds it by the divergence of the full trajectories,
\begin{equation}
\mathrm{KL}\bigl(p_\theta(\mathcal{M}'_0)\,\|\,p_{\mathrm{pre}}(\mathcal{M}'_0)\bigr)
\le \mathrm{KL}\bigl(p_\theta(\mathcal{M}'_{0:T})\,\|\,p_{\mathrm{pre}}(\mathcal{M}'_{0:T})\bigr),
\end{equation}
and the Markov factorization of the reverse process reduces the right-hand side to a sum of per-step
KL divergences,
\begin{equation}
\mathrm{KL}\bigl(p_\theta(\mathcal{M}'_{0:T})\,\|\,p_{\mathrm{pre}}(\mathcal{M}'_{0:T})\bigr)
=\sum_{t=1}^{T}\mathbb{E}_{p_\theta(\mathcal{M}'_t)}\,\mathrm{KL}\bigl(p_\theta(\mathcal{M}'_{t-1}\mid\mathcal{M}'_t)\,\|\,p_{\mathrm{pre}}(\mathcal{M}'_{t-1}\mid\mathcal{M}'_t)\bigr),
\end{equation}
where the shared terminal prior $p_\theta(\mathcal{M}'_T)=p_{\mathrm{pre}}(\mathcal{M}'_T)$ eliminates the
boundary term. Weighting each trajectory's per-step KL by $(\lambda-r(\mathcal{R}_G(\mathcal{M}'_0)))$
yields the reward-weighted regularizer
\begin{equation}
\mathbb{E}_{p_\theta(\mathcal{M}'_{0:T})}\left[\bigl(\lambda-r(\mathcal{R}_G(\mathcal{M}'_0))\bigr)\sum_{t=1}^{T}\mathrm{KL}\bigl(p_\theta(\mathcal{M}'_{t-1}\mid\mathcal{M}'_t)\,\|\,p_{\mathrm{pre}}(\mathcal{M}'_{t-1}\mid\mathcal{M}'_t)\bigr)\right],
\end{equation}
where $\lambda$ is chosen above the maximum reward so that $\lambda-r(\mathcal{R}_G(\mathcal{M}'_0))$ remains positive. The KL penalty decreases as the reward increases. High-reward trajectories are therefore allowed to deviate more from $p_{\mathrm{pre}}$, whereas low-reward trajectories remain more strongly constrained. Combining this regularization term with the policy-gradient contribution using weights $\alpha,\beta>0$ gives the SPARC training gradient reported in the main text.

\section{Property surrogate models}\label{sec:surrogate}
This section documents the three property surrogates used as reward models in the inverse-design loop, a band-gap $E_g$ predictor, an SLME $\eta$ predictor, and a static-limit dielectric-tensor predictor (hereafter the dielectric-tensor surrogate). All three use the same Tensorial Spectra Equivariant Neural Network (TSENN)~\cite{hsuAccuratePredictionTensorial2026} backbone but differ in the output irreducible representations (irreps), training dataset, and task-specific hyperparameters. The output irreps are chosen to match the symmetry of each target property. Although pretrained weights are publicly available for some related tasks, we retrain a TSENN model from scratch on every task so that all three surrogates share a common \texttt{e3nn}-based backbone and a consistent evaluation protocol, allowing us to swap target properties by changing only the output head, without maintaining a separate model definition per objective. The trained surrogate weights are released alongside the code repository.

\subsection{Band gap}
For the band gap, the output head is a single scalar irrep (\texttt{1x0e}), so the network predicts the density functional theory (DFT) band gap $E_{g}$ as one real number per material.

\paragraph{Dataset.}
The training data are drawn from the Materials Project dataset~\cite{jainCommentaryMaterialsProject2013}, which contains $151,330$ entries. The full dataset is not uniformly labeled for every target property, particularly, band-gap labels are available only for a subset of the structures. We therefore restrict the band-gap surrogate to entries with valid band-gap targets. The band-gap labels were obtained from VASP calculations using Materials Project-consistent Perdew--Burke--Ernzerhof (PBE) $+U$ settings. Full details of the underlying DFT protocols are provided in the original dataset references. As these are semilocal PBE(+U) gaps, they underestimate experimental band gaps as expected at this level of theory, and the surrogate predicts on the same PBE scale.

\paragraph{Train/validation/test splits.}
We stratified the dataset into an 8:1:1 ratio, resulting in $121,065$ for training, $15,138$ for validation, and $15,127$ for testing.

\paragraph{Training.}
Learning rate $1.5\times 10^{-3}$, weight decay $5\times 10^{-2}$; remaining hyperparameters in Supplementary Table~\ref{tab:tsenn-bandgap-hparams}.
\begin{table}[h]
\centering
\caption{Hyperparameters for the TSENN band-gap surrogate.}
\label{tab:tsenn-bandgap-hparams}
\setlength{\tabcolsep}{12pt}
\renewcommand{\arraystretch}{1.25}
\begin{tabular}{p{0.55\linewidth} p{0.15\linewidth}}
\toprule
Hyperparameter & Value \\
\midrule
Maximum spherical harmonic order $\ell_{\max}$ & 2 \\
Embedding dimension & 128 \\
Number of layers & 4 \\
Channel multiplicity & 16 \\
Batch size & 128 \\
\bottomrule
\end{tabular}
\end{table}

The surrogate attains $R^{2}=0.902$ with mean absolute error $\mathrm{MAE}=0.223\,\mathrm{eV}$ on the training set, $R^{2}=0.869$ with $\mathrm{MAE}=0.265\,\mathrm{eV}$ on the validation set, and $R^{2}=0.881$ with $\mathrm{MAE}=0.252\,\mathrm{eV}$ on the held-out test set. The corresponding parity plots are shown in Supplementary Figure~\ref{fig:tsenn-bandgap-parity}, where the train, validation, and test predictions closely follow the $y=x$ reference line. The close agreement between validation and test metrics, together with the high $R^{2}$ across this chemically diverse dataset, indicates that the surrogate provides a reliable held-out estimate and is suitable for scoring generated candidates within the chemical regime represented by the training data.

\begin{figure}[H]
    \centering
    \includegraphics[width=1.0\linewidth]{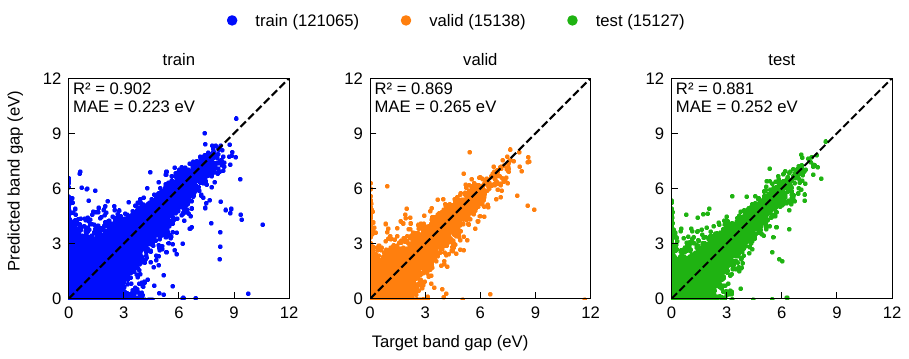}
    \caption[Predicted versus DFT-computed band gap]{
    \textbf{Predicted versus DFT-computed band gap.}
    Parity plots of the predicted band gap $E_g$ against the DFT reference, shown for the train, validation, and test splits (left to right). The dashed line indicates perfect agreement ($y=x$). On the held-out test set, the model achieves $R^{2}=0.881$ with $\mathrm{MAE}=0.252\,\mathrm{eV}$.
    }
    \label{fig:tsenn-bandgap-parity}
\end{figure}

\subsection{Spectroscopic limited maximum efficiency}
The output head is a sequence of scalar irreps: $201\times0e$, predicting the
frequency-dependent trace of the imaginary part of the dielectric tensor,
$\varepsilon_2(\omega)=\tfrac{1}{3}\operatorname{Tr}\boldsymbol{\varepsilon}_{2,ij}(\omega)$, on a
fixed frequency grid from $0$ to $\SI{20}{\electronvolt}$ with spacing
$\Delta E=\SI{0.1}{\electronvolt}$. The corresponding real part $\varepsilon_1(\omega)$ is recovered
from the predicted $\varepsilon_2(\omega)$ via the Kramers--Kronig (KK) relations, and the absorption
coefficient is obtained from the standard expression
\begin{equation}
\alpha(\omega) = \frac{\sqrt{2}\,\omega}{c}\left[\sqrt{\varepsilon_1^2(\omega)+\varepsilon_2^2(\omega)}-\varepsilon_1(\omega)\right]^{1/2}.
\end{equation}
Given $\alpha(E)$, the SLME $\eta$ is evaluated at a fixed film thickness
$L_{\mathrm{abs}}=\SI{0.3}{\micro\meter}$, device temperature $T_{\mathrm{cell}}=\SI{300}{\kelvin}$, and reference
solar spectrum (AM1.5G).

\paragraph{Dataset.}
The training data are drawn from OptiMate~\cite{grunertDeepLearningSpectra2024}, a large-scale DFT dataset of dielectric spectra comprising $21{,}064$ materials sampled from the Alexandria database~\cite{schmidtMachineLearningAssistedDeterminationGlobal2023}. The underlying ground-state calculations were performed with Quantum~ESPRESSO~\cite{giannozziQUANTUMESPRESSOModular2009}, and the frequency-dependent dielectric spectra were obtained with Yambo~\cite{sangalliManybodyPerturbationTheory2019}.

\paragraph{Train/validation/test splits.}
Following the protocol of the original TSENN study~\cite{hsuAccuratePredictionTensorial2026}, we partition the data into $16{,}848$ training, $2{,}106$ validation, and $2{,}110$ test materials --- an $80/10/10$ split stratified by crystal system, so that the symmetry distribution is preserved across the three splits.

\paragraph{Training.}
Learning rate $1\times 10^{-2}$, weight decay $5\times 10^{-2}$; remaining hyperparameters in Supplementary Table~\ref{tab:tsenn-hparams}.

\begin{table}[h]
\centering
\caption{Hyperparameters for the TSENN SLME surrogate.}
\label{tab:tsenn-hparams}
\setlength{\tabcolsep}{12pt}
\renewcommand{\arraystretch}{1.25}
\begin{tabular}{p{0.55\linewidth} p{0.15\linewidth}}
\toprule
Hyperparameter & Value \\
\midrule
Maximum spherical harmonic order $\ell_{\max}$ & 2 \\
Embedding dimension & 128 \\
Number of layers & 4 \\
Channel multiplicity & 64 \\
Batch size & 24 \\
\bottomrule
\end{tabular}
\end{table}

The parity plots in Supplementary Figure \ref{fig:tsenn-slme-parity} compare the predicted SLME values with those computed for the train, validation, and test splits. The surrogate attains $R^2 = 0.943$, $\mathrm{MAE} = 1.15\%$ on the training sets, $R^2 = 0.921$, $\mathrm{MAE} = 1.37\%$ on the validation sets, and $R^2 = 0.925$, $\mathrm{MAE} = 1.35\%$ on the test sets. The high $R^{2}$ on the held-out test set indicates that the predictor remains accurate on held-out materials, which is the most relevant benchmark for inverse design because generated candidates are not part of the training set.

\begin{figure}[H]
    \centering
    \includegraphics[width=1.0\linewidth]{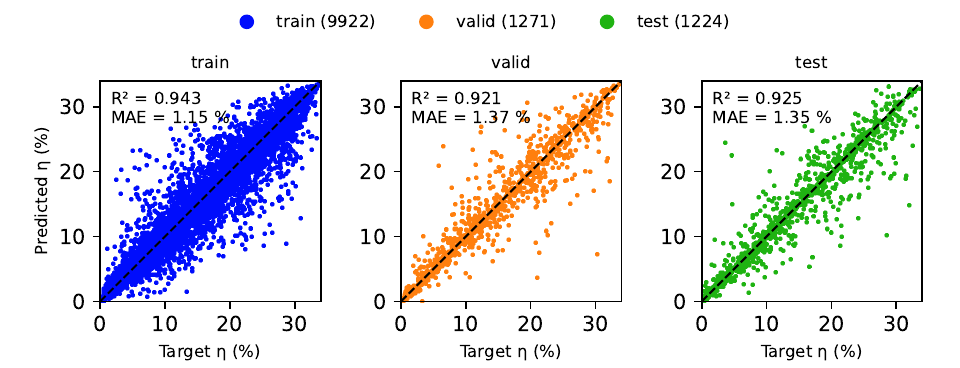}
    \caption[Predicted versus DFT-computed SLME]{
    \textbf{Predicted versus DFT-computed SLME.}
    Parity plots of the predicted SLME $\eta$ against the value computed from the DFT dielectric spectrum, shown for the train, validation, and test splits (left to right). The dashed line indicates perfect agreement ($y=x$). On the held-out test set, the model achieves $R^{2}=0.925$ with $\mathrm{MAE}=1.35\%$ (absolute efficiency percentage points).
    }
    \label{fig:tsenn-slme-parity}
\end{figure}

\subsection{Static-limit dielectric tensor}
The output head emits two irreps of $O(3)$: a scalar (\texttt{1x0e}) and a symmetric rank-2 traceless tensor (\texttt{1x2e}), giving $1+5=6$ independent coefficients. These coefficients are mapped back to the Cartesian symmetric dielectric tensor $\varepsilon_{ij}$ via the spherical-to-Cartesian transformation, recovering the six independent entries $\{\varepsilon_{xx},\varepsilon_{yy},\varepsilon_{zz},\varepsilon_{xy},\varepsilon_{xz},\varepsilon_{yz}\}$. This decomposition separates the isotropic trace from the anisotropic traceless response. This is important because the diagonal entries are typically much larger than the off-diagonal entries; separating the two scales prevents the loss from being dominated by the diagonal components~\cite{hsuAccuratePredictionTensorial2026}.

\paragraph{Dataset.}
The training data comprise $6{,}706$ materials drawn from the Materials Project database~\cite{jainCommentaryMaterialsProject2013}, with dielectric tensors computed by density-functional perturbation theory (DFPT) in VASP~\cite{petousisBenchmarkingDensityFunctional2016a}.

\paragraph{Train/validation/test splits.}
We partition the data into $5{,}363$ training, $671$ validation, and $672$ test materials, an $\approx 80/10/10$ split stratified by crystal system, so that the symmetry distribution is preserved across the three splits.

\paragraph{Training.}
Learning rate $1\times 10^{-2}$, weight decay $5\times 10^{-2}$; remaining hyperparameters in Table~\ref{tab:tsenn-tensors-hparams}.

\begin{table}[h]
\centering
\caption{Hyperparameters for the TSENN dielectric-tensor surrogate.}
\label{tab:tsenn-tensors-hparams}
\setlength{\tabcolsep}{12pt}
\renewcommand{\arraystretch}{1.25}
\begin{tabular}{p{0.55\linewidth} p{0.15\linewidth}}
\toprule
Hyperparameter & Value \\
\midrule
Maximum spherical harmonic order $\ell_{\max}$ & 2 \\
Embedding dimension & 64 \\
Number of layers & 2 \\
Channel multiplicity & 32 \\
Batch size & 16 \\
\bottomrule
\end{tabular}
\end{table}

Per-component test-set metrics are reported in Supplementary Table~\ref{tab:tsenn-tensor-percomp}, and the corresponding parity plots across all splits are shown in Supplementary Figure~\ref{fig:tsenn-dielectric-tensors-parity}. The off-diagonal components have substantially smaller absolute errors than the diagonal components, consistent with their smaller magnitudes in the DFT data. Despite this scale separation, the model maintains high $R^{2}$ values for both diagonal and off-diagonal components. These results indicate that the scalar and rank-2 irrep decomposition enables the surrogate to learn the full symmetric dielectric tensor without letting the large diagonal components dominate the prediction. The resulting model is therefore suitable as a tensorial scorer for inverse-design tasks, including those targeting materials with finite off-diagonal dielectric responses.

\begin{table}[h]
\centering
\caption{Per-component performance of the TSENN dielectric-tensor surrogate on the held-out test split. MAE is reported in units of the relative dielectric constant.}
\label{tab:tsenn-tensor-percomp}
\renewcommand{\arraystretch}{1.2}
\begin{tabular*}{0.75\linewidth}{@{\extracolsep{\fill}} lcc}
\toprule
Component & $R^{2}$ & MAE \\
\midrule
$\varepsilon_{xx}$ & 0.993 & 0.087 \\
$\varepsilon_{yy}$ & 0.992 & 0.083 \\
$\varepsilon_{zz}$ & 0.991 & 0.088 \\
\midrule
$\varepsilon_{xy}$ & 0.986 & 0.005 \\
$\varepsilon_{xz}$ & 0.976 & 0.006 \\
$\varepsilon_{yz}$ & 0.956 & 0.009 \\
\bottomrule
\end{tabular*}
\end{table}

\begin{figure}[H]
    \centering
    \includegraphics[width=1.0\linewidth]{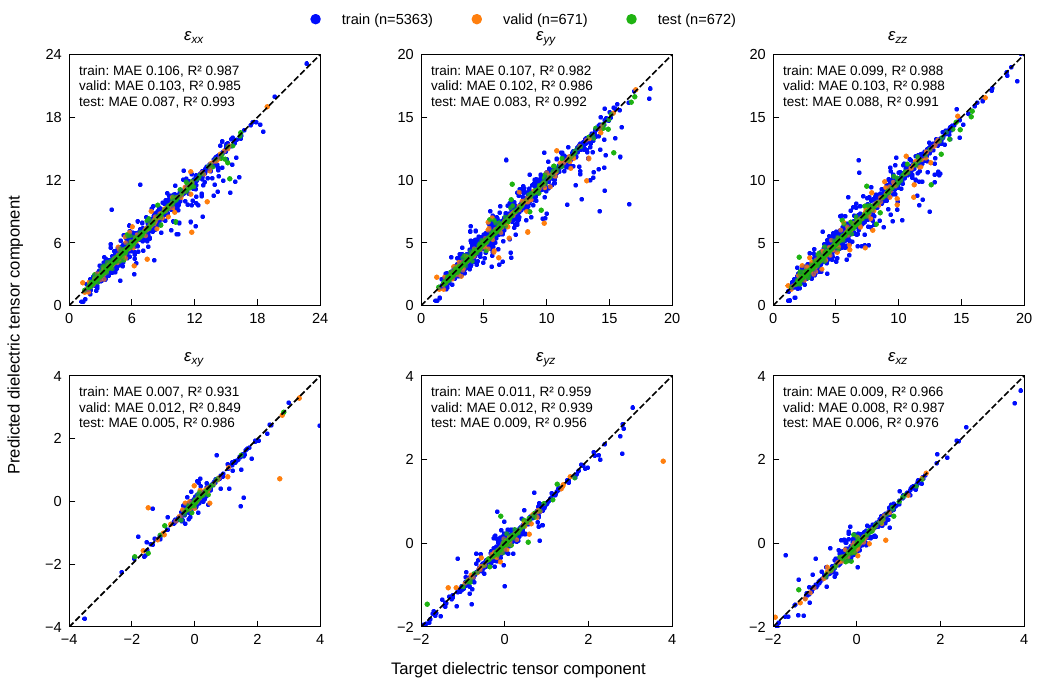}
    \caption[Predicted versus DFT-computed static-limit dielectric tensors across all three splits]{
    \textbf{Predicted versus DFT-computed static-limit dielectric tensors across all three splits.}
    Parity plots of each independent Cartesian component $\varepsilon_{ij}$ ($i,j\in\{x,y,z\}$) of the dielectric tensor, evaluated on the train, validation, and test splits. The dashed line indicates perfect agreement ($y=x$). The inset showing the MAE for per-component performance, and all the metrics are also reported in Supplementary Table~\ref{tab:tsenn-tensor-percomp}.
    }
    \label{fig:tsenn-dielectric-tensors-parity}
\end{figure}

\section{Reward shaping}

For both inverse-design tasks, the reward depends on a task-specific property and the predicted band gap. Selected quantities are mapped to $[0,1]$ using the clamped ascending linear ramp
\begin{equation}
\Phi(x;a,b)=\mathrm{clip}\left(\frac{x-a}{b-a},\,0,\,1\right),
\end{equation}
which is zero for $x\leq a$, increases linearly between $a$ and $b$, and saturates at one for $x\geq b$. The dielectric and SLME tasks combine their property and band-gap terms differently. The dielectric reward uses the smaller of the two terms, while the SLME reward uses a weighted sum that favors band gaps near the single-junction optimum.

\paragraph{Dielectric task.}

The tensor-based uniaxial dielectric score is
\begin{equation}
q_{\mathrm{uni}}=g_z(1-m)\in[0,1],
\end{equation}
where $g_z$ measures the contrast between the out-of-plane and mean in-plane dielectric responses, and $m$ measures the residual mismatch between $\varepsilon_{xx}$ and $\varepsilon_{yy}$. These quantities are defined in the main Methods.
It is convenient to name the
in-plane isotropy factor separately,
\begin{equation}
q_{\mathrm{iso}}\equiv 1-m\in[0,1],\qquad q_{\mathrm{uni}}=g_z\,q_{\mathrm{iso}},
\end{equation}
since the two factors suppress different failure modes, where $g_z$ rejects near-isotropic
tensors, while $q_{\mathrm{iso}}$ rejects biaxial ones. The uniaxial score is mapped to $[0,1]$ using
\begin{equation}
r_q=\Phi(q_{\mathrm{uni}};0,0.20),
\end{equation}
while the band-gap contribution is
\begin{equation}
r_{g,\mathrm{uni}}(E_g)=\Phi(E_g;0.3,0.8).
\end{equation}
The final dielectric reward is
\begin{equation}
r_{\mathrm{uni}}
=
\min\left[
r_q,\,
r_{g,\mathrm{uni}}(E_g)
\right]
=
\min\left[
\Phi(q_{\mathrm{uni}};0,0.20),\,
\Phi(E_g;0.3,0.8)
\right].
\end{equation}
The minimum operator makes the smaller term the limiting contribution. A large uniaxial score cannot compensate for a small predicted band gap, and a large band gap cannot compensate for a poor uniaxial response. The reward is therefore nonzero only when $q_{\mathrm{uni}}>0$ and $E_g>0.3$~eV. It reaches one when $q_{\mathrm{uni}}\geq0.20$ and $E_g\geq0.8$~eV.
Because the ramp saturates at $q_{\mathrm{uni}}=0.20$, the reward does not discriminate
among candidates with $q_{\mathrm{uni}}>0.20$. We therefore apply an additional in-plane
isotropy criterion, $q_{\mathrm{iso}}\geq0.95$, when selecting candidates for
first-principles validation.

This construction prevents near-metallic candidates from receiving a high reward solely because of their predicted dielectric response. It also keeps the search within the finite-gap regime represented by the dielectric surrogate and used for subsequent DFPT validation. The resulting reward surface is shown in Supplementary Figure~\ref{fig:reward}(a).

\paragraph{SLME property term.}

For the photovoltaic task, the property term is the SLME $\eta$. We evaluate $\eta$ under the AM1.5G illumination spectrum using a film thickness $L_{\mathrm{abs}}=\SI{0.3}{\micro\meter}$, a cell temperature $T_{\mathrm{cell}}=\SI{300}{\kelvin}$, and a radiative fraction $f_r=1$~\cite{yuIdentificationPotentialPhotovoltaic2012,hungUniversalEnsembleEmbeddingGraph2024}. The efficiency is defined as
\begin{equation}
\eta=\frac{P_{\max}}{P_{\mathrm{solar}}},
\end{equation}
where the incident solar power is
\begin{equation}
P_{\mathrm{solar}}
=
\int_0^{\infty}
I_{\mathrm{solar}}(E)\,\mathrm{d}E.
\end{equation}
Here, $I_{\mathrm{solar}}(E)$ is the AM1.5G energy irradiance. The maximum electrical output power is obtained from the diode $J\text{--}V$ curve,
\begin{equation}
P_{\max}
=\max_V\{J(V)V\},\qquad J(V) = J_{\mathrm{sc}} - J_0 \left( e^{qV/(k_B T_{\mathrm{cell}})}-1 \right),
\end{equation}
where $q$ is the elementary charge and $k_B$ is the Boltzmann constant.

The material absorptance is calculated from the predicted absorption coefficient as
\begin{equation}
A(E) = 1-e^{-2\alpha(E)L_{\mathrm{abs}}},
\end{equation}
where $E=\hbar\omega$. The short-circuit current density is
\begin{equation}
J_{\mathrm{sc}} = q\int_{E_g}^{\infty} A(E)\, \phi_{\mathrm{solar}}(E)\, \mathrm{d}E,
\end{equation}
where $E_g$ defines the absorption edge and $\phi_{\mathrm{solar}}(E)$ is the incident solar photon flux. When the photon energy $E$ is expressed in electronvolts, the photon flux is obtained from the energy irradiance as
\begin{equation}
\phi_{\mathrm{solar}}(E) = \frac{I_{\mathrm{solar}}(E)}{qE},
\end{equation}
where $qE$ converts the photon energy from electronvolts to joules.

The dark saturation current is modeled as radiative-recombination limited,
\begin{equation}
J_0 = \frac{J_0^{\mathrm{rad}}}{f_r} = \frac{q\pi}{f_r} \int_{E_g}^{\infty} A(E)\, \phi_{\mathrm{bb}}(E,T_{\mathrm{cell}}) \,\mathrm{d}E,
\end{equation}
where $\phi_{\mathrm{bb}}(E,T_{\mathrm{cell}})$ is the blackbody photon radiance per unit solid angle,
\begin{equation}
\phi_{\mathrm{bb}}(E,T_{\mathrm{cell}}) = \frac{2E^2}{h^3c^2} \frac{1}{ \exp\left(E/k_B T_{\mathrm{cell}}\right)-1}.
\end{equation}
Here, $h$ is Planck's constant and $c$ is the speed of light. The factor $\pi$ performs the hemispherical angular integration that converts the radiance into an emitted photon flux. Because $\eta$ depends on both the band gap and the full absorption spectrum $\alpha(E)$, it provides the task-specific photovoltaic property term.

\paragraph{SLME reward.}

Although $\eta$ already depends on the band gap, we include an additional band-gap term to guide the generated population toward the favorable region for single-junction solar cells. The final SLME reward is
\begin{equation}
r_{\mathrm{SLME}}=w_{\eta}r_{\eta}+w_g r_{g,\mathrm{SLME}},
\end{equation}
where the normalized efficiency term is
\begin{equation}
r_{\eta}
=
\Phi(\eta;0,0.35),
\end{equation}
and the band-gap localization term is
\begin{equation}
r_{g,\mathrm{SLME}}(E_g)
=
\mathrm{clip}\left(
1-\frac{\left|E_g-E^*\right|}{\Delta},
0,
1
\right).
\end{equation}
We use $w_{\eta}=0.8$, $w_g=0.2$, $E^*=1.3$~eV, and $\Delta=1.0$~eV. Because both terms lie in $[0,1]$ and the weights sum to one, the final reward also satisfies $r_{\mathrm{SLME}}\in[0,1]$.

The band-gap term forms a symmetric tent centered at $E^*=1.3$~eV. It reaches one at the center and decreases linearly to zero at $E_g=0.3$ and $2.3$~eV. Outside this interval, the band-gap contribution remains zero. This term does not act as a hard gate. A candidate outside the preferred interval can still receive reward through its predicted efficiency, but it does not receive the additional contribution from band-gap localization.

The larger weight $w_{\eta}=0.8$ keeps SLME as the dominant optimization objective. The smaller band-gap term discourages the search from drifting toward very small-gap or overly wide-gap regions. As shown in Supplementary Figure~\ref{fig:reward}(b), the reward increases with SLME and contains an additional ridge centered near $E_g=1.3$~eV.

For reference, the dashed curve in Supplementary Figure~\ref{fig:reward}(b) shows the Shockley--Queisser detailed-balance limit $\eta_{\mathrm{SQ}}(E_g)$ evaluated using the same AM1.5G spectrum. In our implementation, the curve reaches a maximum efficiency of $33.7\%$ at $E_g=1.34$~eV. This curve provides a theoretical reference and is not included in the reward.

\begin{figure}[H]
    \centering
    \includegraphics[width=1.0\linewidth]{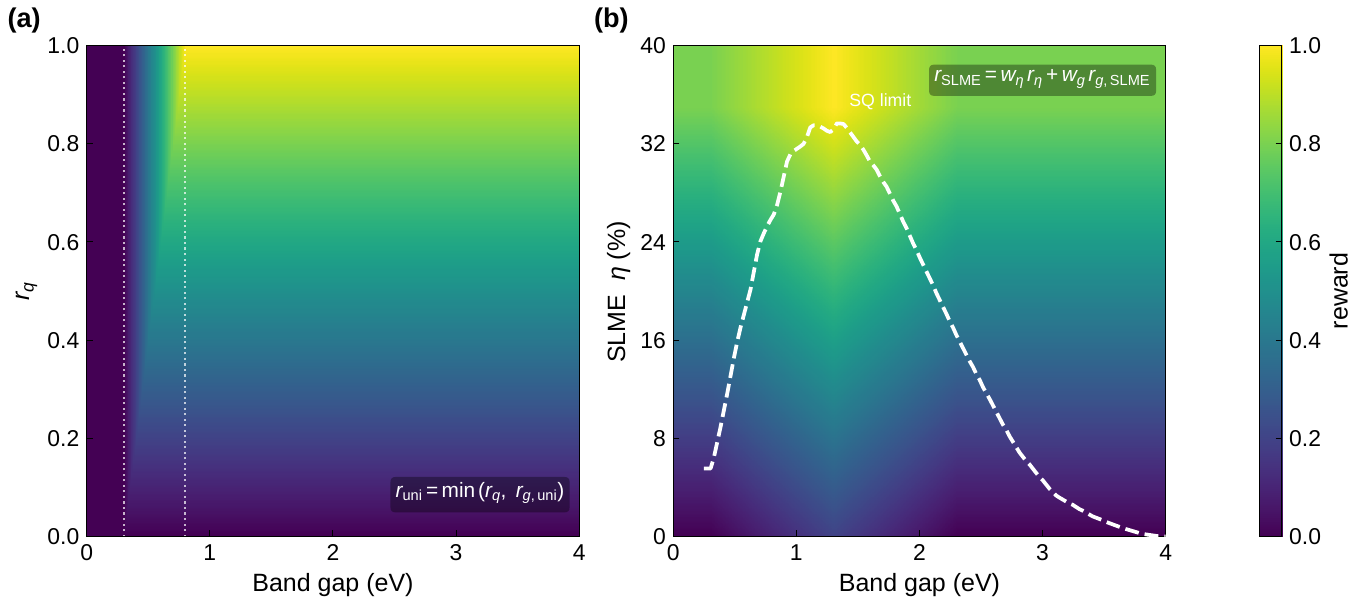}
    \caption{\textbf{Reward surfaces.} Reward as a function of the predicted band gap $E_g$ and the task-specific property for (a) the uniaxial dielectric objective and (b) the SLME objective. In (a), the tensor-based property score is $q_{\mathrm{uni}}=g_z(1-m)$, and the final reward is $r_{\mathrm{uni}}=\min[\Phi(q_{\mathrm{uni}};0,0.20),\Phi(E_g;0.3,0.8)]$. The vertical dotted line marks the onset of the band-gap contribution at $E_g=0.3$~eV. In (b), the final reward is $r_{\mathrm{SLME}}=0.8r_{\eta}+0.2r_{g,\mathrm{SLME}}$, where $r_{\eta}=\Phi(\eta;0,0.35)$ and $r_{g,\mathrm{SLME}}$ is a symmetric tent centered at $E^*=1.3$~eV. The dashed curve shows the Shockley--Queisser detailed-balance limit $\eta_{\mathrm{SQ}}(E_g)$ evaluated using the same AM1.5G spectrum.}
    \label{fig:reward}
\end{figure}

\section{First-principles validation of generated candidates across crystal systems}

\subsection{Uniaxial dielectric responses}

To test whether the dielectric responses identified by the surrogate remain valid at the first-principles level, we relaxed selected generated structures with DFT and calculated their dielectric tensors using DFPT. Candidates were selected from the generated population by requiring a uniaxial dielectric reward of $r_{\mathrm{uni}}\geq0.5$ and a predicted $E_{\mathrm{hull}}\leq0.10$~eV/atom. The reward threshold ensures that the selected candidates satisfy both the dielectric-anisotropy and finite-gap objectives defined in the reward function.

The high-throughput calculations were managed using \texttt{atomate2}~\texttt{0.1.5} and performed with VASP~\texttt{6.4.2}. Each candidate underwent two successive full cell-and-ion relaxations followed by a static calculation using the Materials Project-compatible PBE($+U$) protocol implemented through \texttt{MPRelaxSet} and \texttt{MPStaticSet}. These calculations used projector augmented-wave (PAW) PBE potentials, a plane-wave cutoff of $E_{\mathrm{cut}}=520$~eV, and the standard Materials Project Hubbard-$U$ corrections for transition-metal oxides and fluorides. The relaxed geometry was then evaluated using the DFPT dielectric workflow with \texttt{LEPSILON = .TRUE.}, \texttt{IBRION = 8}, the PBEsol exchange-correlation functional, and a plane-wave cutoff of $E_{\mathrm{cut}}=680$~eV. The resulting dielectric tensor is denoted by $\boldsymbol{\varepsilon}_0$.

\begin{figure}[H]
    \centering
    \includegraphics[width=1.0\linewidth]{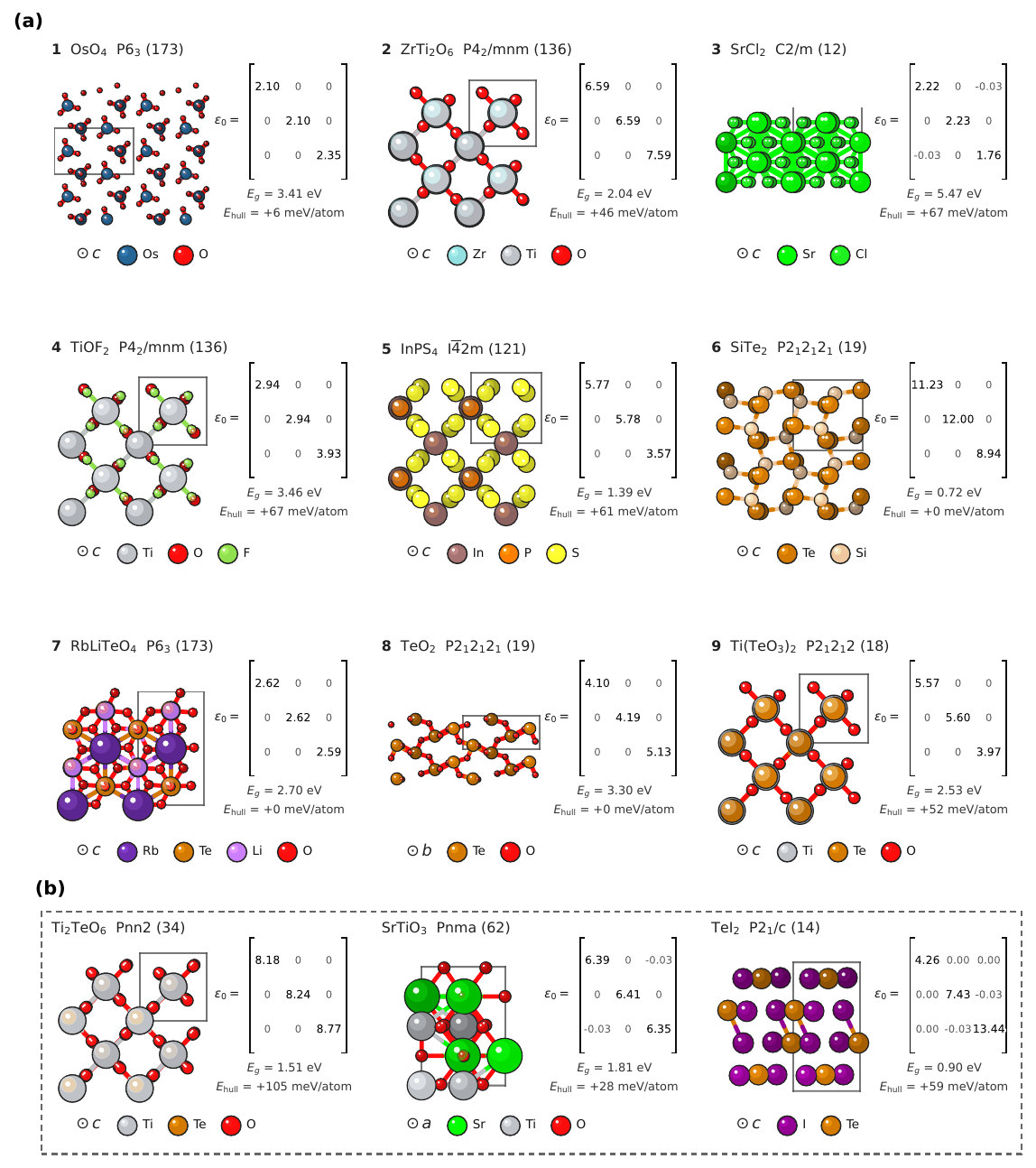}
    \caption{\textbf{First-principles validation of generated uniaxial dielectric candidates.} \textbf{(a)} Nine generated structures whose dielectric tensors $\boldsymbol{\varepsilon}_{0}$ retain the targeted relation $\varepsilon_{xx}\approx\varepsilon_{yy}$ with a distinct out-of-plane component. Each entry shows the DFT-relaxed structure, realized space group, calculated dielectric tensor, DFT band gap $E_g$, and $E_{\mathrm{hull}}$. The viewing direction is indicated below each structure, and $\odot$ denotes an axis pointing out of the page. The hexagonal and tetragonal structures exhibit symmetry-enforced in-plane equality, while the monoclinic and orthorhombic structures exhibit approximate equality that is not required by symmetry. \textbf{(b)} Comparison cases illustrating three rejection mechanisms. \ce{Ti2TeO6} retains an approximately uniaxial tensor but exceeds the stability cutoff, \ce{SrTiO3} has a nearly isotropic response and is suppressed by the out-of-plane contrast term, and \ce{TeI2} has a biaxial response that fails the in-plane isotropy requirement.}
    \label{fig:dfpt_verified_structures}
\end{figure}

Supplementary Figure~\ref{fig:dfpt_verified_structures}(a) shows nine generated candidates whose first-principles dielectric tensors retain the targeted relation
\begin{equation}
\varepsilon_{0,xx}\approx\varepsilon_{0,yy},
\qquad
\varepsilon_{0,zz}\neq
\frac{\varepsilon_{0,xx}+\varepsilon_{0,yy}}{2}.
\end{equation}
For simplicity, the subscript $0$ is omitted from the individual tensor components below. The validated candidates span the hexagonal, tetragonal, orthorhombic, and monoclinic crystal systems. Their DFT band gaps range from $0.72$ to $5.47$~eV, and their $E_{\mathrm{hull}}$ range from $0$ to $67$~meV/atom. 

For the hexagonal and tetragonal candidates, equality between the two in-plane components is imposed by crystal symmetry. These structures include \ce{OsO4} in $P6_3$, \ce{ZrTi2O6} and \ce{TiOF2} in $P4_2/mnm$, \ce{InPS4} in $I\bar{4}2m$, and \ce{RbLiTeO4} in $P6_3$. The tetragonal $P4_2/mnm$ phase of \ce{ZrTi2O6} shown here is a distinct polymorph from the trigonal $P\bar{3}1m$ phase highlighted in the main text. The occurrence of both polymorphs further illustrates that the targeted dielectric response is controlled by the crystal structure and symmetry, rather than by composition alone. Their DFPT tensors satisfy $\varepsilon_{xx}=\varepsilon_{yy}$ within the reported numerical precision, while $\varepsilon_{zz}$ remains distinct. Among these candidates, \ce{InPS4} exhibits the largest contrast, with $\varepsilon_{xx}=5.77$, $\varepsilon_{yy}=5.78$, and $\varepsilon_{zz}=3.57$.

The lower-symmetry candidates show that the same numerical tensor relation can also arise without symmetry protection. Monoclinic \ce{SrCl2} in $C2/m$ and the orthorhombic structures \ce{SiTe2} in $P2_12_12_1$, \ce{TeO2} in $P2_12_12_1$, and \ce{Ti(TeO3)2} in $P2_12_12$ exhibit approximately equal in-plane components and a distinct response along the selected out-of-plane direction. Their space groups do not require $\varepsilon_{xx}=\varepsilon_{yy}$, so the approximate equality arises from the particular relaxed structures. These responses may therefore be more sensitive to strain, defects, temperature, or further structural perturbations than the symmetry-enforced responses of the hexagonal and tetragonal candidates.

Supplementary Figure~\ref{fig:dfpt_verified_structures}(b) presents three comparison cases that illustrate distinct rejection mechanisms. Orthorhombic \ce{Ti2TeO6} in $Pnn2$ retains an approximately uniaxial tensor, with diagonal components $8.18$, $8.24$, and $8.77$, but its $E_{\mathrm{hull}}$ is $105$~meV/atom, slightly exceeding the $0.10$~eV/atom stability cutoff. Orthorhombic \ce{SrTiO3} in $Pnma$ has an almost isotropic tensor, with diagonal components $6.39$, $6.41$, and $6.35$, and is therefore suppressed by the out-of-plane contrast term. Monoclinic \ce{TeI2} in $P2_1/c$ has strongly unequal diagonal components, $4.26$, $7.43$, and $13.44$, and is rejected by the in-plane isotropy requirement. These examples show how stability, out-of-plane contrast, and in-plane equality contribute separately to candidate selection.

The first-principles calculations confirm that the surrogate-guided search identifies finite-gap structures whose dielectric tensors reproduce the desired relation between the in-plane and out-of-plane responses. They also distinguish symmetry-enforced uniaxiality from approximate uniaxiality that occurs only for a specific relaxed configuration. The comparison cases further demonstrate that a favorable tensor contrast alone is not sufficient. A candidate must also satisfy the stability, in-plane isotropy, and finite-gap requirements of the complete selection procedure.

\subsection{SLME}
To verify that the generative loop discovers genuinely diverse semiconductors rather than repeatedly exploiting a single low-symmetry motif, we relaxed a representative set of generated structures with DFT and computed their optical responses. Because the TSENN surrogate was trained on the OptiMate dataset, the validation calculations were performed using the same high-throughput PBE independent-particle workflow used to construct the training data~\cite{grunertDeepLearningSpectra2024}. Briefly, ground-state calculations were performed using Quantum ESPRESSO with SG15 norm-conserving pseudopotentials, and the frequency-dependent dielectric functions were calculated using Yambo over the 0--20~eV range with an energy spacing of 0.01~eV and a broadening of 0.10~eV. The plane-wave cutoff, k-point density, and number of unoccupied bands were converged following the OptiMate protocol.

Candidate materials were selected according to the surrogate predictions, requiring a TSENN-predicted band gap of 0.8--1.9~eV and an SLME exceeding 25\%. Of the 576 structures satisfying these criteria, 241 (42\%) relaxed to gapped semiconductors with well-defined absorption spectra, whereas the remainder became metallic or nearly metallic after DFT relaxation. Because SLME depends strongly on the absorption onset, uncertainty in the predicted band gap is the dominant source of error, rather than inaccuracies in the predicted optical spectra.  The eight materials shown in Supplementary Figure~\ref{fig:slme_dft_structure_spectra} are drawn from this DFT-confirmed semiconducting pool, and their first-principles $\varepsilon_2(\omega)$ spectra are compared with the TSENN surrogate that guided the search.

The selected materials span six of the seven crystal systems, including cubic \ce{K2Te} ($Fm\bar{3}m$) and \ce{NaHfCu3S4} ($P\bar{4}3m$), trigonal \ce{CsGeBr3} ($R3m$), tetragonal \ce{SnBr2} ($P4_2/mnm$), orthorhombic \ce{BiSI} ($Pnma$) and \ce{PbS} ($Pmn2_1$), monoclinic \ce{InGaS2} ($P2_1/c$), and triclinic \ce{AsS4} ($P\bar{1}$). This diversity shows that the search discovers promising photovoltaic absorbers across a broad range of crystal symmetries rather than collapsing onto the triclinic $P1$/$P\bar{1}$ structures that dominate the raw efficiency ranking.

Across this representative set, the candidates exhibit PBE indirect band gaps of 1.15--2.12~eV and SLME values of 18--32\%, placing most within the desired range for single-junction photovoltaic absorbers. The TSENN surrogate closely reproduces the DFT-computed $\varepsilon_2(\omega)$ spectra, including the absorption onset, dominant absorption peak, and higher-energy spectral features across all represented crystal systems. Spectral agreement was quantified using the spectral similarity coefficient (SC),
\begin{equation}
    \operatorname{SC}[\varepsilon_2^{\mathrm{TSENN}}(\omega) ; \varepsilon_2^{\mathrm{DFT}}(\omega)]=1-\frac{\int|\varepsilon_2^{\mathrm{DFT}}(\omega)-\varepsilon_2^{\mathrm{TSENN}}(\omega)| d \omega}{\int|\varepsilon_2^{\mathrm{DFT}}(\omega)| d \omega}
\end{equation}
following Ref.~\cite{grunertDeepLearningSpectra2024}, where $\mathrm{SC}=1$ indicates identical spectra. The layered orthorhombic \ce{PbS} shows the closest agreement, with an SC of 0.97.

The reported band gaps and optical spectra were calculated at the PBE independent-particle level and should therefore be interpreted as internally consistent screening quantities rather than quantitative quasiparticle or optical excitation energies. PBE generally underestimates band gaps, while the independent-particle approximation neglects electron--hole interactions and microscopic local-field effects. Nevertheless, the close agreement between the surrogate and the corresponding DFT calculations confirms that the reward model reproduces the level of optical physics represented in its training data across a structurally diverse set of semiconductors, supporting its use in the generative search.

\begin{figure}[H]
    \centering
    \includegraphics[width=1.0\linewidth]{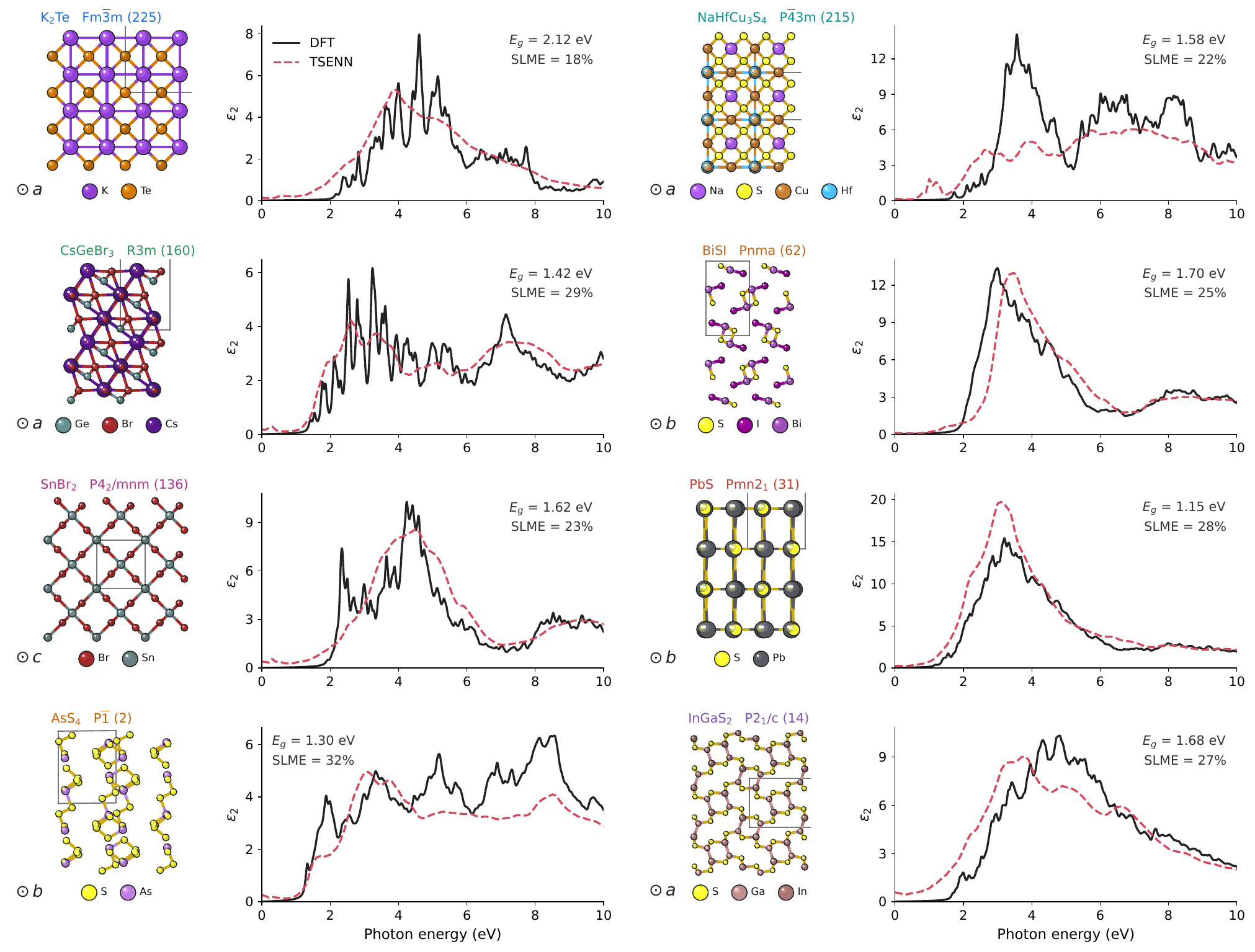}
    \caption{\textbf{First-principles optical response of eight generated candidates.} For each material the left panel shows the relaxed crystal structure (ball-and-stick, unit cell outlined) viewed down the crystallographic axis marked in the lower-left corner. The $\odot$ denotes a viewing axis pointing out of the page. Each structure panel associated to a spectra panel comparing the imaginary dielectric function $\varepsilon_2(\omega)$ from first-principles DFT (QE, black solid) with the TSENN prediction (red dashed) over $0$--$10$~eV. Annotations give the PBE indirect band gap $E_g$ and the SLME evaluated following the parameters throughout the work describe in main text.}
    \label{fig:slme_dft_structure_spectra}
\end{figure}

\bibliography{sn-bibliography}